\PassOptionsToPackage{pdfpagelabels=false}{hyperref}

\documentclass[fleqn,usenatbib]{mnras}

\usepackage[T1]{fontenc}
\usepackage{ae,aecompl}

\usepackage{float}
\usepackage{hyperref}
\usepackage{apjfonts}
\usepackage{amssymb}
\usepackage{amsmath}
\usepackage{graphicx}	
\usepackage[normalem]{ulem}
\usepackage{multirow}
\usepackage{mathtools}
\usepackage{dutchcal}
\usepackage[all]{hypcap}
\hypersetup{colorlinks,citecolor=blue}

\usepackage{etoolbox}
\makeatletter
\patchcmd\@combinedblfloats{\box\@outputbox}{\unvbox\@outputbox}{}{%
  \errmessage{\noexpand\@combinedblfloats could not be patched}%
}%
\makeatother

\DeclarePairedDelimiter\abs{\lvert}{\rvert}%

\newcommand{\mystar}{\ast}        
\newcommand\run[1]{\texttt{#1}}   

\newcommand{\mvir}{M_\mathrm{vir}}
\newcommand{\rvir}{R_\mathrm{vir}}

\newcommand{\rtwoh}{R_{\rm 200}}

\newcommand{\msun}{\textnormal{M}_\odot}
\newcommand{\Msun}{\textnormal{M}_\odot}
\newcommand{\mstar}{M^\mystar_\mathrm{galaxy}}
\newcommand{\mgas}{M^\mathrm{gas}_\mathrm{galaxy}}

\newcommand{\mpc}{\mathrm{Mpc}}

\newcommand{\kpc}{\mathrm{kpc}}
\newcommand{\pc}{\mathrm{pc}}

\newcommand{\kms}{{\rm km} \ {\rm s}^{-1}}
\newcommand{\lcdm}{$\Lambda$CDM}

\newcommand{\mcoldgas}{M^\mathrm{gas}_\mathrm{cold}}

\newcommand{\rspear}{r_\mathrm{sp}}
\newcommand{\fgdisk}{f_\mathrm{gas}^\mathrm{disk}}
\newcommand{\fsdisk}{f^\mystar_\mathrm{disk}}
\newcommand{\fsdbirth}{f_{\mystar,\,{\rm birth}}^\mathrm{disk}}

\newcommand{\Rstar}{R^{\mystar}_{90}}
\newcommand{\rstar}{r^{\mystar}_{90,\,3D}}
\newcommand{\Zstar}{Z^{\mystar}_{90}}
\newcommand{\Rshalf}{R^\mystar_{50}}
\newcommand{\rshalf}{r^{\mystar}_{50,\,3D}}
\newcommand{\Zshalf}{Z^\mystar_{50}}
\newcommand{\Rgas}{R_\mathrm{gas}}
\newcommand{\Zgas}{Z_\mathrm{gas}}

\newcommand{\etal}{et al.}
\newcommand{\degree}{^\circ}
\newcommand{\gyr}{\mathrm{Gyr}}
\newcommand{\myr}{\mathrm{Myr}}
\newcommand{\jd}{\mathcal{j}_{\rm d}}
\newcommand{\md}{\mathcal{m}_{\rm d}}

\defcitealias{MoMaoWhite}{MMW98}

\newcommand\altaffilmark[1]{$^{#1}$}
\newcommand\altaffiltext[1]{$^{#1}$}

\title[Morphological diversity in the FIRE-2 MW-mass galaxies]{The origin of the diverse morphologies and kinematics of Milky Way-mass galaxies in the FIRE-2 simulations
\vspace{-0.5cm}}

\vspace{-0.2cm}
\author[S. Garrison-Kimmel \etal]{
\parbox[t]{\textwidth}{ 
Shea Garrison-Kimmel\thanks{$\!$sheagk@caltech.edu}\thanks{Einstein Fellow}\altaffilmark{1},
Philip F.~Hopkins\altaffilmark{1},
Andrew Wetzel\thanks{Caltech-Carnegie Fellow}\altaffilmark{1,2,3},
Kareem El-Badry\altaffilmark{4},
Robyn E. Sanderson\altaffilmark{1,5},
James S.~Bullock\altaffilmark{6},
Xiangcheng Ma\altaffilmark{1},
Freeke van de Voort\altaffilmark{7,8},
Zachary Hafen\altaffilmark{9},
Claude-Andr{\'e} Faucher-Gigu{\`e}re\altaffilmark{9},
Christopher C. Hayward\altaffilmark{10},
Eliot Quataert\altaffilmark{4},
Du{\v s}an Kere{\v s}\altaffilmark{11},
Michael Boylan-Kolchin\altaffilmark{12}
}
\vspace*{6pt} \\
\altaffiltext{1}{TAPIR, Mailcode 350-17, California Institute of Technology, Pasadena, CA 91125, USA} \\
\altaffiltext{2}{The Observatories of the Carnegie Institution for Science, Pasadena, CA 91101, USA} \\
\altaffiltext{3}{Department of Physics, University of California, Davis, CA 95616, USA} \\
\altaffiltext{4}{Department of Astronomy and Theoretical Astrophysics Center, University of California Berkeley, Berkeley, CA 94720} \\ 
\altaffiltext{5}{Columbia University Department of Astronomy, 550 W 120th St, Mail Code 5246, New York, NY, 10027} \\
\altaffiltext{6}{Center for Cosmology, Department of Physics and Astronomy, University of California, Irvine, CA 92697, USA} \\
\altaffiltext{7}{Heidelberg Institute for Theoretical Studies, Schloss-Wolfsbrunnenweg 35, 69118, Heidelberg, Germany} \\
\altaffiltext{8}{Astronomy Department, Yale University, PO Box 208101, New Haven, CT 06520-8101, USA} \\
\altaffiltext{9}{Department of Physics and Astronomy and CIERA, Northwestern University, 2145 Sheridan Road, Evanston, IL 60208, USA} \\ 
\altaffiltext{10}{Center for Computational Astrophysics, Flatiron Institute, 162 Fifth Avenue, New York, NY 10010, USA} \\
\altaffiltext{11}{Department of Physics, Center for Astrophysics and Space Science, University of California at San Diego, 9500 Gilman Drive, La Jolla, CA 92093} \\
\altaffiltext{12}{Department of Astronomy, The University of Texas at Austin, 2515 Speedway, Stop C1400, Austin, TX 78712, USA} \\
\vspace{-0.5cm}
}

\date{Accepted XXX. Received YYY; in original form ZZZ}

\pubyear{2017}

\begin{document}
\label{firstpage}
\pagerange{\pageref{firstpage}--\pageref{lastpage}}

\maketitle

\begin{abstract}
We use hydrodynamic cosmological zoom-in simulations from the FIRE project 
to explore the morphologies and kinematics of fifteen Milky Way (MW)-mass galaxies.
Our sample ranges from compact, bulge-dominated systems with 90\% of their stellar 
mass within $2.5~\kpc$ to well-ordered disks that reach $\gtrsim15~\kpc$.  The gas 
in our galaxies always forms a thin, rotation-supported disk at $z=0$, with sizes 
primarily determined by the gas mass.  For stars, we quantify kinematics and 
morphology both via the fraction of stars on disk-like orbits and with the radial 
extent of the stellar disk.  In this mass range, stellar morphology and kinematics 
are poorly correlated with the properties of the halo available from dark 
matter-only simulations (halo merger history, spin, or formation time).  They 
more strongly correlate with the gaseous histories of the galaxies:  those that 
maintain a high gas mass in the disk after $z\sim1$ develop well-ordered stellar disks.  
The best predictor of morphology we identify is the spin of the gas in the halo at 
the time the galaxy formed $1/2$ of its stars (i.e. the gas that builds the galaxy).  
High-$z$ mergers, before a hot halo emerges, produce some of the most massive bulges 
in the sample (from compact disks in gas-rich mergers), while later-forming bulges 
typically originate from internal processes, as satellites are stripped of gas before 
the galaxies merge.  Moreover, most stars in $z=0$ MW-mass galaxies (even $z=0$ bulge 
stars) form in a disk:  $\gtrsim60-90\%$ of stars begin their lives rotationally 
supported.
\end{abstract}




\section{Introduction}
\label{sec:intro}

Galactic morphologies vary widely. Broadly speaking, galaxies range
from elliptical, dispersion-supported systems to disk-dominated 
structures where the majority of stars are on well-ordered circular 
orbits \citep[e.g.][]{Hubble1926,Huertas-Company2011}.  The former 
dominate at both the high-mass end \citep[e.g.][]{Bamford2009} and 
at the low-mass end \citep[e.g.][]{Wheeler2017}, with disky galaxies 
emerging primarily at intermediate stellar masses of $\sim10^{9}-10^{11}\msun$
\citep[e.g.][]{Simons2015}.  The preponderance of ellipticals at the 
high-mass end is typically associated with these galaxies growing primarily 
through dry mergers \citep{vanDokkum2005,vanDokkum2010,Rodriguez-Puebla2017}, 
which scramble stellar orbits and promote bulge formation 
\citep[e.g.][]{White78,Hopkins2009b,Stewart2009,Hopkins2010}.  At the 
low-mass end, stars are both born out of gas with a high degree of 
pressure support (rather than rotational support), and they are then 
dynamically heated by the repeated cycles of gas blowouts that continue 
to $z\sim0$ in $\lesssim10^{11}\msun$ halos \citep[][]{Kaufmann2007,
Pontzen2012,Governato2012,diCintio2014a,DiCintio2014b,Onorbe2015,Chan2015,
Wheeler2017,ElBadry2016,KEB2017,AnglesAcazar2017,Sparre2017,CAFG2017}.

At intermediate masses, however, the exact properties of a galaxy and/or 
halo that drive the morphology of that system remain relatively poorly 
understood.  \citet*[][hereafter MMW98]{MoMaoWhite} reproduced both the 
$z=0$ population of disk galaxies and the properties of $z\sim2.5$ damped 
Ly$\alpha$ systems in semi-analytic models by assuming (1) galaxy sizes 
are determined by their angular momentum, (2) the baryons in a galaxy 
acquire their angular momentum from the host dark matter (DM) halo, (3) DM 
halos respond adiabatically to the growth of galaxies, and (4) baryons 
initially have the same density profile as DM \citep[also see ][]{Fall1980,
Fall1983,Romanowsky2012,Fall2013}.  This model therefore predicts that 
the size of a galactic disk (relative to the radius of the halo) depends 
primarily on the spin of the host DM halo, such that elliptical galaxies 
reside in low angular momentum halos.


Though the \citetalias{MoMaoWhite} paradigm broadly reproduces the 
galactic population, it has not been possible to 
directly test it against hydrodynamic simulations that include 
star formation and feedback, the latter of which appears to be 
particularly important for regulating the angular momentum (and 
therefore shapes) of galaxies.  Such simulations typically fall 
into two categories: large-volumes simulations such as Illustris 
\citep{Illustris1,Illustris2}, Illustris-TNG \citep{IllustrisTNG2,IllustrisTNG1},
and EAGLE \citep{EAGLE}; and ``zoom-in'' simulations 
\citep{Katz1993,Onorbe2014} that focus on individual systems.  
While the former contain huge populations of galaxies in a given 
mass bin ($\gg10^3$), each galaxy typically contains $\lesssim10^3$ 
resolution elements, with spatial resolutions $\gtrsim1~\kpc$, such 
that it is impossible to fully resolve the vertical scale lengths 
of MW-like disks.  However, recent work with this style of simulations 
have managed to broadly reproduce the observed Hubble sequence of galaxy types 
\citep[e.g.][]{Pedrosa2014,Pedrosa2015,Genel2015,Teklu2015,Zavala2016,Genel2017}.  
In particular, \citet{Rodriguez-Gomez2017} found that the morphologies of 
massive systems ($\mstar\geq10^{11}\msun$) in the Illustris simulation are 
determined by their merger histories, while the morphologies of low mass 
galaxies ($\mstar\leq10^{10}\msun$) correlate with their host halo spin.  
However, they found that neither spin nor merger history could individually 
explain morphologies at the intermediate mass scale occupied by the MW.



Conversely, zoom-in simulations excel at resolving the structure of the 
galaxy (or galaxies) that they target, but each additional galaxy incurs 
a significant CPU cost, such that many suites of zoom-in simulations only 
include a few galaxies at a given mass simulated with a given physical 
model.  There are thus only a few suites of hydrodynamic zoom-in runs 
(e.g. GIMIC, \citealp{Crain2009}; MAGICC, \citealp{Stinson2012}; NIHAO,
\citealp{Wang2015}; Auriga, \citealp{Grand2017}) that have the sample 
size to test and explore even basic correlations between morphology 
and halo properties (such as the \citetalias{MoMaoWhite} model).  However, 
some trends have emerged across a number of analyses of various zoom-in 
simulations, which have generally become successful in recent years at 
producing realistic disk galaxies \citep{Governato2007,Governato2009,Scannapieco2009,Guedes2011,Aumer2013,
Marinacci2014,Fiacconi2015,Murante2015,Colin2016}.  A wide variety of 
authors using different simulation codes agree that stellar feedback is 
crucial for regulating star formation in low angular momentum material, 
which otherwise quickly collapses to form overly-massive bulge 
components \citep{Okamoto2005,Scannapieco2008,Agertz2011,Roskar2014,
Agertz2016,Brooks2016}.

Some of these authors have examined the conditions that lead to disk 
formation.  For example, \citet{SpringelHernquist2005} and \citet{Robertson2006} 
found that mergers of gas-rich galaxies can result in an 
extended star-forming disk, rather than a bulge-dominated system 
\citep[also see][]{Robertson2008}.  Similarly, \citet{Governato2007} 
found that a substantial disk formed following a gas-rich major merger 
in a cosmological simulation.  \citet{Governato2009} also examined the 
distribution of \emph{light} at $z=0$ in a galaxy that experienced a 
major merger at $z=0.8$, and found that this violent merger primarily 
grows the disk, rather the bulge.  Combined with the passive evolution 
of the older stars in the bulge, this fresh star formation results in a 
bright, blue stellar disk.  Together, these results suggest that gas-rich 
major mergers can lead to extended stellar disks \citep{Hopkins2009a}, 
particularly if they occur at late times when the potential is deep enough 
to prevent the burst-quench cycles that occur at higher redshift 
\citep{Muratov2015,Sparre2017,CAFG2017}, which heat stellar orbits and 
generally inhibit disk formation.  

Other works have used suites (of varying sizes) of zoom-in simulations 
to attempt to uncover the underlying drivers of stellar morphology.  
\citet{Scannapieco2009}, for example, argued that the fraction of mass 
in the disk does not depend on the spin parameter of the halo, but 
instead that the individual formation history of each galaxy is crucial 
to predicting its $z=0$ morphology.  They also showed that spheroidal 
(bulge) components typically form earlier, while disks tend to form 
at later times from the inside-out \citep[also see][]{Aumer2013,Sokolowska2017},
in general agreement with observations tracing the evolution of
the kinematics of gas in galaxies \citep{Simons2017}.  Using a set of 
100 MW-mass halos in high-resolution regions embedded within the
Millennium \citep{millennium} simulation volume, \citet{Sales2012} 
similarly found that galaxy morphology was not correlated with the 
spin of the halo.  They then further showed that it also does not monotonically 
depend on either the halo formation time \citep[which scales with the 
concentration of a halo; e.g.][]{Ludlow2014} or the merger history: even 
halos that grow significantly through major mergers can host either a 
disk-dominated or a bulge-dominated system at $z=0$.  Instead, they argued 
that the star formation history is key:  disks tend to form gradually and 
at late times, while spheroidal components assemble in episodic bursts of 
star formation that occur following the accretion of gas that is misaligned 
from the existing galaxy.  More recently, \citet{Grand2017} used 30 
galaxies from the Auriga Project to argue (1) that disk size \emph{does} 
correlate with halo spin (though the kinematic disk fraction, which we 
define below, does not) and (2) that well-aligned mergers of gas-rich
satellites promote disk growth.  


Collectively, the results from large-volume and zoom-in simulations suggest 
that a picture where stellar morphology is regulated by angular momentum 
is not necessarily wrong, but that it is likely incomplete.  However, the 
majority of these studies have focused on simulations that adopt a stiff 
equation of state for the interstellar medium, which could plausibly introduce 
artifacts into, e.g., the behavior of the gas during galactic mergers, 
motivating a study with an more physical description of the interstellar 
medium.  Here, we use a sample of fifteen MW-mass galaxies, seven of which are 
isolated and eight of which are in Local Group-like pairs, from high resolution 
zoom-in simulations, run with physically-motivated and identical models and 
parameters for star formation and feedback, to explore correlations and drivers 
of (primarily) stellar morphology.  We first test the \citetalias{MoMaoWhite} 
predictions against the sizes of our galaxies, then search for physically meaningful 
correlations between stellar morphology at $z=0$ and various properties of the 
host halo, including their evolutionary histories.  We then explore the evolution 
of the stellar morphologies and the fraction of stars born in a disk at any 
given time to better understand the impact of dynamical interactions and the 
instantaneous state of the star-forming gas at any given time.  Finally, we examine 
the morphology of the gas at $z=0$ to understand the morphologies of stars being 
born today.

Throughout this work, we assume flat \lcdm\ cosmologies, with 
$h = 0.68$~--~$0.71$, $\Omega_\mathrm{m} = 0.266$~--~$0.31$, 
$\Omega_\mathrm{b} = 0.0455$~--~$0.048$, and 
$\Omega_\Lambda = 1 - \Omega_\mathrm{m}$ 
\citep[e.g.][]{Larson2011,Planck15}.\footnote{The differences in 
average halo properties due to variances in the cosmological parameters 
are smaller than the typical halo-to-halo variance within a given 
cosmology, and, moreover, any systematic variations would be automatically 
included in the physical parameters we explore here.}  We adopt the 
\citet{Bryan1998} definition of $\mvir$ and $\rvir$ throughout, except 
when computing the \citetalias{MoMaoWhite} predictions, which depend 
on the properties of the halo within $\rtwoh$, the radius at 
which the density is $200$ times the critical density.  For all stellar 
images and properties presented herein, we use a coordinate system where 
the $z$-axis is aligned with the shortest principal axis of the moment of 
inertia tensor of all star particles within $20~\kpc$.  For the gas, we
align our coordinate system with the shortest principal axis of the gas 
within $10~\kpc$; we select a smaller radius for the gas because the gas 
moment of inertia tensor at $20~\kpc$ is occasionally dominated by gas 
outside the galaxy.  We sometimes refer to halo properties in the corresponding 
dark matter-only simulation; such properties will be indicated as ``DMO.''

We explicitly opt not to make comparisons with observations in this work
because our goal is not to demonstrate the ``reasonableness'' of our 
galactic disks, but rather to understand why and how they came to have
their $z=0$ morphologies.  However, we note that the FIRE/FIRE-2 physics
are broadly successful at reproducing observed galactic properties over
a range of galaxy masses, including the stellar mass \emph{vs} halo mass
relation \citep{FIRE,FIRE2}, the normalization and scatter of the star 
formation rate \emph{vs} stellar mass relationship \citep{Sparre2017},
the Kennicutt-Schmidt law \citep{Orr2017}, the mass-metallicity 
relationship \citep{MaMassMetallicity}, and even the vertical and radial 
structure (including stellar ages and metallicities) of the MW disk
\citep{Ma2016b}.  Sanderson et al. (in prep) also show that the masses
of the stellar halos around the FIRE-2 MW-mass galaxies are in relative
agreement with those measured by \citet{Merritt2016}.  Moreover, proper
comparisons to observations requires a careful conversion from the stellar 
mass to observed light to make a fair comparison with observables, 
including the effects of dust attenuation and stellar evolution 
\citep[e.g. radial variations in the mass-to-light ratio;][]{Wuyts2010}.
\citet{Scannapieco2010}, for example, used mock observations to show that 
photometric bulge/disk decompositions typically overestimate the true disk
fractions by at least a factor of two.  A detailed comparison of observer-space 
disk indicators will be the focus of subsequent work(s).

This paper is organized as follows.  In \S~\ref{sec:sims}, we describe 
the simulations and briefly review the star formation and feedback models.  
\S~\ref{sec:quant} presents our measures of morphology, $\fsdisk$ and $\Rstar$, 
and compares them to other (primarily theoretical) quantifiers.  
\S~\ref{sec:drivers} compares the actual morphologies to those predicted 
by the \citetalias{MoMaoWhite} model, then presents correlations between 
$z=0$ morphologies and various properties of the galaxy and their host halos, 
while \S~\ref{sec:evolution} explores the evolution of the stellar 
morphologies and the birth properties of stars.  \S~\ref{sec:gas_morph} 
presents the morphologies of the gas disks in our sample.  We summarize 
our results and conclusions in \S~\ref{sec:conclusions}.

\section{Simulations}
\label{sec:sims}

We analyze hydrodynamic, cosmological zoom-in \citep{Katz1993,Onorbe2014} 
simulations from the Feedback in Realistic Environments 
(FIRE)\footnote{\url{http://fire.northwestern.edu}} project, specifically 
with the improved ``FIRE-2'' version of the code from \citet{FIRE2}.  In order 
to maximize our sample size, we include simulations with varying resolutions, 
which we discuss below, but the numerical methods and primary physical models 
are identical across all of the simulations.  All of the simulations were run 
using \texttt{GIZMO} 
\citep{GIZMO},\footnote{\url{http://www.tapir.caltech.edu/~phopkins/Site/GIZMO.html}} 
a multi-method gravity plus hydrodynamics code, in meshless finite-mass 
(``MFM'') mode. This is a mesh-free Lagrangian finite-volume Godunov 
method which automatically provides adaptive spatial resolution while 
maintaining conservation of mass, energy, and momentum (for extensive 
tests, see \citealt{GIZMO}). Gravity is solved with an improved 
version of the Tree-PM solver from GADGET-3 
\citep{Springel2005}, with fully-adaptive (and fully-conservative) 
gravitational force softenings for gas (so hydrodynamic and force 
softenings are always self-consistently matched), following 
\citet{Price2007}. 

The FIRE physics and source code are {\em exactly} identical to those in
previous FIRE-2 simulations; these are described in detail in the papers 
above but we briefly review them here. Radiative heating and cooling is 
treated (from $10-10^{10}\,$K), including free-free, 
photo-ionization/recombination, Compton, photoelectric \& dust collisional, 
cosmic ray, molecular, and metal-line \& fine-structure processes (following 
each of 11 tracked species independently), and accounting for photo-heating 
both by a UV background \citep{FaucherGiguere2009} and an approximate model
for local sources, and self-shielding. Star formation occurs only in gas 
identified as self-gravitating according to the \citet{Hopkins2013sf_criteria} 
criterion, which is also molecular and self-shielding (following 
\citealt{Krumholz2011}), Jeans unstable, and exceeds a minimum density threshold 
$n_{\rm min}=1000\,{\rm cm^{-3}}$. Once a star particle forms, the simulations 
explicitly follow several different stellar feedback mechanisms, including 
(1) local and long-range momentum flux from radiation pressure (in the initial 
UV/optical single-scattering, and re-radiated light in the IR), (2) energy, 
momentum, mass and metal injection from SNe (Types Ia and II) and stellar mass loss 
(both OB and AGB), and (3) photo-ionization and photo-electric heating. Every star 
particle is treated as a single stellar population with known mass, age, and 
metallicity, and then all feedback event rates, luminosities and energies, mass-loss 
rates, and all other quantities are tabulated directly from stellar evolution models 
({\sc starburst99}; \citealt{Leitherer1999}), assuming a \citet{Kroupa2001} IMF.  
We emphasize that the FIRE physics were not tuned to reproduce galaxy sizes or
morphologies.  One of the pairs, \run{Romulus} \run{\&} \run{Remus}, was simulated 
with subgrid turbulent metal diffusion \citep{Hopkinsmetaldiff,Escala2017}; however, 
\citet{Su2016} showed metal diffusion has a small impact on the morphology of a 
MW-mass galaxy.

\begin{table*}
\centering

\begin{tabular}{lcccccccccccc}
\vspace*{0.125em}       
Galaxy        & $\mvir$           & $\mstar$          & $\mgas$           & $\fsdisk$ & $\Rstar$ & $\Zstar$ & $\Rgas$ & $f^\mystar_{\geq0.7}$ & $m_i,\,\mathrm{gas}$ & $m_{\rm DM}$  & Reference \\           
              & [$10^{12} \Msun$] & [$10^{10} \Msun$] & [$10^{10} \Msun$] &           & [kpc]    & [kpc]    & [kpc]   &                       & [$10^3\msun$]        & [$10^4\msun$] & \\             
\hline\hline
\run{Romeo}   & 1.28              & 6.98              & 3.45              & 0.79      & 17.4     & 1.95     & 30.5    & 0.65                  & 28                   & 15            & A \\             
\run{Juliet}  & 1.06              & 5.26              & 3.16              & 0.76      & 13.7     & 1.67     & 20.8    & 0.59                  & 28                   & 15            & A \\             
\run{Louise}  & 1.10              & 6.39              & 3.23              & 0.69      & 12.2     & 1.5      & 24.2    & 0.56                  & 32                   & 16            & A \\             
\run{Robin}   & 1.56              & 5.99              & 2.90              & 0.66      & 9.5      & 1.65     & 20.8    & 0.51                  & 57                   & 31            & A \\             
\run{Thelma}  & 1.44              & 11.58             & 2.56              & 0.65      & 11.6     & 2.13     & 11.2    & 0.5                   & 32                   & 16            & A \\             
\run{m12f}    & 1.58              & 7.53              & 2.85              & 0.64      & 11.1     & 2.39     & 20.8    & 0.48                  & 7.1                  & 3.5           & B \\             
\run{Romulus} & 1.95              & 13.46             & 3.55              & 0.61      & 11.6     & 2.55     & 22.4    & 0.48                  & 32                   & 16            & E \\             
\run{m12i}    & 1.14              & 6.16              & 2.23              & 0.58      & 9.9      & 2.07     & 17.8    & 0.44                  & 7.1                  & 3.5           & C \\             
\run{m12z}    & 0.86              & 3.5               & 1.82              & 0.57      & 11.4     & 3.23     & 8.3     & 0.4                   & 33                   & 17            & D \\             
\run{m12c}    & 1.27              & 8.09              & 0.92              & 0.56      & 4.3      & 1.08     & 3.6     & 0.42                  & 57                   & 28            & A \\             
\run{Remus}   & 1.23              & 10.05             & 0.90              & 0.53      & 7.7      & 1.71     & 8.3     & 0.45                  & 32                   & 16            & E \\             
\run{m12m}    & 1.47              & 10.88             & 1.41              & 0.53      & 13.3     & 2.75     & 12.1    & 0.34                  & 7.1                  & 3.5           & A \\           
\run{m12b}    & 1.36              & 9.13              & 2.32              & 0.33      & 5.2      & 1.16     & 12.1    & 0.27                  & 57                   & 28            & A \\           
\run{m12q}    & 1.61              & 11.23             & 0.56              & 0.21      & 5.4      & 1.57     & 0.9     & 0.11                  & 57                   & 28            & A \\           
\run{Batman}  & 1.89              & 10.21             & 1.96              & 0.20      & 2.4      & 0.98     & 11.2    & 0.08                  & 57                   & 31            & A \\           
\end{tabular}
\caption{
Properties of the central galaxies and their host halos, sorted by decreasing 
$\fsdisk$.  In order, columns indicate the host halo virial mass, the stellar 
and gas mass of the central galaxy (defined in \S\ref{ssec:simprops}), the 
fraction of stars in the central galaxy on ``disk-like'' orbits 
($\epsilon\geq0.5$), and the sizes of the stellar and gas disks (see 
\S\ref{ssec:defns} for details).  To give an estimate of how sensitive the 
disk fractions/ordering are to our $\epsilon\geq0.5$ cut, the following 
column lists the fraction of stellar mass with $\epsilon\geq0.7$.
The remaining columns list the resolution of each simulation, given by the 
initial gas particle mass and the mass of the DM particles in the high 
resolution region.  The final column lists the publication each run first 
appeared in:   A: \citet{FIRE2}, B: \citet{GKDisk}, C: \citet{Wetzel2016}, 
D: \citet{Hafen2016}, E:  this work.  Galaxies beginning with 
``\run{m12}'' are isolated MW-mass analogues, while those with names of 
individuals are in Local Group-like pairs.  \run{Romulus} \run{\&} \run{Remus} 
and \run{Thelma} \run{\&} \run{Louise} are hydrodynamic re-simulations of the 
same pairs originally presented (as DMO simulations) in \citet{ELVIS}.  
Figures~\ref{fig:morph_v_mass}~and~\ref{fig:gas_drivers} plot the 
relationships between several of these properties.
}

\label{tab:sims}
\end{table*}

We focus on the roughly MW-mass galaxies simulated with FIRE-2.   
Therefore, we combine the Latte halo (here referred to as 
\run{m12i}) from \citet{Wetzel2016}; five additional isolated halos 
simulated with an identical pipeline, two at the same resolution and 
three with a factor of $8$ higher mass particles; one isolated halo from 
\citet{Hafen2016}; three pairs of halos in Local Group-like 
configurations \citep[first reported in][but analyzed in detail here 
for the first time]{FIRE2}, and one additional pair that has not yet 
been reported elsewhere.  Hosts in Local Group-like pairs were selected 
with the same criteria as \citet{ELVIS}:  isolated pairs with $\mvir\sim10^{12}\msun$ 
that are approaching one another.  All other hosts were selected purely on 
the basis of their mass and isolation from other massive halos.  The mass 
resolution of each galaxy is listed in Table~\ref{tab:sims}.\footnote{We 
list the initial mass of a gas particle in each simulation, but note that 
due to deposition onto gas particles from stellar mass loss, baryonic 
particle masses fluctuate slightly about their initial value.}  Softening 
lengths for the gas are fully adaptive, typically down to $1~\pc$, with
fixed stellar and DM softening lengths set according to the typical 
inter-particle spacing.  \citet{FIRE2} list the exact values for our 
runs, but all are sufficient to resolve the disk heights.   For each galaxy, 
we analyze the highest resolution simulation available that has been 
completed to $z=0$.  We demonstrate the stability of our morphologies
and sizes with numerical resolution in Appendix~\ref{sec:resolution}:
the general trends are robust to resolution, but we caution that 
quantitative values do change slightly with resolution.

Movies showing the formation and evolution of each galaxy in our sample, 
created using identical pipelines, may be found at 
\url{http://www.tapir.caltech.edu/~sheagk/firemovies.html}.

\section{Quantifying morphology of the FIRE-2 galaxies}
\label{sec:quant}

There are a wide variety of reasonable definitions for galactic morphology 
that one can adopt.  Broadly speaking, they range from kinematic distinctions 
(e.g.\ the fraction of stars on circular orbits) to visual quantifiers 
(e.g.\ photometric bulge-to-disk ratios, \citealp{Sersic1963} indices, and 
half-light radii).  Though the former are straightforward to measure in 
simulations, they are difficult to determine with observations.  The latter, 
however, are relatively straightforward to extract with photometry, but can 
only be measured for simulated galaxies if one assumes models for stellar 
evolution and dust attenuation.  Though the relationship between observable 
morphological measures and kinematic quantifiers is extremely interesting, 
a full study requires ``mock observations'' of the simulated galaxies (including 
radiative transfer) and subsequent fitting of those images with the tools 
typically used by observers.  We consider these steps to be beyond the scope 
of this paper, which instead focuses on the physical drivers of those 
morphologies, but plan to investigate this question in greater detail in 
future work.

\begin{figure*}
\centering
\includegraphics[width=\textwidth]{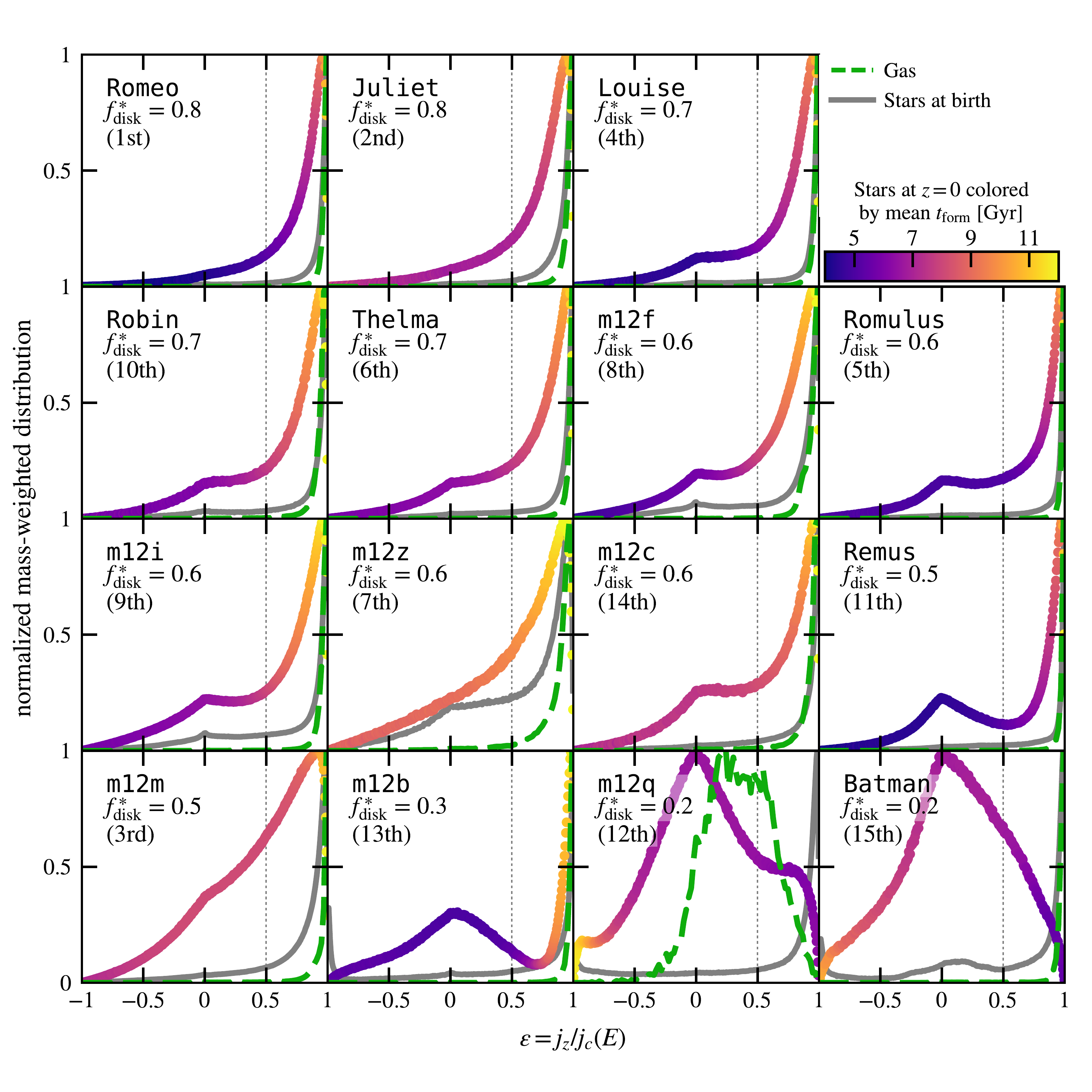}
\vspace{-3em}
\caption{Mass-weighted PDFs (normalized to a maximum of one) of the 
circularity $\epsilon = j_z/j_\mathrm{c}(E)$ for the stars (colored histograms) 
and gas (dashed green curves) within the MW-mass FIRE-2 galaxies at $z=0$.  
The stellar distributions are colored by the average cosmic formation time
(where $z=0$ corresponds to $t_\mathrm{form}\simeq13$--$14~\gyr$) of the stars 
in each bin.  The vast majority of the galaxies transition to younger
ages at higher circularities; the exceptions are \run{m12q} and \run{Batman},
which form counter-rotating disks at late times.  The gray curves show 
the kinematics measured when stars are born; we discuss them in detail in 
\S~\ref{sec:evolution}, but emphasize here that almost all stars are born 
in $z=0$ MW-mass galaxies form on disk-like orbits (i.e. with very high 
circularities).  We quantify the ``diskiness'' of a galaxy by $\fsdisk$, 
defined here as the fraction of stars with $\epsilon \geq 0.5$, indicated 
by the dashed vertical line.  The panels are sorted by decreasing $\fsdisk$, 
with the numbers in parentheses indicating the rank they would have if the 
panels were instead sorted from largest to smallest $\Rstar$, the 2D radial 
extent of the stars.  We also demonstrate in 
Figures~\ref{fig:mymorh_vs_othermorph}~and~\ref{fig:morph_vs_sig1} that 
both $\fsdisk$ and $\Rstar$ correlate strongly with other kinematic and 
spatial measures of morphology. All but one of the galaxies has a 
well-ordered, rotating gas disk at $z=0$; the exception is \run{m12q}, 
which is nearly gas-free and is in the process of expelling what little 
gas remains by $z=0$.  The stars display a range of kinematics 
ranging from well-ordered disks (\run{Romeo}) to dispersion supported bulges 
(\run{Batman}).  
}
\label{fig:joverjz}
\end{figure*}

\subsection{Definitions}
\label{ssec:defns}
Here, we focus primarily on morphological measures that do not rely on 
specific profiles or on assumptions regarding the luminosities/colors of 
individual star particles.  We primarily adopt two independent measures 
of galactic morphology, $\fsdisk$ and $\Rstar$.  The latter, $\Rstar$, 
is the radial extent of the disk.  It is defined together with $\Zstar$, 
the height of each galaxy, such that 90\% of the stellar \emph{mass} 
within $30~\kpc$ of the galactic center is contained within a 2D radius 
$\Rstar$ and a height above/below the disk $\Zstar$ when the stars are 
aligned with their principal axes. 
We then define $\mstar$ as the stellar mass within a radial distance $\Rstar$ 
and a height above/below the disk $\Zstar$.\footnote{We note that this 
definition differs from the stellar masses listed in \citet{FIRE2}, who 
quoted total stellar masses within $3\times\rshalf$.}.  For the purposes of 
comparing with semi-analytic models (\S\ref{ssec:spin}), we identically define 
$\Rshalf$, the 2D radius that encloses 50\% of the stellar mass. We 
similarly define 3D stellar radii $\rstar$ and $\rshalf$ as the radii 
that contain 90\% and 50\% of the stellar mass within $30~\kpc$.  Though 
the same process typically yields accurate results for the gas, it 
artificially inflates the sizes of extremely gas-poor galaxies (e.g. \run{m12c} 
and \run{m12q}; see Figure~\ref{fig:gas_viz}).  Therefore, we define the 
radial and vertical extents of the gas disk by first taking the peak of 
the face-on mass profile, $\mathrm{d}M_\mathrm{gas}(R)/\mathrm{d}\ln R$, as 
$\Rgas$, then defining $\Zgas$ as the break in the vertical 1D mass profile 
of all the gas with a projected radius $R<\Rgas$.  $\mgas$ is then defined 
as the total gas mass within ($\Rgas, \Zgas$).  $\mgas$ typically changes 
by only $\sim10-20\%$ between this method and the approach we adopt for
the stars, with the technique we adopt for the gas yielding a slightly 
lower $\mgas$ in all but two cases.  All properties are based on centers 
calculated via a shrinking spheres approach \citep{Power2003}.  

Our kinematic morphological definition, $\fsdisk$, measures the fraction of 
stars on circular orbits that are aligned with the angular momentum of the 
galaxy as a whole.  Specifically, for each particle within $\rstar$, we 
compute the circularity $\epsilon = j_z/j_\mathrm{circ}(E)$ following the 
method of \citet{Abadi2003} and described in detail in \citet{KEB2017}.  For 
a given mass element, the circularity relates the component of the specific 
angular momentum that is aligned with the average angular momentum vector of 
the galaxy, $j_z$, to the specific angular momentum of a circular orbit with 
the same energy, $j_\mathrm{circ}(E)$.  Stars (or gas) with $\epsilon = 1$ 
are therefore on perfectly circular orbits in the plane of the galaxy, 
those with $\epsilon = 0$ have orbits that are exactly perpendicular to 
the galaxy, and those with $\epsilon = -1$ are perfectly counter-rotating.  
We adopt a cut of $\epsilon \geq 0.5$ to distinguish disk stars, and define 
$\fsdisk$ as the mass fraction of stars that meet this cut within $\rstar$.  
We find nearly identical disk fractions if we consider all stars within 
$30~\kpc$:  the fractional difference is typically $<5\%$.

\begin{figure*}
\centering
\includegraphics[width=0.95\textwidth]{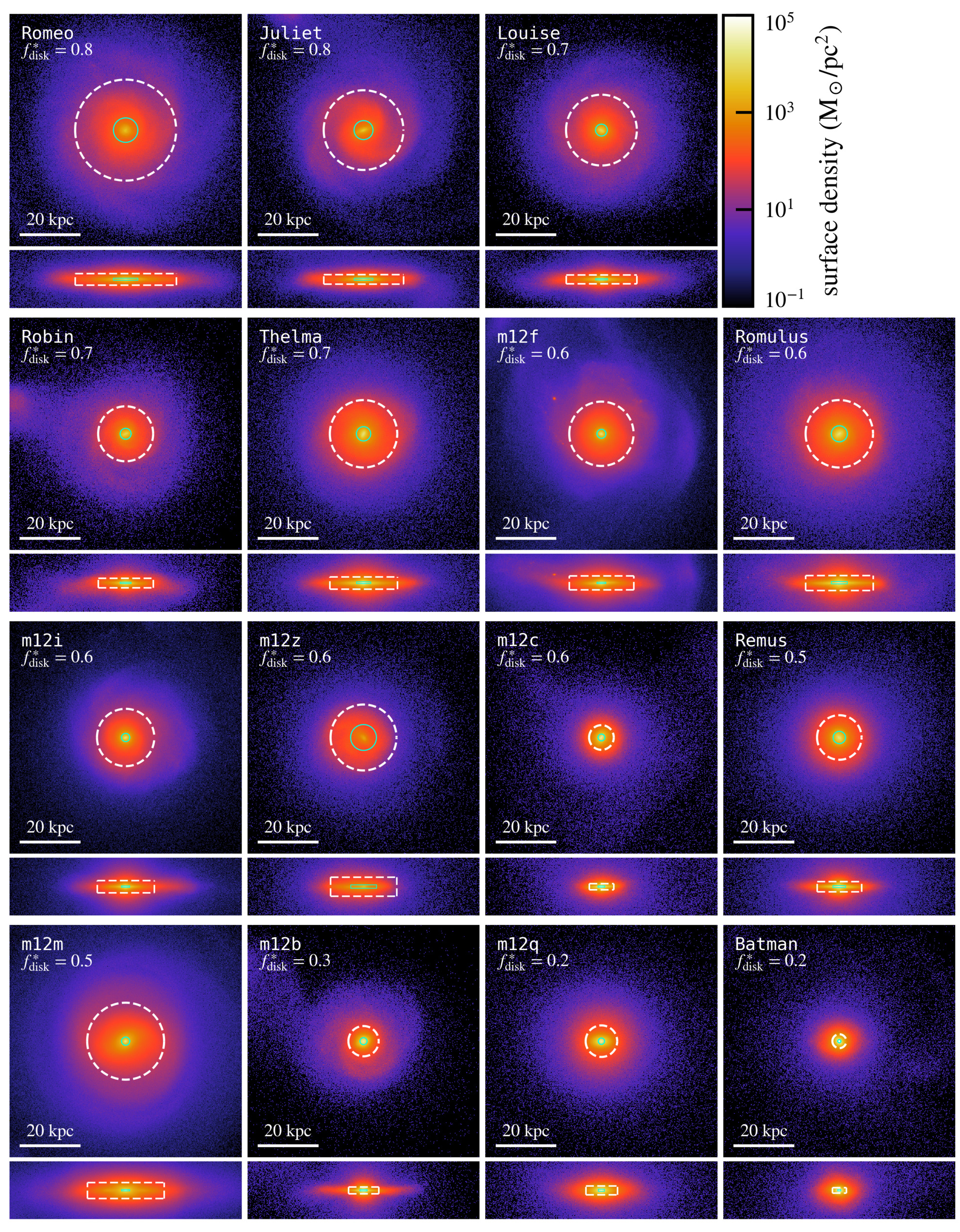}
\caption{Face-on (top panels) and edge-on (bottom panels) projections 
of the stars in the FIRE-2 galaxies, again sorted by decreasing $\fsdisk$, 
with the radius $\Rstar$ and height $\Zstar$ that contain 90\% of the 
mass indicated by the white circles/rectangles; the green lines show 
the equivalent half-mass quantities.  Each panel is $80~\kpc$ across; 
the edge-on projections are $20~\kpc$ tall.  Though there is not a direct 
correspondence between $\fsdisk$ and disk size, they are clearly 
correlated (see Figure~\ref{fig:mymorh_vs_othermorph}).  
}
\label{fig:stellarviz}
\end{figure*}

\subsection{Simulation Properties}
\label{ssec:simprops}

\begin{figure*}
\centering
\includegraphics[width=\textwidth]{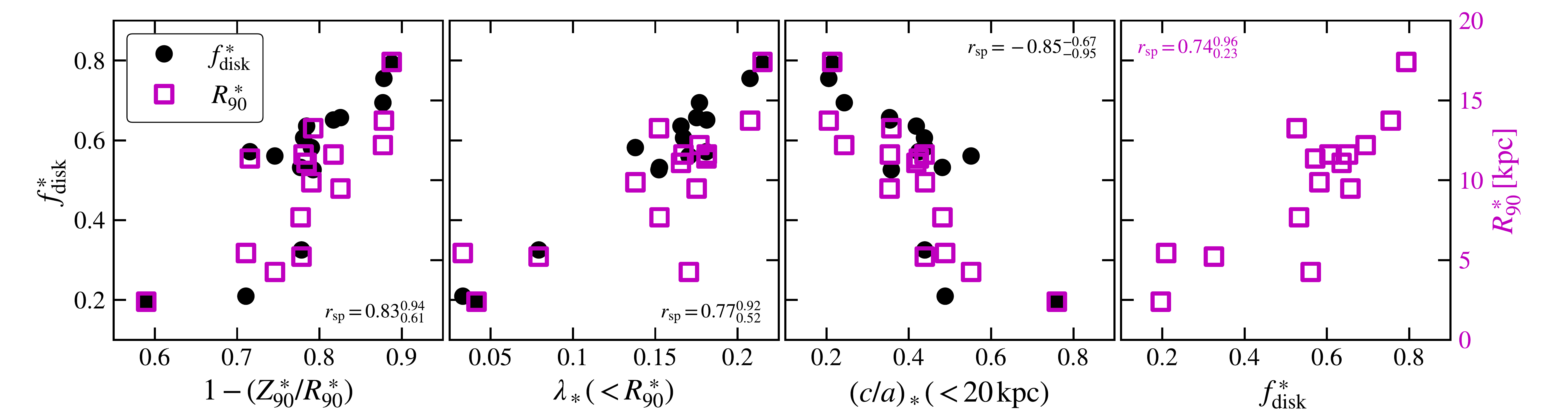}
\caption{
Comparing our adopted measures of morphology, $\fsdisk$ (black points; 
left axis) and $\Rstar$  (open magenta squares; right axis), to other 
spatial and kinematic measures of morphology, all measured at $z=0$.  The 
panels show $1-(\Zstar/\Rstar)$, a measure of the flatness of the stellar 
distribution; $\lambda_\mystar(\Rstar)$, the \citet{Bullock2001} spin 
parameter of the stars in the galaxy; $(c/a)_\mystar$, the ratio of the 
shortest to longest principal axes of the stars, measured within $20~\kpc$; 
and the disk fraction $\fsdisk$ \emph{vs} $\Rstar$.  The $\rspear$ values indicate 
the median Spearman $r$-coefficient obtained over 100,000 bootstrapping 
trials; the upper and lower values give the full 95\% confidence interval.
In all but the final column (and in the remainder of this work), we compute 
the $r$-coefficient by combining the magenta squares and black circles;
i.e. $\rspear$ represents the joint correlation between the property on the $x$
axis and both $\fsdisk$ and $\Rstar$ (see $\S\ref{ssec:morphvmorph}$ for
details).
}
\label{fig:mymorh_vs_othermorph}
\end{figure*}

The distributions of $\epsilon$ for both the stars and gas in each MW-mass 
FIRE-2 galaxy (i.e.\ within $\rstar$) are shown in Figure~\ref{fig:joverjz}.  
Each panel represents an individual galaxy; they are ordered by decreasing 
$\fsdisk$, which is indicated for each galaxy.  The number in parentheses 
below $\fsdisk$ indicates the rank that each galaxy \emph{would} have if 
they were instead ordered by decreasing $\Rstar$; we compare $\fsdisk$ and 
$\Rstar$ explicitly in Figure~\ref{fig:mymorh_vs_othermorph}.  We will retain 
this sorting by $\fsdisk$ in other figures to ease comparison.

The stellar distributions, which are plotted as the colored histograms 
in Figure~\ref{fig:joverjz}, vary widely even in our relatively small 
sample.  Without pre-selecting for expected morphology, the MW-mass 
FIRE-2 sample includes nearly bulge-less disk galaxies (e.g. \run{Romeo} 
and \run{Juliet}), galaxies with clear bulge and disk components (e.g. 
\run{Remus} and \run{m12b}), and almost entirely dispersion-supported 
galaxies (\run{Batman} and \run{m12q}).\footnote{We note that \run{Batman} 
and \run{m12q} are very compact, with $\Rshalf\simeq0.5$~and~$1~\kpc$ 
respectively, and may be outliers in observations \citep[e.g.][]{Shen2003}.  
As noted, though, we caution against direct comparisons with observations 
without mock-observing the sample.}  The color of each curve at a 
given $\epsilon$ indicates the average formation time of stars with 
that $\epsilon$.  Other than \run{Batman} and \run{m12q}, which have 
formed roughly counter-rotating disks at late times, the disk 
($\epsilon \geq 0.5$) is almost always composed of younger stars on 
average, in agreement with previous results that disks in MW-mass 
galaxies begin to appear at $z\lesssim1$ \citep[e.g.][]{Ma2016b,Mametalgrad}.  
In some cases, such as \run{m12b} and \run{Remus}, the average 
ages of the bulge and disk components differ dramatically, while 
the transition is much smoother in other systems (e.g. \run{m12m} 
and \run{m12z}).

In contrast with the diversity in the kinematics of the stars, the gas 
distributions (green dashed curves) are almost uniform across this 
mass-selected sample.  Specifically, every galaxy except \run{m12q}, 
(which has not experienced any significant gas accretion since $z\sim0.1$)
hosts a thin, primarily rotation-supported gas disk.  The gray curves 
in Figure~\ref{fig:joverjz}, which show the circularity distributions 
of the stars formed in the galaxy \emph{at birth} (i.e.\ stacking over 
all snapshots) are similarly uniform, with the vast majority of stars 
forming with $\abs{\epsilon} \geq 0.5$.  We will discuss the kinematics of 
stars at birth along with the evolution of those kinematics in \S~\ref{sec:evolution}, 
and we will explore the characteristics of the gas disks in greater detail in 
\S~\ref{sec:gas_morph}, but we first focus on the $z=0$ stellar morphologies.

Visualizations of the stars in all fifteen galaxies are shown in 
Figure~\ref{fig:stellarviz}, again sorted by $\fsdisk$.  The top panels show 
face-on views of each galaxy, while the lower panels visualize the galaxy 
edge-on. There is a clear trend for galaxies to become more elliptical, less 
disky, and typically more spatially compact as $\fsdisk$ decreases.  The thick 
dashed and thin solid circles (rectangles) in the upper (lower) panels of 
Figure~\ref{fig:stellarviz} indicate ($\Rstar$, $\Zstar$) and ($\Rshalf$, 
$\Zshalf$), respectively.  As intended, the former captures roughly the full 
extent of the stellar populations.  We also plot circular velocity 
profiles for the full sample in Appendix~\ref{sec:rotcurves}:  galaxies
with higher disk fractions tend to have flatter, more extended circular
velocity curves and, conversely, the bulge-dominated systems have rotation 
curves that peak at small radii, but there is some scatter about that 
trend.

We summarize several basic properties of each galaxy in Table~\ref{tab:sims}, 
including the host virial mass $\mvir$, the galaxy stellar mass $\mstar$, and 
the mass in gas within the galaxy $\mgas$, along with the fraction of stars in 
the galaxy on circular orbits $\fsdisk$ and the radial extent of the stars and 
gas in each galaxy, $\Rstar$ and $\Rgas$.  To give an indication of how sensitive
our results are to our definition of ``disk'' stars having $\epsilon\geq0.5$, we 
also list the fraction of stellar mass with $\epsilon\geq0.7$.

While the FIRE-2 physics successfully reproduce observed relationships over a 
wide range of masses \citep[see][and \S\ref{sec:intro}]{FIRE2}, our 
mass-selected sample does face some tension with observations.  First, our 
galaxies are overly massive for their halo masses:  our stellar mass definition 
places our sample between $0.2$--$0.55$~dex above the \citet{Behroozi2013} 
stellar mass \emph{vs} halo mass relation.  Second, at these stellar masses, a 
non-negligible fraction of observed galaxies are quenched, with strongly 
suppressed star formation rates \citep[e.g.][]{Salim2007}.  However, none of 
the galaxies in our sample fall into this category:  our lowest 100~Myr averaged 
specific star formation rate at $z=0$ is $\sim10^{-11.5}$~yr$^{-1}$ (possibly 
because these simulations do not include AGN feedback; \citealp{Bower2006,
Cattaneo2006,Croton2006,Somerville2008}).  Though our sample includes only 
fifteen galaxies, we caution that we may overproduce (or at least over-represent) 
late-type galaxies, which could potentially alter the correlations we present 
herein.  Furthermore, if quenching correlates with properties of either the 
galaxy or the halo \citep[e.g. the mass of the DM halo at fixed stellar 
mass;][]{Woo2013} in a way not captured by the FIRE-2 models, then our 
analysis will miss those relationships.

\subsection{Comparing morphological measures}
\label{ssec:morphvmorph}

Before examining correlations between various halo/galaxy properties, 
$\fsdisk$, and the radial extents of our galaxies, we briefly explore the 
relationship between our morphological measures ($\fsdisk$ and $\Rstar$) 
and other potential measures of morphology.  As discussed in \S\ref{sec:intro} 
and \ref{ssec:defns}, we do not explicitly compare with observational measures, 
as that lies beyond the scope of this work.  However, we do note that the stellar 
radii that we adopt in this paper scale closely with the half-mass radii derived 
from fitting two-component S\'{e}rsic profiles to these same galaxies (Sanderson 
et al., in prep), though the bulge-to-disk ratios of those profiles do not 
correlate particularly well with the true kinematic disk fraction $\fsdisk$.  

In addition to the properties we discuss above, there are a number of 
viable morphological definitions we could adopt, such as the angular 
momentum of the stars, the thickness of the stellar disk, or the shape 
of the stellar mass distribution.  We examine how these properties correlate 
with $\fsdisk$ and $\Rstar$ in Figure~\ref{fig:mymorh_vs_othermorph}.   
In the first three panels, the filled black circles plot each quantity 
against $\fsdisk$ (left $y$ axis), while the open magenta squares correspond 
to the right $y$ axis and indicate $\Rstar$.  The final panel shows $\fsdisk$ 
and $\Rstar$ against one another and therefore omits the black points.

\begin{figure}
\centering
\includegraphics[width=\columnwidth]{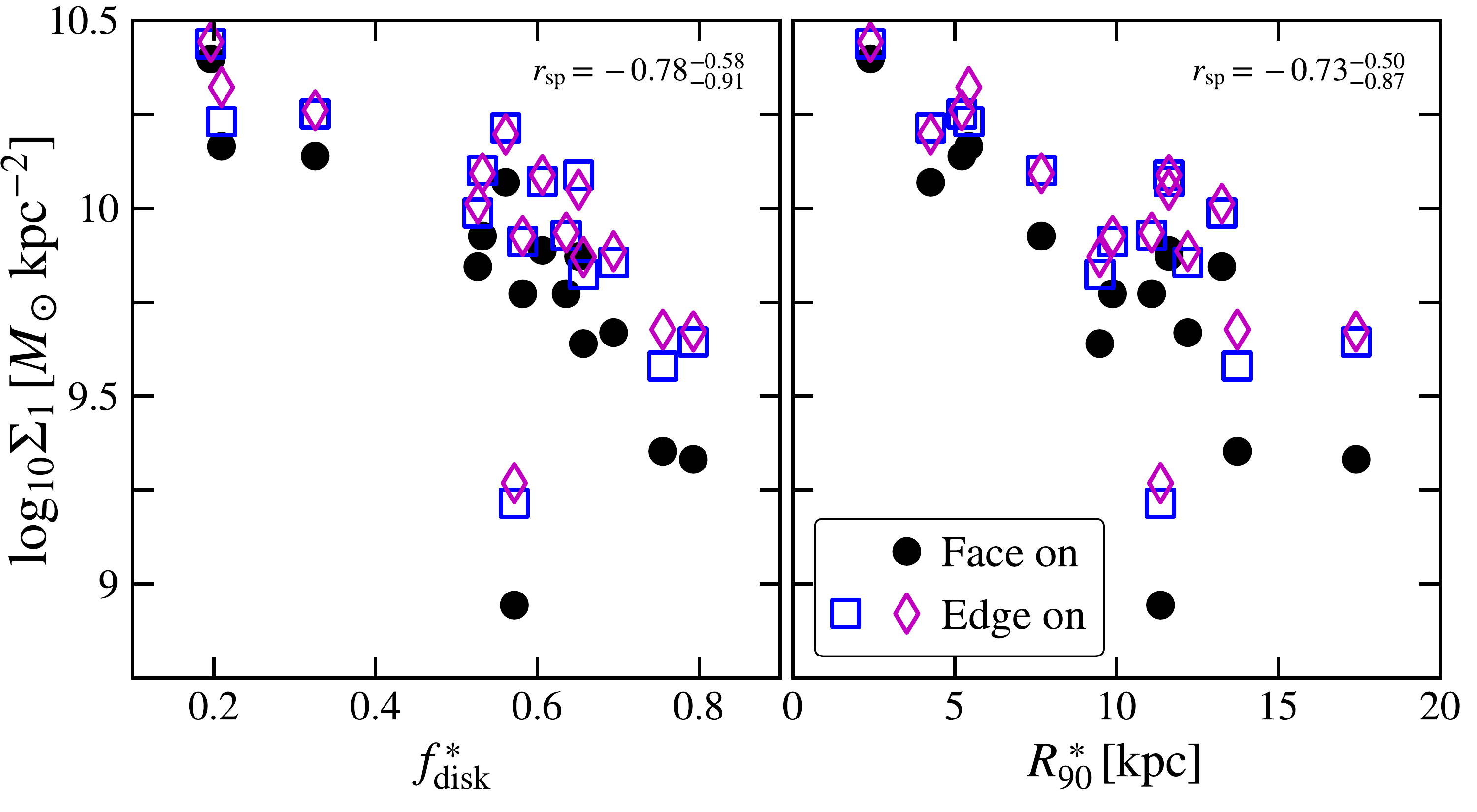}
\caption{The stellar surface density within $1$ projected kpc of the 
center of the galaxy $\Sigma_1$ \emph{vs} our adopted morphological 
measures, $\fsdisk$ on the left and $\Rstar$ on the right.  Black circles 
show $\Sigma_1$ as measured from a perfectly face-on view, while the open 
symbols indicate $\Sigma_1$ measured along orthogonal edge-on projections.  
Regardless of viewing angle, the projected central density of the 
galaxy correlates strongly with both $\fsdisk$ and $\Rstar$.  The 
anti-correlation arises because higher central densities necessarily 
imply more compact (and therefore less disky) galaxies at fixed mass.}
\label{fig:morph_vs_sig1}
\end{figure}

Any of these properties shown in Figure~\ref{fig:mymorh_vs_othermorph} (along 
with other measures that we do not plot here, such as the stellar radius scaled 
by the virial radius, the specific angular momentum of the stars, the radius where 
the log-slope of the stellar density profile equals $-3$, or the kinematic 
bulge-to-disk ratio) are viable alternatives to $\fsdisk$ and $\Rstar$.  The 
correlations are unsurprising:  at roughly fixed mass, galaxies that are 
radially extended are also flatter, have larger stellar spin parameters, 
and have a greater fraction of rotation support.  The final panel in 
Figure~\ref{fig:mymorh_vs_othermorph} indicates the relationship between 
$\fsdisk$ and $\Rstar$.  As suggested by the visualizations in 
Figure~\ref{fig:stellarviz}, the radial extent of the stars correlates with 
the degree of order in the disk, but with non-trivial scatter, motivating 
our analysis of both properties throughout.

\begin{figure*}
\centering
\includegraphics[width=\textwidth]{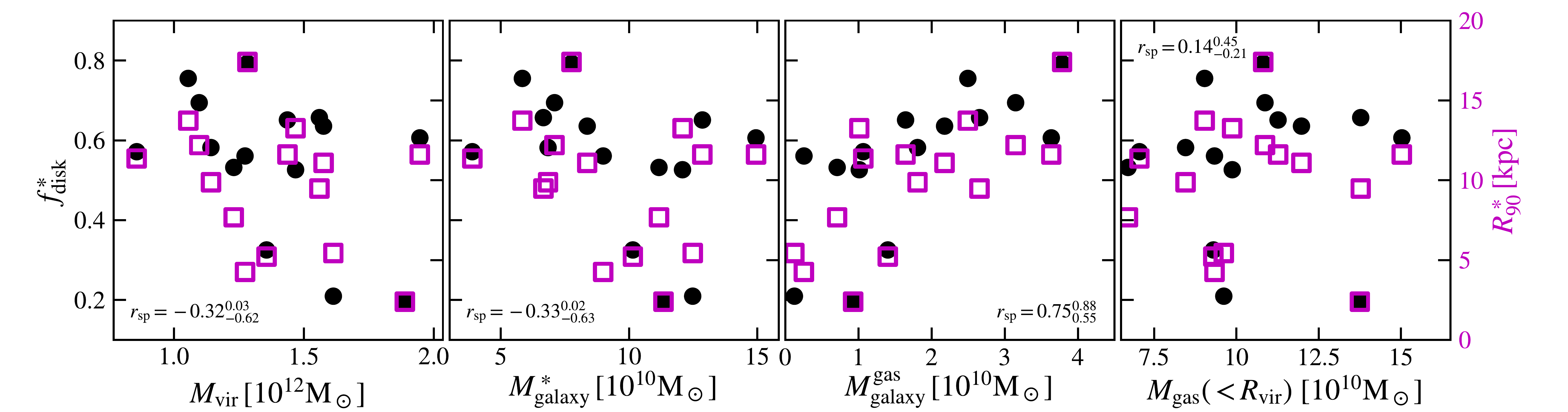}
\caption{The $z=0$ relationship between our morphological measures and, 
(\textit{a}) the virial mass of the halo $\mvir$, (\textit{b}) the 
stellar mass of the galaxy $\mstar$, (\textit{c}) the gas mass within 
the galaxy $\mgas$, and (\textit{d}) the total gas mass within $\rvir$.  
Morphology in the FIRE-2 simulations is not correlated with either 
$\mvir$, $\mstar$, or the total gas mass within $\rvir$, across this 
narrow mass range, but gas rich galaxies are more likely to be disky 
today; we discuss this correlation in more detail below.}
\label{fig:morph_v_mass}
\end{figure*}

The text in each panel (and in similar Figures below) indicates the 
Spearman $r$-coefficient, $\rspear$, which quantifies the monotonicity of each 
relationship.  We compute $\rspear$ on the joint relationship with $\fsdisk$ 
and $\Rstar$: we assign each galaxy a rank based on $\fsdisk$ and a rank 
based on $\Rstar$, then combine those ranked datasets and compute
$\rspear$ against two copies of the ranked $x$ values of each plot.  Our 
qualitative conclusions are unchanged if we compute $\rspear$ against 
$\fsdisk$ and $\Rstar$ independently.  For each relationship, we perform 
100,000 bootstrapping trials (randomly drawing $N$ points, with replacement).  
We report the median $\rspear$ of those trials, and the values in superscripts 
and subscripts indicate the full 95\% confidence interval for those trials.  
We provide identical statistics throughout the remainder of this work.  Based 
on the correlations reported in Figure~\ref{fig:mymorh_vs_othermorph}, which 
plots correlations between morphological properties that we expect to be 
reasonably well correlated, we adopt a rough criterion of $\abs{\rspear}\gtrsim 0.8$,
with a lower 95\% bound on the confidence interval of $\abs{\rspear}\gtrsim 0.6$, 
as a ``tight'' correlation.

Before turning to the drivers of stellar morphology, we briefly examine 
one non-parametric morphological measure that is relatively easy to measure 
in both simulations and observations: $\Sigma_1$, the stellar surface density 
within the central $1~\kpc$ \citep[e.g.][]{Cheung2012,Bell2012,Fang2013}.  
For this mass-selected sample, we find a tight relationship between 
$\Sigma_1$ and morphology at $z=0$: Figure~\ref{fig:morph_vs_sig1} shows 
$\Sigma_1$ as measured edge-on in open symbols and face-on in black circles.  
The viewing angle has a small impact, though edge-on projections are always 
higher, as expected.  The anti-correlation between $\Sigma_1$ and the true 
morphology of a galaxy is striking, though somewhat unsurprising:  for a 
roughly fixed stellar mass, galaxies with high central densities must be 
more compact, and Figure~\ref{fig:mymorh_vs_othermorph} demonstrated that 
radial extent and degree of order in the stellar orbits are well correlated, 
again at fixed galaxy mass.  We therefore conclude that $\Sigma_1$ is a reliable 
morphological measure, at least for roughly MW-mass galaxies.  However, 
we caution that the low-lying outlier from the trend is \run{m12z}, our
lowest mass galaxy, suggesting the possible emergence of a mass trend.  
Moreover, while some analyses have associated high $\Sigma_1$ with galactic 
quenching \citep[e.g.][]{Woo2015,Woo2017}, all of our galaxies show some 
level of continued star formation to $z=0$ (as noted above).

\section{Drivers of Stellar Morphology}
\label{sec:drivers}

We now turn to correlations between stellar morphology, quantified 
primarily by $\fsdisk$ and $\Rstar$, and various properties of the 
galaxy and the host halo, both in the hydrodynamic simulation and 
in the analogous dark matter-only (DMO) run.  In short, we search 
for physical drivers of and explanations for the $z=0$ morphologies 
of each of the galaxies in our sample.

\subsection{Mass (around MW masses)}
\label{ssec:mass}

We begin by checking whether the morphologies of the FIRE-2 MW-mass 
galaxies are driven by either the halo or galaxy mass.  
Figure~\ref{fig:morph_v_mass} indicates the virial mass $\mvir$, the 
stellar mass $\mstar$, the gas mass $\mgas$, and the total gas mass 
within $\rvir$, all at $z = 0$.  As in Figure~\ref{fig:mymorh_vs_othermorph},  
black points correspond to the left axis and plot $\fsdisk$, while 
magenta squares indicate $\Rstar$ (right axis).  Of the masses shown 
in Figure~\ref{fig:morph_v_mass}, only $\mgas$ displays evidence for 
a correlation with the $z=0$ stellar morphology.  Though we do not plot 
it, we also find no correlation between the total baryonic mass within 
$\rvir$ and kinematics/morphology ($\rspear = -0.6$ -- 0.06).  There is 
evidence for a correlation with the total mass in cold gas (defined 
as $T<10^5$ K) within $\rvir$ ($\rspear = 0.52$--$0.85$), but because the 
cold gas is predominantly in the galaxy, this correlation is driven 
by $\mgas$.  We will return to the correlation with $\mgas$ below, 
but here we emphasize that the morphologies of the MW-mass FIRE-2 
galaxies do not correlate with either the halo mass, the stellar 
mass of the galaxy, or the total baryonic mass within $\rvir$.  Note 
that over a large dynamic range, however, there is a strong mass 
dependence (e.g. \citealt{KEB2017} showed that the FIRE-2 dwarfs 
are spherical and dispersion dominated).

\subsection{Spin (and other DM properties)}
\label{ssec:spin}

As discussed in \S\ref{sec:intro}, many authors have pointed out 
that, if baryons acquire their angular momentum from their dark matter 
halos and begin with the same density profile as those halos, then the 
size of the stellar disk should be predicted by a combination of the 
\citet{PeeblesSpin} spin parameter $\lambda_{\rm Peebles}$, the size 
of the host halo, the fraction of angular momentum in the halo that 
resides in the disk $\jd$, and the fraction of halo mass that resides in the 
disk $\md$.  In the simpler model of \citetalias{MoMaoWhite}, wherein the 
galaxy is hosted by a static isothermal sphere,
\begin{equation}
R_\mathrm{d} = 2^{-1/2} (\jd/\md) \lambda_{\rm Peebles} \rtwoh.
\label{eqn:mmwRd}
\end{equation}
In their more complete model, where the disk grows adiabatically within an 
initially NFW \citep{Navarro1996} halo, the disk radius is modified by two
multiplicative functions; the first arises from the change in the total 
energy of the NFW profile relative to an isothermal sphere and the second 
from the (assumed) adiabatic contraction of the halo in response to the 
growth of the disk.  If such a relationship is borne out by the FIRE-2 
simulations, and if $\jd/\md = j_{\rm disk}/j_{\rm halo}$, the ratio of 
the specific angular momentum of the disk to the halo, is roughly constant 
(i.e. if the baryons acquire their angular momentum from the halo, as 
assumed in \citetalias{MoMaoWhite}), then one can accurately populate halos 
in DMO simulations with galaxies of the proper size and, by virtue of the 
correlation between $\Rstar$ and $\fsdisk$, roughly the proper disk 
fraction.  Moreover, validation of the model would provide evidence for 
the overall theory of angular momentum-regulated disk growth.

\begin{figure}
\centering
\includegraphics[width=\columnwidth]{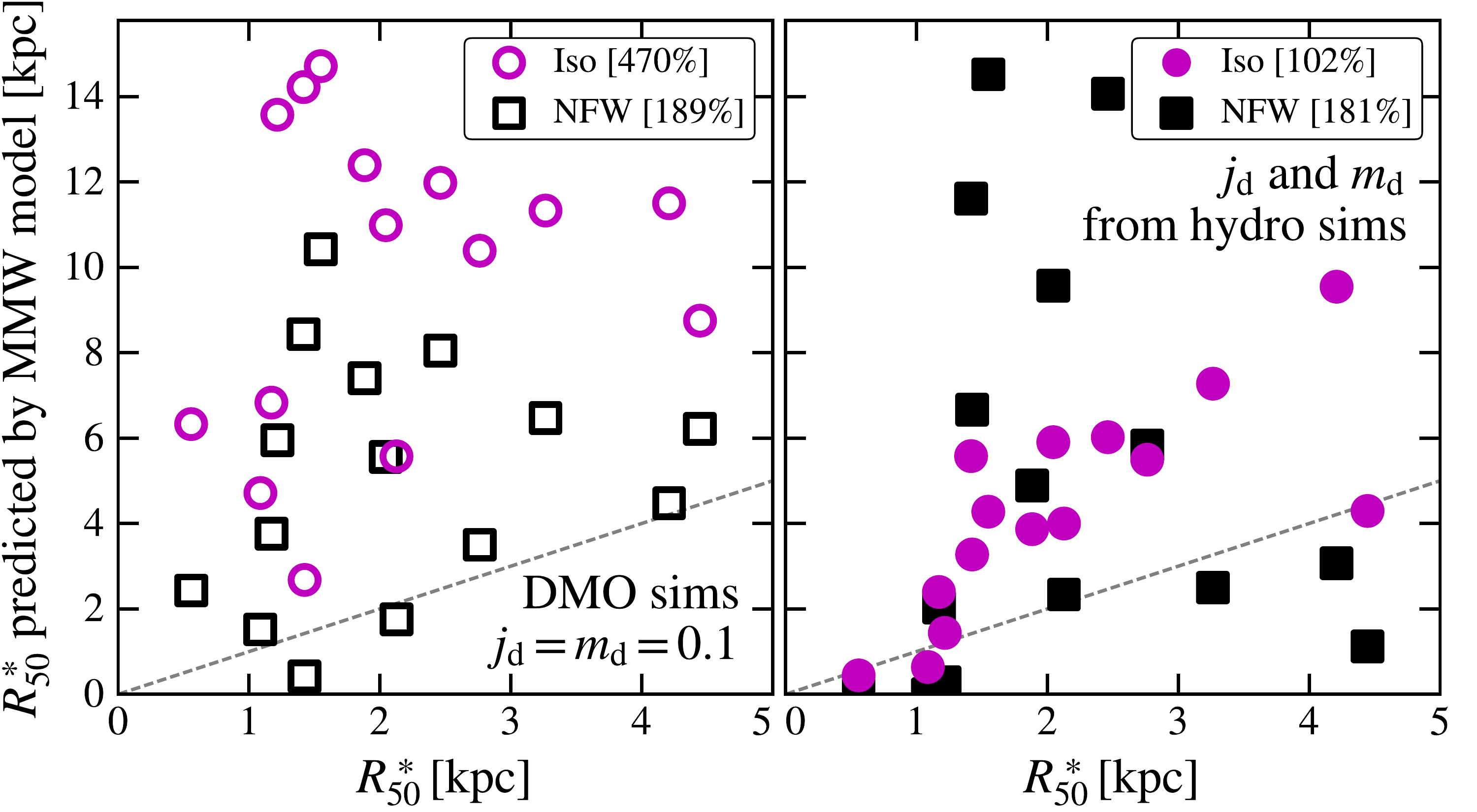}
\caption{The half-mass radius of the FIRE-2 galaxies \emph{vs} the disk 
radius predicted by the \citetalias{MoMaoWhite} model for each galaxy
(see Eq.~\ref{eqn:mmwRd}).  Circles show results assuming an isothermal 
potential, and squares indicate the full model (an adiabatically contracted 
NFW profile).  The left panel uses the properties of the halo 
available in the DMO simulation (fixing $\jd = \md$), and therefore 
tests the assumption that the baryons acquire their angular momentum 
from the DM halo (given the small variations in $\rtwoh$ in our
sample).  The right panel frees this assumption by adopting $\jd$
and $\md$ from the hydrodynamic simulations, and instead tests 
whether the galaxies are well-described by a rotationally-supported
disk in a fixed potential.  The numbers in brackets indicate
the average fractional error of the model relative to the 
simulations.}
\label{fig:mmw_comparison}
\end{figure}

Figure~\ref{fig:mmw_comparison} tests this picture by comparing the half-mass 
radius predicted by the models of \citetalias{MoMaoWhite} to the 
half-mass radius of each simulated galaxy.  Circles show the results 
of the isothermal model (Equation~\ref{eqn:mmwRd}), and squares plot the 
full model assuming an adiabatically contracted NFW halo (Equation 28 of 
\citetalias{MoMaoWhite}).  In order to test the assumption that galaxies
acquire their angular momentum from the dark matter, the left panel uses 
properties available from the DMO simulations and fixes $\jd=\md$.\footnote{We 
adopt $\jd = \md = 0.1$, but our overall results are insensitive to the 
chosen value.}  Given the relatively small variations in $\rtwoh$ within 
our sample, the left panel implicitly tests whether disk size is driven
by the spin of the halo at $z=0$.  Neither model is able to reproduce the 
actual size of our galaxies, in line with the general results of zoom-in 
simulations discussed in \S~\ref{sec:intro}.  The bracketed numbers in the 
legends indicate the average fractional error of each set of points relative 
to the simulations: the isothermal \citetalias{MoMaoWhite} model dramatically 
over-predicts the size of the galaxies when assuming $\jd=\md$.  The 
contracted-NFW halo model produces a reasonable order-of-magnitude 
estimate of $\Rshalf$, but the actual predictive value is quite poor.

The right panel frees the assumption that the angular momentum of the 
galaxy is correlated with the spin of the halo and instead fits the
galactic angular momentum independently by adopting $\jd$ and $\md$ 
(along with the remainder of the halo properties) from the hydrodynamic
simulations.  We calculate $\jd$ ($\md$) from the simulations as the 
ratio of the stellar angular momentum (mass) within $\Rstar$ to the 
total angular momentum (mass) within $\rtwoh$.  By doing so, we measure 
the true angular momentum of the galaxy (i.e., independent of the spin 
of the halo) and therefore test the assumption that a rotationally supported 
disk in a fixed gravitational potential (determined by a simple NFW or 
isothermal model) provides a reasonable approximation.  Even under this 
assumption, the predictions are only moderately accurate, though we do 
find order-of-magnitude agreement across this mass range,  in line with 
observational results that show a correlation between virial radius 
(i.e. halo mass) and galaxy size \citep[e.g.][]{Kravtsov2013}.  
The relative success of the isothermal model (compared to the NFW 
model) may suggest that the density profiles are closer to isothermal 
spheres at their centers, but we see no strong evidence in the 
actual profiles (though see \citealp{Chan2015}, who found that the
total density profiles at the centers of MW-mass FIRE-1 galaxies
are well-fit by an isothermal sphere).

\begin{figure*}
\includegraphics[width=\textwidth]{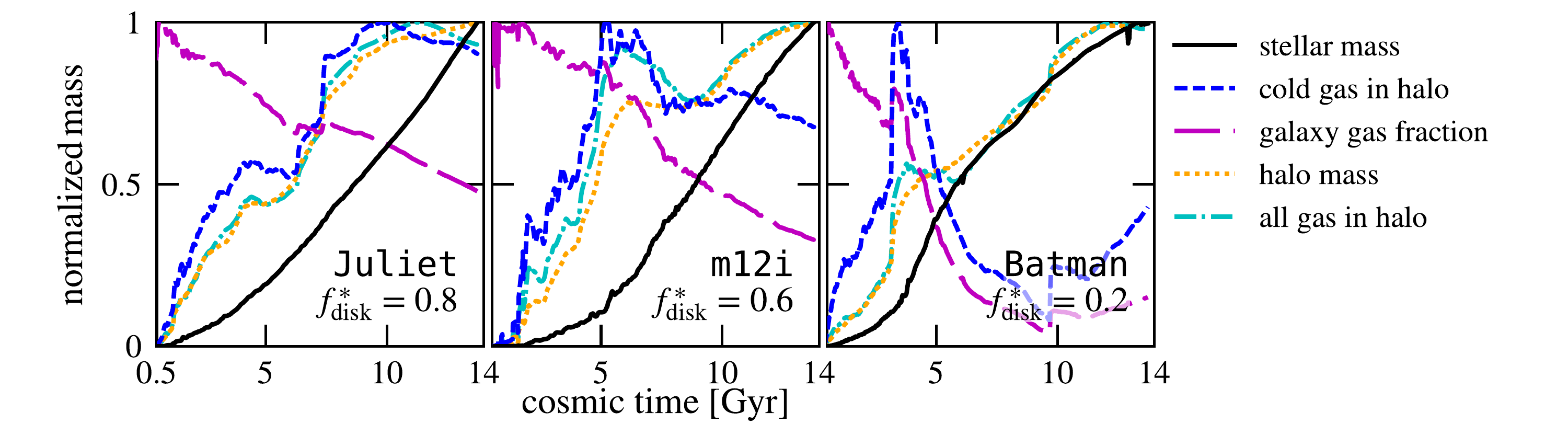}
\vspace{-1.5em}
\caption{Evolutionary histories of three representative galaxies; 
the entire sample is shown in Appendix~\ref{sec:allgrowth}.  Each 
curve is normalized to its maximum value.  The clearest trend, which 
holds generically for our sample, is that galaxies that have higher 
cold gas fractions and more gas available to form stars at late times 
(relative to their mass at early times) form the majority of their stars 
in disky configurations.  This follows directly from the fact that 
star formation is chaotic and bursty at high redshift, but settles 
into an ordered, disk-like configuration after $z\sim1$ (for most
galaxies that will be MW-mass at $z=0$).  We quantify these trends 
for the full sample in Figure~\ref{fig:morph_vs_gashistories}.}
\label{fig:growth_hist}
\end{figure*}

Though we adopt all of the halo parameters in the hydrodynamic 
simulation in the right panel ($\rtwoh$, $\lambda_\mathrm{Peebles}$, 
$c$, $\jd$ and $\md$), the majority of the changes are driven by 
allowing $\jd$ and $\md$ to vary freely and independently:  even 
for our sample of fifteen galaxies, $\jd$, $\md$, and their ratio 
vary by nearly an order of magnitude:  $0.005\lesssim\jd\lesssim0.07$, 
$0.04\lesssim\md\lesssim0.09$, and $0.1\lesssim\jd/\md\lesssim0.75$.  
Galaxies acquire a broad range of the specific angular momentum available 
in their hosts, and one must know the true $\jd$ and $\md$ in order to 
even roughly predict the radial extent of a given galaxy with the 
\citetalias{MoMaoWhite} model.  We are unable to recover a tight 
correlation with a single value of $\jd$ and $\md$ for all galaxies 
(even when $\jd\neq\md$).  
There is some evidence for a correlation between $\jd$ and the 
$1~\mpc$ environment:  the median $\jd$ of the galaxies in Local
Group-like pairs is twice that of the isolated sample.  Accordingly, 
six of the seven diskiest galaxies in our sample are in Local Groups.  
However, our sample size is too small to make definitive statements.

In our parameter exploration, we have generally found that, of the 
properties of the $z=0$ DMO halo, $\lambda$ (or $\lambda_{\rm Peebles}$) 
correlates most tightly with morphology ($\rspear = 0.05$ -- 0.7 for the latter)
though the correlation is weak with a large degree of scatter about the 
average relationship:  our largest, most ordered galaxy has an average spin 
parameter.  While $\lambda(z=0)$ is relatively stable between the DMO and 
baryonic simulations,\footnote{The average fractional difference in our 
sample is $\sim1\%$.}  $\lambda_{\rm DMO}$ alone is insufficient to 
predict morphologies without alleviating the scatter by multiplying by 
the true values of $\jd$ and $\md$.  Given the difficulty of predicting 
the morphology of a galaxy with only the information available about the host 
halo in a DMO simulation, we therefore turn our attention to identifying 
physical drivers of the morphology in the hydrodynamic simulations.  That is, 
we do not attempt to predict morphologies, but rather to explain them through 
galactic/halo properties at all redshifts.

\subsection{Gas fraction and accretion history}
\label{ssec:growthhist}

\begin{figure*}
\centering
\includegraphics[width=\textwidth]{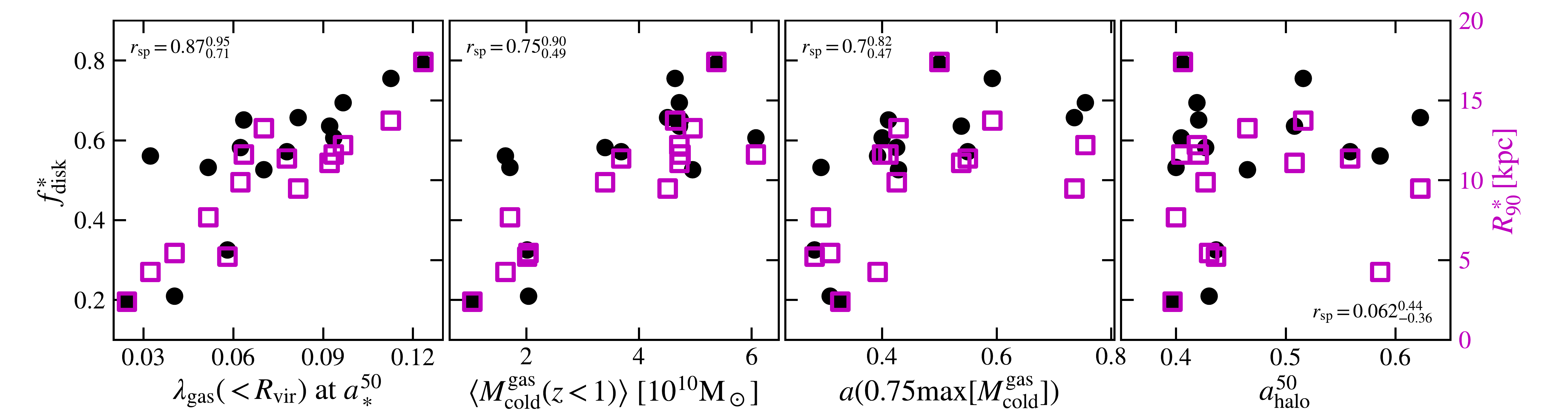}
\caption{The morphologies of our galaxies as a function of several 
parameterizations of the gas accretion histories of our galaxies 
and their host halos.  In order, the columns plot the spin parameter
$\lambda$ of the gas within $\rvir$ at the scale factor when half of 
stars in the galaxy at $z=0$ formed $a_{\mystar}^{50}$, the average 
amount of cold gas in the halo after $z=1$, the scale factor at which 
the cold gas in the halo first reaches 75\% of its peak, and the scale 
factor at which the halo reaches 50\% of its mass $a_{\rm halo}^{50}$.  
The first three, which contain information about the accretion history
and buildup either of the galaxy itself or of material that helps to 
build the galaxy, all correlate reasonably tightly with $z=0$ morphology:  
the first panel is actually the tightest correlation we have identified.  
The final column, however, indicates that the history of the DM is less 
meaningful:  the DM accretion history and halo merger history contains 
little information about the $z=0$ morphology (also see 
\S\ref{ssec:dmopred}).}
\label{fig:morph_vs_gashistories}
\end{figure*}

Figures~\ref{fig:growth_hist}~and~\ref{fig:morph_vs_gashistories} represent
the culmination of these searches.  The former, Figure~\ref{fig:growth_hist},
shows the normalized mass accretion histories of three representative 
galaxies, \run{Batman}, \run{m12i}, and \run{Juliet} (growth histories for the 
full sample are plotted in Appendix~\ref{sec:allgrowth}).  The black curves 
indicate the stellar mass within $\Rstar$, 
the blue curves show the total cold gas within $\rvir$ (where ``cold'' is again defined as $T<10^5$ K),
and the magenta curves indicate the ratio of the cold gas mass to the stellar mass of the galaxy (i.e. the ratio of the black and blue curves without normalizing).
Finally, the cyan and orange curves indicate the total gas mass 
within $\rvir$ and the total halo mass, respectively.  Each curve 
is normalized to its maximum value.  
We find qualitatively identical results measuring the total gas mass near the galactic center via the same iterative process we adopt for the stars:  the vast majority of the cold gas in the halo at any given time is in the galactic disk.  However, we opt to use the total cold gas mass within the virial radius, $\mcoldgas$, because this iterative process can falsely capture hot gas in the halo, as discussed earlier.

Of course, every galaxy has a unique evolutionary history, and our 
results suggest that history is instrumental in shaping the $z=0$ 
galaxy.  However, there are trends that hold generically across our 
sample, which are exemplified by the three panels in 
Figure~\ref{fig:growth_hist}. First, while the total gas in the halo
closely tracks the total halo mass for all galaxies, the behavior of 
the cold gas in the halo, i.e. the fuel for star formation, varies 
strongly with $\fsdisk$.  Galaxies similar to \run{Batman} with low 
$\fsdisk$ tend to reach their maximum cold gas mass at early times
(both in absolute terms and in comparison to the growth of their dark
matter halos) when star formation is chaotic and bursty, and quickly 
exhaust (or heat) that gas.  They therefore form more stars with 
bulge-like configurations, and stars that are formed in a disk are 
subject to greater dynamical disruption from the powerful feedback 
events.  Galaxies that reach their maximum cold gas mass at early times 
but maintain a relatively large reservoir for star formation until 
late times, either through mergers or accretion from the circumgalactic 
medium (CGM), form a similar fractions of stars during the bursty 
period and at late times ($z\lesssim1$), when the star-forming gas 
has settled into a rotation-supported disk, as is the case with 
\run{m12i}.  Finally, galaxies such as \run{Juliet} that are 
disk-dominated tend to have relatively little gas and form relatively 
few stars during the bursty phase.  These trends are also evident 
in the gas fractions:  bulge-dominated systems tend to reach gas 
fractions $\lesssim0.25$ at or before $z\sim1$, while disk dominated 
systems maintain high gas fractions until late times.  

\subsection{Galaxy mergers}
\label{ssec:galmergers}

Second, comparing the evolution of \run{Batman} and \run{Juliet} 
reveals the varying impacts of mergers on the $z=0$ morphologies.  
\run{Batman}, which experiences a double merger at $z\sim2$ 
($t\sim3~\gyr$; revealed by the sharp up-tick in $\mvir$) when 
the halo mass is relatively low, has a large amount of cold gas 
dumped into the halo.  That gas then forms nearly half of the 
$z=0$ stellar mass over the next $\sim1-2~\gyr$, the majority of 
which ends up as a compact, dispersion supported system.  Other 
bulge-dominated galaxies in our sample typically experience 
similarly large mergers at early times (when the systems have 
$\mvir\ll10^{12}\msun$ and therefore no extended ``hot halos'' 
of gas; see, e.g., \citealp{Keres2005}). Those mergers then tend 
to funnel their gas into the center of the galaxy relatively rapidly 
\citep{Barnes1991,Bournaud2011,Zolotov2015}.  Mergers that occur
later (when the hot halo is in place), however, tend to have their 
gas gradually stripped off and incorporated into the central galaxy 
more gradually.  \run{Juliet}, for example, has a gas-rich halo fall 
inside the virial radius at $z\sim1$, but the gas in that subhalo is 
slowly stripped off and accreted onto the central disk over the course 
of several pericentric passages, feeding an extended star forming disk.  

Overall, visual inspection of the movies indicates that large 
\emph{galactic} mergers do typically lead to bulges in our sample.
This is particularly true those mergers occur on first infall (i.e.\ 
with low angular momentum), before the gas in the merging system can
be gradually stripped and mixed with the halo, then more gently added 
on to the host (in agreement with the results of \citealp{Sales2012} 
regarding morphology as a function of the dominant gas accretion mode).  
The prominent bulges of \run{Batman}, \run{m12q}, \run{m12b}, and 
\run{Remus}, for example, were all created by such events.  However, 
the sizes of the bulges built by these events varies:  \run{Robin} 
experiences such a merger at $z\sim2$, but the overall masses were 
low at that time, leading to a small bulge relative to the disk 
that grows later.

Because a small fraction of the stellar mass in the central galaxy 
is formed ex situ (i.e. brought in by mergers; typically less than
5\%), they do not significantly contribute directly to $\fsdisk$
\citep[also see][]{AnglesAcazar2017}.  The actual ex situ fraction 
is also not correlated with morphology ($\rspear=-0.3$--0.55), and in the 
majority of our sample, the deposited ex situ stars are typically 
dispersion supported.  However, a few galaxies in our sample have 
accreted stars that contribute to the disk:  three galaxies have 
their ex situ circularity distributions peak at $\epsilon\sim0.6$, 
and in particular, our two diskiest galaxies have even their accreted 
stars on disky orbits with $\epsilon$ peaking at $\sim0.9$ and $1$.

\subsection{Secular evolution and bulges}
\label{ssec:secevol}
Not all bulges are built by mergers, however:  neither \run{m12m}, 
\run{m12i}, nor \run{Louise} experience a major head-on galactic 
merger, but all host dispersion-supported stars today.  The bulge 
in \run{m12m} is built by a secular bar-buckling event (Sanderson 
\etal, in prep.).  Meanwhile, \run{m12i} hosts a compact gas disk
that initially loses angular momentum in a series of mergers, but
then slowly builds up a larger disk at late times.  Therefore, systems 
that have undergone direct galactic mergers are more likely to host 
a bulge than compared to those evolving just under secular 
evolution (internal effects), but the exact morphology depends 
on the interplay between the merger history, star formation history, 
and angular momentum of the gas that builds the disk.

\subsection{Clump sinking/migration}
\label{ssec:clumps}
Both observations \citep[e.g.][and references therein]{ForsterSchreiber2011}
and simulations \citep[e.g.][]{Mandelker2017} of star-forming, disky galaxies 
at $z\sim2$ have found evidence for large ($\sim10^7$--$10^9\msun$) gas clumps 
that may migrate to the centers of their host disks to form secular bulges.  
However, the galaxies we study here are low enough mass at $z\sim2$ 
that we simply do not expect or see this channel of bulge formation.  
Moreover, we note that \citet{Oklopcic2017} showed that while giant clumps
do form in massive galaxies at $z\sim2$ in the FIRE simulations, there
was no evidence that these clumps have a net inward migration inwards that 
build a bulge, even at higher masses than we study here.

\subsection{Misalignments and counter-rotating disks}
\label{ssec:misalignment}
By examining the angular momentum of the material that end up in
the galaxy and halo at turn-around ($z\sim3.5$), \citet{Sales2012} 
argued that disk-dominated galaxies are typically formed out of 
well-aligned material, while bulge-dominated systems are more likely 
to experience misaligned accretion events.  We see some evidence 
for this picture in our sample:  \run{m12q}, in particular, is 
formed out of the merger of two counter-rotating disks at $z\sim0.8$,
and \run{Batman} and \run{m12b} also experience large, misaligned
galactic mergers.  

However, in our sample, this effect manifests primarily through 
mergers, and it therefore has either a dramatic impact or a 
nearly negligible one:  the fraction of counter-rotating stellar 
mass ($\epsilon\leq - 0.5$) at $z=0$ is less than $4\%$ in the 
remaining twelve galaxies and, as we show in \S\ref{ssec:formdisky}, 
the fraction that \emph{forms} counter-rotating is even smaller.
Moreover, \S\ref{ssec:formdisky} demonstrates that the fraction
of stars that form in a bulgy configuration ($\abs{\epsilon<0.5}$)
is relatively smooth across our sample, suggesting a minor (or 
relatively constant) contribution from misaligned gas that forms
stars before integrating with the disk.  However, our results do 
not preclude the possibility of misaligned accretion contributing 
to torquing the disk and shifting stars to lower circularities.  
Together with the large scatter in the trend identified by 
\citet{Sales2012}, we conclude that our results are in overall 
agreement with theirs.

\subsection{Summary:  the evolution of the gas mass and spin}
\label{ssec:driversum}
We quantify these trends in Figure~\ref{fig:morph_vs_gashistories}.  As
in Figure~\ref{fig:mymorh_vs_othermorph}~and~\ref{fig:morph_v_mass}, 
the black circles show $\fsdisk$ and the magenta squares indicate 
$\Rstar$.  From left to right, the $x$-axes plot the spin parameter of 
the gas in the halo at the scale factor $a_\mystar^{50}$ when half of the 
$z=0$ stellar mass had formed,
the average cold gas mass within the halo after $z=1$, the scale factor when $\mcoldgas$ first reaches 75\% of its peak,
and the scale factor when the halo mass reaches half of the $z=0$ value, 
$a_{\rm halo}^{50}$.  The first three are positively, and relatively strongly, 
correlated with morphology:  the spin parameter of the gas at $a_\mystar^{50}$ 
is actually the tightest (non-morphological) correlation we have identified, 
and $\langle\mcoldgas(z<1)\rangle$ displays the tightest relationship outside
of other related spin parameters.  In fact, $\langle\mcoldgas(z<1)\rangle$ is
even more tightly correlated with morphology than the spin parameter of the 
gas in the halo at $z=0$, which has $\rspear = 0.31$ -- $0.81$.  We find similar 
correlations for other descriptions of the gas accretion history of the halo, 
such as the scale factor when the total gas mass within $30~\kpc$ first 
reaches 75\% of its maximum.

\begin{figure*}
\centering
\includegraphics[width=\textwidth]{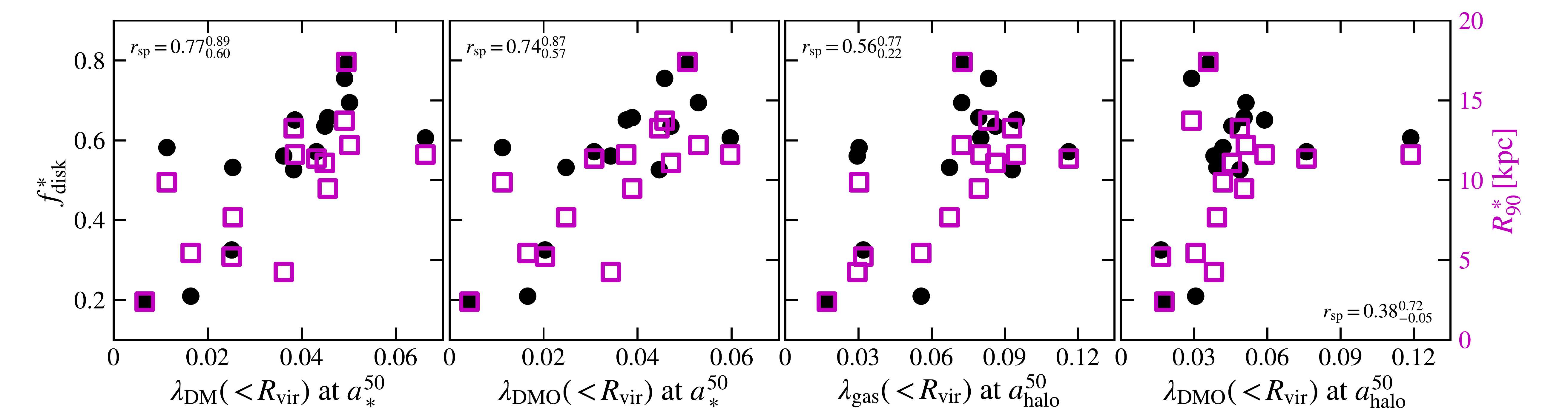}
\caption{Galactic kinematics and morphology as a function of the
spin parameter, measured at the time (scale factor) when half of 
the stars in the galaxy have formed, $a_\mystar^{50}$ (left two 
panels, as in the left panel of Figure~\ref{fig:morph_vs_gashistories}), 
and at the time (scale factor) when the halo reached half of its 
$z=0$ mass, $^{50}_{\rm halo}$ (right two panels).  The relationships 
with the spin parameter of the dark matter at $a = a_\mystar^{50}$, 
either in the baryonic or DMO run, shows more scatter than with the 
gas, but only marginally so.  However, the spin parameters at 
$a=a_{\rm halo}^{50}$ (whether gas or DMO) are only weakly correlated 
with morphology, at best, emphasizing the difficulty of predicting 
galactic morphology given only a DMO simulation.}
\label{fig:spins}
\end{figure*}

Together, Figures~\ref{fig:growth_hist}~and~\ref{fig:morph_vs_gashistories} 
suggest that disk-dominated galaxies are formed in systems that maximize
their star forming reservoir at late times (often via fresh gas delivered
by infalling subhalos), and where the gas that will turn into stars at late 
times has a high spin.  These conditions often coincide -- gas that 
infalls at late times, whether through mergers or smooth accretion, tends 
to have a higher impact parameter and therefore more angular momentum than 
similar interactions at early times 
\citep{White1984,Peirani2004,Keres2009,Stewart2011,Pichon2011,Stewart2013,Lagos2017,Stewart2017}.  
This picture is also largely consistent with the results of previous theoretical 
and observational works, which have generally found that stellar disks form 
inside-out and are composed of young stars, while the oldest stars reside 
in the galactic bulge \citep{Kepner1999,Pilkington2012,Howes2015}.  It is 
also relatively unsurprising based on the coloring in Figure~\ref{fig:joverjz}, 
which indicates that the youngest stars in each galaxy have disk-like kinematics.  
Unfortunately, it further reinforces the discussion above that it is difficult 
to accurately predict the $z=0$ morphology of a galaxy that will form in a 
given halo based on a DMO simulation, as the morphology is primarily driven 
by the gas dynamics.  The ubiquity of galactic winds and gas recycling in the
FIRE simulations further complicates efforts to connect DMO simulations to the 
galactic morphology/kinematics \citep{AnglesAlcazar2014,Muratov2015,AnglesAcazar2017}.

Any individual bulge, meanwhile, can be sourced by either mergers 
or secular processes, and both contribute significantly.  However, 
the former are more important at early times, while the latter 
typically lead to later bulge formation.  In fact, when mergers 
happen at later times in our sample, they tend to be smaller, 
gas-rich galaxies merging onto the central host and depositing 
more gas at larger radii, enhancing the chance of disk survival 
\citep[consistent with][]{Hopkins2009a}.  In our limited sample, 
however, there is no obvious way to attribute all morphological 
trends to ``merger history'' or to ``bar formation.''

\subsection{Predicting morphology from DM-only properties}
\label{ssec:dmopred}

To emphasize the difficulty of using a DMO simulation to 
\emph{a priori} estimate the morphology of a galaxy in light
of the tight correlation between $\lambda_\mathrm{gas}$ at 
$a_\mystar^{50}$ (which suggests that a similarly tight 
correlation might exist for $\lambda_\mathrm{DMO}$ at some
$z>0$), Figure~\ref{fig:spins} shows galactic morphology 
against various spin parameters at two scale factors: 
$a_\mystar^{50}$ and $a_\mathrm{halo}^{50}$, the half-mass 
time of the total halo mass; only the latter is available
in a DMO simulation.  The first panel shows the spin parameter 
of the dark matter in the baryonic simulation at $a_\mystar^{50}$.  
While it is less tightly correlated with morphology than the 
spin parameter of the gas in the halo at the same scale factor 
(Figure~\ref{fig:morph_vs_gashistories}), the relationship remains 
relatively tight.  The second panel demonstrates that the correlation 
between the DM spin (at $a_\mystar^{50}$) and morphology is not 
driven by interactions between the baryons and the DM -- the spin 
parameter in the DMO simulation at the same scale factor also correlates 
with morphology, though again less strongly.\footnote{Though we 
have not identified any direct correlations between DMO halo 
properties and $a_\mystar^{50}$, the relationship in the second 
panel suggests that, if one \emph{could} predict the galaxy 
half-mass time from a DMO simulation alone, there may be a path 
from the properties of the halo in the DMO simulation to the 
galactic morphology.}  However, those correlations have not yet 
appeared at the (earlier) $a_\mathrm{halo}^{50}$:  the third 
panel shows that the spin parameter of the gas at $a_{\rm halo}^{50}$ 
is only weakly correlated with morphology, and the final panel 
illustrates that the spin parameter in the DMO simulation contains 
little information at this time (as it does at $z=0$).  We also 
note that $\lambda_{\rm gas}$ is typically $2$--$3$ times the spin 
of the dark matter (both at high $z$ and at $z=0$; first pointed 
out by \citealp{Stewart2011}), emphasizing the disconnect 
between the angular momentum of the baryons (particularly those 
that eventually form the galaxy) and the halo.   Moreover,
while there is a reasonably tight correlation between $a_\mystar^{50}$
and $a_{\rm halo,\,DMO}^{50}$ ($\rspear = 0.4$--0.92), a
direct route from DMO halo properties to galaxy morphology
would require a similar correlation between
$\lambda_{\rm gas}(a = a_\mystar^{50}$) and 
$\lambda_{\rm DMO}(a = a_{\rm halo,\,DMO}^{50})$, which we see
no strong evidence for ($\rspear =-0.22$--0.85).

\begin{figure}
\includegraphics[width=\columnwidth]{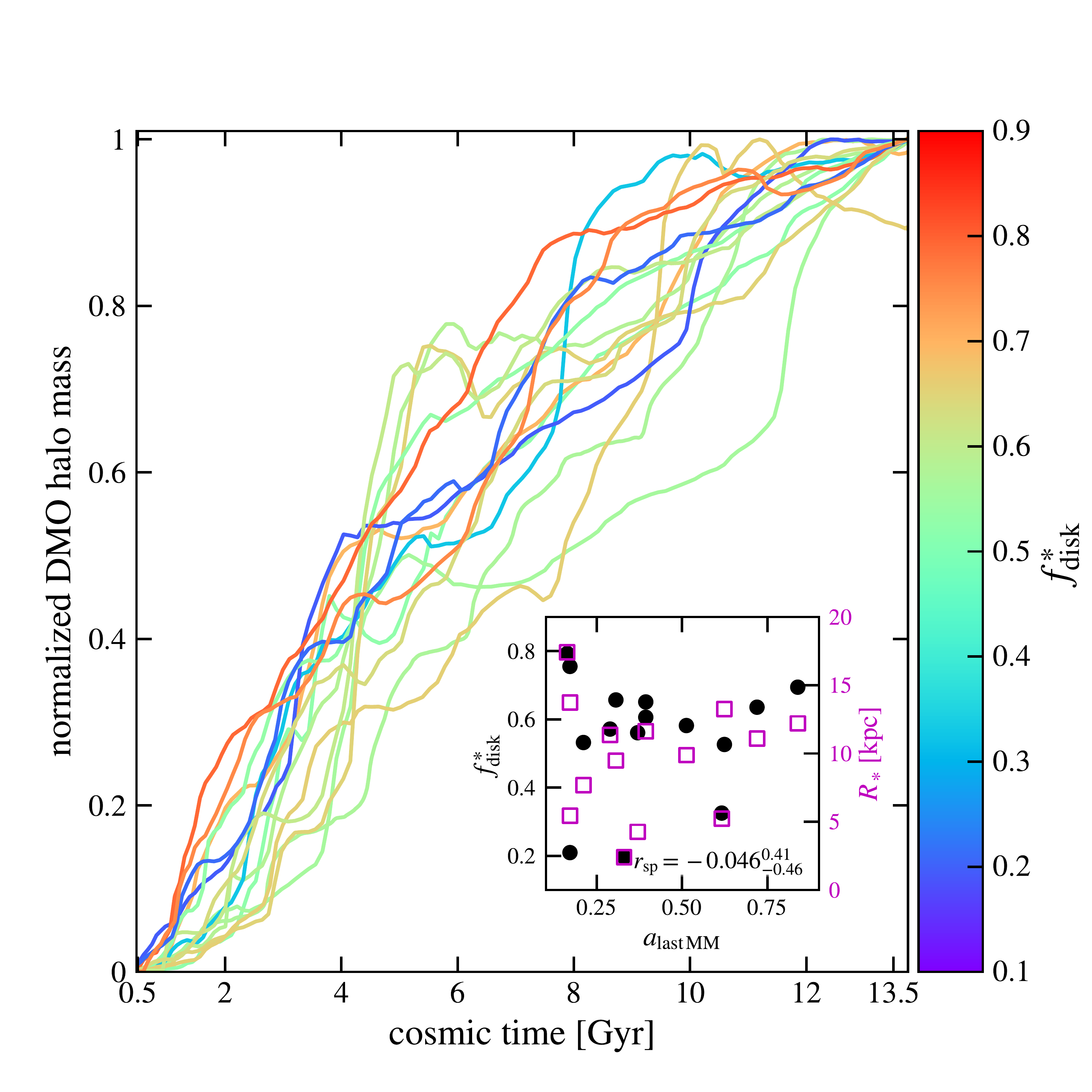}
\vspace{-1.5em}
\caption{
The normalized mass accretion histories of our host halos in the
DMO simulation.  The inset shows the scale factor of the last 
major merger (defined as a mass ratio of $\geq0.3$) in the DMO 
simulation against galactic morphology.  While the galaxies with
the largest bulges do tend to reside in halos that form early, 
the most disk dominated systems have actually accreted a greater 
fraction of their mass by $z\sim1$ ($t\sim6~\gyr$).  Moreover, 
there is effectively no correlation between the timing of the last 
major merger and morphology.
}
\label{fig:dmo_mah}
\end{figure}

We also explore trends with the accretion history of the
main branch of the DMO halo in Figure~\ref{fig:dmo_mah}.  The inset
panel shows the scale factor of the last major merger in the DMO
run against the galactic morphology.\footnote{Merger times and mass 
accretion histories are drawn from merger trees built with 
\texttt{consistent-trees} \citep{ctrees} using \texttt{rockstar}
\citep{rockstar} halo catalogs.} The curves are colored by 
$\fsdisk$; the most bulge-dominated galaxies tend to have 
higher masses at early times, but the halos that host the 
galaxies with the highest $z=0$ disk fractions in our sample 
typically have even higher (normalized) masses at any 
$z\gtrsim1$.  This is similar to the result in 
Figure~\ref{fig:morph_vs_gashistories}:  the evolution (and 
spin parameter) of the halo contains relatively little 
information about the galactic morphology compared to 
the evolution (and spin parameter) of the galaxy itself.

\subsection{Other}
\label{ssec:other}
In addition to the properties shown in Figures~\ref{fig:morph_v_mass},
~\ref{fig:morph_vs_gashistories},~\ref{fig:spins},~and~\ref{fig:dmo_mah}
and discussed above, we have also checked for correlations with numerous
other parameters of the galaxy, halo, or DMO halo both at $z=0$ and at 
higher redshifts, including their growth histories.  Examples of those
that correlate with the $z=0$ morphology, but less strongly than those we 
present above, include 
properties associated with the star formation history, such as the 
amount of time that the galaxy maintains a (200 Myr averaged) star formation 
rate (SFR) of at least 50\% of its peak value and the fraction of stars
formed during that time.  Similarly, the actual peak SFR shows a weak
anti-correlation:  relatively constant, extended star formation is
more likely to create a well-ordered disk \citep[as discussed in][]{Muratov2015}.  
However, the scale factor when the galaxy reaches its peak SFR is uncorrelated 
with morphology today ($\rspear = -0.33$ -- 0.54). The fraction of specific 
angular momentum in the disk, $\jd/\md$, is also weakly correlated with 
$z=0$ morphology, as is the spin of the gas/halo at $z=1$.  

Finally, a non-exhaustive list of properties that show no statistically
significant signs of correlation with $\fsdisk$ or $\Rstar$ include (along 
with their associated bootstrapped 95\% CI on $\rspear$) includes:
\begin{itemize}
    \item $\mstar/\mvir$ ($\rspear = -0.49$ -- $0.11$), 
    \item the total angular momentum in the DMO halo at $z=0$ ($\rspear = 0.09$ -- $0.48$), 
    \item the NFW scale radius of the DMO halo at $z=0$ ($\rspear = -0.52$ -- $0.14$), 
    \item the $z=0$ shape of the DMO halo at various radii ($\rspear = 0.14$ -- $0.5$ at 10~kpc and $\rspear = 0.08$ -- $0.64$ at 300~kpc),
    \item the fraction of $\mvir$ in bound subhalos at $z=0$ ($\rspear = -0.18$ -- $0.54$), 
    \item the scale factor at which the SFR peaks ($\rspear = -0.32$ -- $0.52$),
    \item the $z=0$, 100-Myr-averaged SFR ($\rspear = -0.34$ -- $0.52$) and specific SFR ($\rspear = -0.27$ -- $0.52$),  
    \item the fraction of stellar mass formed after $z = 1$ ($\rspear = -0.25$ -- $0.44$),
    \item the fraction of halo mass accreted after $z = 1$ ($\rspear = -0.27$ -- $0.31$),
    \item the fraction of in-situ stars within 30~kpc ($\rspear = -0.3$ -- $0.55$),
    \item the mass of the stellar halo, whether selected by $z=0$ distance ($\rspear = -0.41$ -- $0.23$) or formation distance ($\rspear = -0.58$ -- $0.16$),
    \item the maximum gas mass within $\rvir$ over cosmic time ($\rspear = -0.21$ -- $0.44$)
    \item the mean stellar age ($\rspear = -0.46$ -- $0.23$).
\end{itemize}
The final point appears contradictory to the picture that we
describe above at first glance:  if disks form late, then 
one would naively expect the mean stellar age, or $a_\mystar^{50}$,
to correlate with morphology.  However, by close inspection of
Figure~\ref{fig:joverjz} (and as we will show further in 
Section~\ref{sec:evolution}), one can see that while the disk of a given 
galaxy is (almost) always younger than the bulge of that galaxy,
disks emerge at different times in different galaxies.  For 
example, the disk of \run{Romeo} is composed of stars with an 
average age of $\sim6~\gyr$, while \run{m12f}, which hosts a disk 
and a bulge, formed its disk much more recently.

\begin{figure*}
\includegraphics[width=\textwidth]{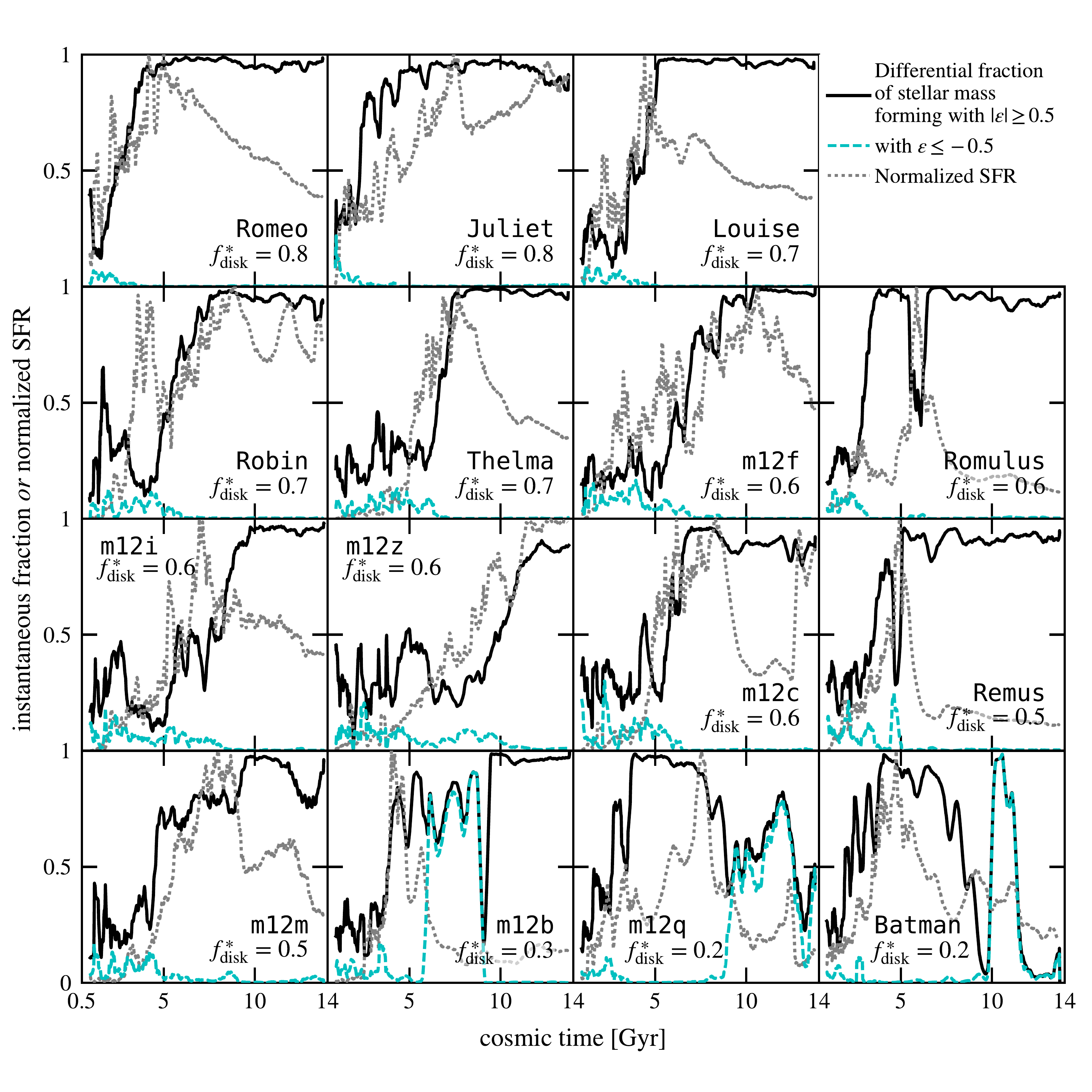}
\vspace{-3em}
\caption{The instantaneous fraction of stars born in either a prograde 
or retrograde disk, i.e. with $\abs{\epsilon}\geq0.5$ at formation 
(black); only in a counter-rotating disk, i.e. with $\epsilon\leq - 0.5$ 
at formation (dashed cyan); and the normalized star formation rate (gray 
dotted curves).  Most stars with high circularity are formed at late 
times, though in \run{Batman} and \run{m12q} most of the young star 
formation occurs in a counter-rotating disk.  The three galaxies with 
the lowest $\fsdisk(z=0)$ all experience some level of counter-rotating 
star formation; the remainder experience almost none.  Galaxies with 
high $z=0$ disk fractions have more prolonged disk-like star formation, 
but mergers sometimes destroy existing disks and scramble the correlation.
}
\label{fig:fsdisk_evol}
\end{figure*}

\section{The evolution of the stellar morphology}
\label{sec:evolution}

\subsection{Overview}
As the previous section showed, the $z=0$ morphology is driven 
primarily by a combination of the accretion histories, the degree
of rotation support in the halo at the half-mass time of the galaxy, 
and the relative amount of mass and angular momentum from the halo 
that end up in the disk.  However, the $z=0$ morphology is also 
the culmination of star formation in the galaxy, stars being 
deposited onto the galaxy through mergers, and dynamical interactions 
altering the orbits of existing stars.  In this section, we explore the 
birth morphologies of stars and the extent to which their orbits are 
shifted to lower circularities over time, and we demonstrate that while 
the $z=0$ morphologies do correlate with the spin of the gas that they 
are born out of, the full picture depends on the mutual evolution of the 
gas kinematics and star formation rate, and the impact of dynamical heating 
on the galactic disk.  We do not explicitly investigate the radial or 
vertical structure of the disk as a function of time, but we refer the 
reader to \citet{Ma2016b} for a detailed discussion of the evolution of 
the disk of \run{m12i} simulated with FIRE-1.

Figure~\ref{fig:fsdisk_evol} shows, as a function of time, the 
instantaneous fraction of stars forming with circularities 
$\abs{\epsilon}\geq0.5$ (measured at the time of formation) and
with $\epsilon\leq0.5$, i.e. in a disk that is counter-rotating 
relative to the overall angular momentum axis of the existing 
stars in the galaxy, and the normalized star formation rate (SFR).  
We define the instantaneous birth disk fraction $\fsdbirth(t)$ from 
the first snapshot that each particle appears in, capturing the 
kinematics of stars that are at most $\sim20~\myr$ old.  
Circularities, and therefore disk fractions, are defined relative 
to the evolving $\hat{z}$ axis of the angular momenta of all the stars 
in the galaxy at a given time.  The curves indicate running averages 
smoothed over $\sim300~\myr$, but the qualitative conclusions are 
insensitive to the size of this window.  We count stars within $\Rstar(t)$, 
but we find similar results using all stars within a fixed cut of 
$30$~physical~kpc.

\begin{figure}
\centering
\includegraphics[width=\columnwidth]{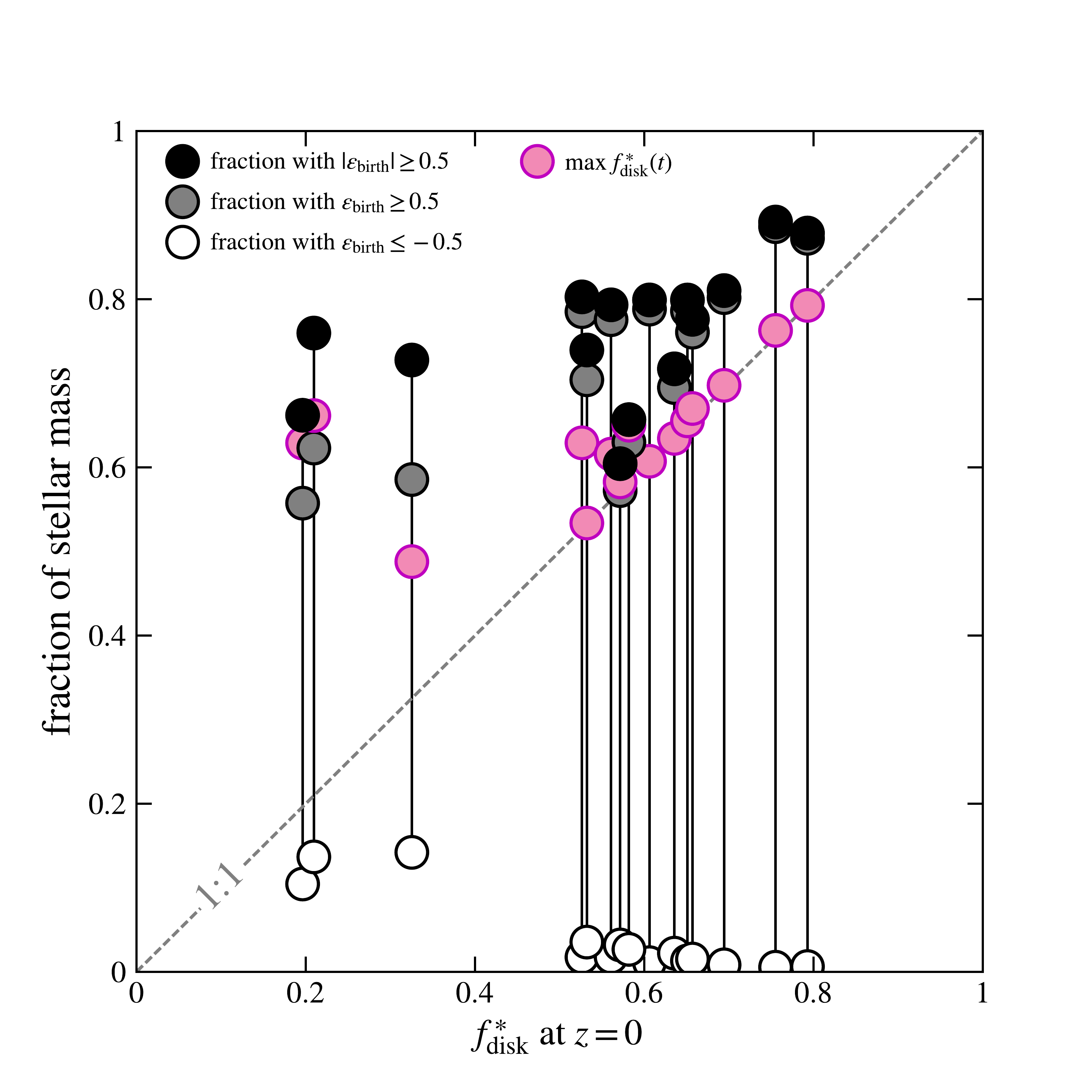}
\vspace{-1.5em}
\caption{The cumulative fraction of stars born in a disk (i.e. with 
$\abs{\epsilon} \geq 0.5$, $\epsilon\geq0.5$, and $\epsilon\leq - 0.5$ at 
birth), and the maximum instantaneous fraction of the stellar mass in 
the galaxy at any given time with $\epsilon\geq0.5$ (i.e. $\fsdisk(t)$),
all as a function of the disk fraction today.
The fraction of stars born in counter-rotating disks (open black circles) 
is $<4\%$ in all but the most bulge-dominated galaxies.  The spread in maximum 
disk fraction and in the birth disk fraction is surprisingly small 
($0.5$--$0.9$): though $z=0$ bulge-dominated systems do tend to form fewer 
stars with high $\abs{\epsilon}$ overall, and more stars in a counter-rotating 
disk, they are primarily differentiated by their subsequent evolution.  Disk 
dominated systems at $z=0$ are more likely to be at or near their maximum 
disk fractions.  
}
\label{fig:birth_v_now}
\end{figure}

\subsection{Most stars form in disky configurations}
\label{ssec:formdisky}

Figure~\ref{fig:fsdisk_evol} illustrates several points about the 
evolution of the disk morphology.  First, at late times, most stars
forming in MW-mass galaxies (black curves) do so with disk-like 
kinematics. This does not preclude them from forming in bulges, 
however, since they can be compact, rotationally-supported 
pseudo-bulge components.  Even in our most bulge-dominated system 
($\fsdisk\sim0.2$), the ``birth disk fraction'' is high at late 
times -- only \run{m12q} does not have $\fsdbirth\sim1$ at some 
point after $t \sim 10$~Gyr.  The three galaxies with the lowest 
$\fsdisk$ at $z=0$ are also the only three to experience a significant 
fraction of counter-rotating star formation.  In \run{m12b}, that 
star formation eventually builds enough of a disk to flip the overall 
angular momentum axis of the galaxy (which occurs at $t\sim8$~Gyr 
when the cyan curve goes to zero), but in \run{Batman} and \run{m12q} 
it only decreases the $z=0$ disk fraction by adding stars opposite
to the predominant $\hat{z}$.  Therefore, even though SFRs typically 
peak around $z\sim1$, and dynamical interactions shift stars to lower 
circularity as time passes, MW-mass galaxies usually increase their 
disk fractions at late times through fresh star formation.  Though it 
is not shown here, the disk size $\Rstar$ also tends to grow smoothly 
after $z\sim1$.

The remainder of the sample demonstrates that the interplay between
the SFR and the fraction of stars forming in the disk as a function
of time is also instrumental to the $z=0$ morphology.  During the 
very early of stages of their growth ($t\lesssim2$~Gyr), the hosts
are dwarf-size systems and experience chaotic, bursty star formation
in clumpy, gas-rich dIrr-type progenitor galaxies.  As the gray curves 
show, though, star formation rates are typically relatively low at 
these early times.  The transition to ordered star formation (which 
is strongly correlated with the emergence of an ordered gas disk;
\citealp{Ma2016b,Simons2017,Ceverino2017}) occurs at different times 
and at a different rate in each system, but it often coincides with a 
peak in the star formation.  

Both the timing of the transition and the behavior of the SFR following 
it strongly influences the $z=0$ morphology:  galaxies with 
$\fsdisk(z=0)\gtrsim0.7$ shift to ordered star formation relatively early 
and maintain a relatively high SFR until $z=0$.  Lower $\fsdisk$ galaxies 
either change over later (e.g. \run{m12z}) or have a relatively low SFR 
following the switch (e.g. \run{Remus}).  

\subsection{Disruption and disordering of stellar disks}

Disks built by ordered star formation can also be heated and 
destroyed.  Figure~\ref{fig:birth_v_now} plots the 
disk fraction at $z=0$ against both the cumulative fraction of 
stellar mass born in the galaxy with high $\epsilon$ and the 
maximum instantaneous fraction of stellar mass in the galaxy at
any given time with high circularities.  The scatter in $\fsdbirth$ 
is relatively small: all of the galaxies in our sample have 
$0.6<\fsdbirth<0.9$.  Accounting for counter-rotating star formation, 
all of our systems form $\geq75\%$ of their stars in a disk.  In 
other words, most stars form in disky configurations, as argued 
above.  Furthermore, counter-rotating young stars (which would be
formed out of retrograde star-forming gas) are typically too rare to 
have a significant impact on the galaxy (\S\ref{ssec:misalignment}).  
Though we do not show it here, we also note that the fraction of 
counter-rotating ($\epsilon\leq - 0.5$) stellar mass at $z=0$ is 
extremely small in all but the most bulge-dominated systems.  
Moreover, while the maximum counter-rotating fraction at any given 
time is $\sim15$--$25\%$ across our sample, these maxima occur at 
very early times ($z\gtrsim3$--$4$) when the galaxies were at the 
dwarf mass scale.  The exception is \run{m12b}, which builds a
disk that is initially counter-rotating relative to the existing
bulge but becomes large enough to dominate the angular momentum
of the galaxy and flip the overall $j_z$ vector; \run{m12b} 
therefore maximizes its counter-rotating fraction immediately
before this transition at $z\sim0.5$.
   

Because the galaxy masses and the degree of order in the galaxy 
build up over time, most of our galaxies have $\fsdisk(z=0)\simeq
\max\left[\fsdisk\right]$,  i.e., the majority of our sample is at 
its ``most disk-dominated'' today. 



\begin{figure*}
\centering
\includegraphics[width=0.95\textwidth]{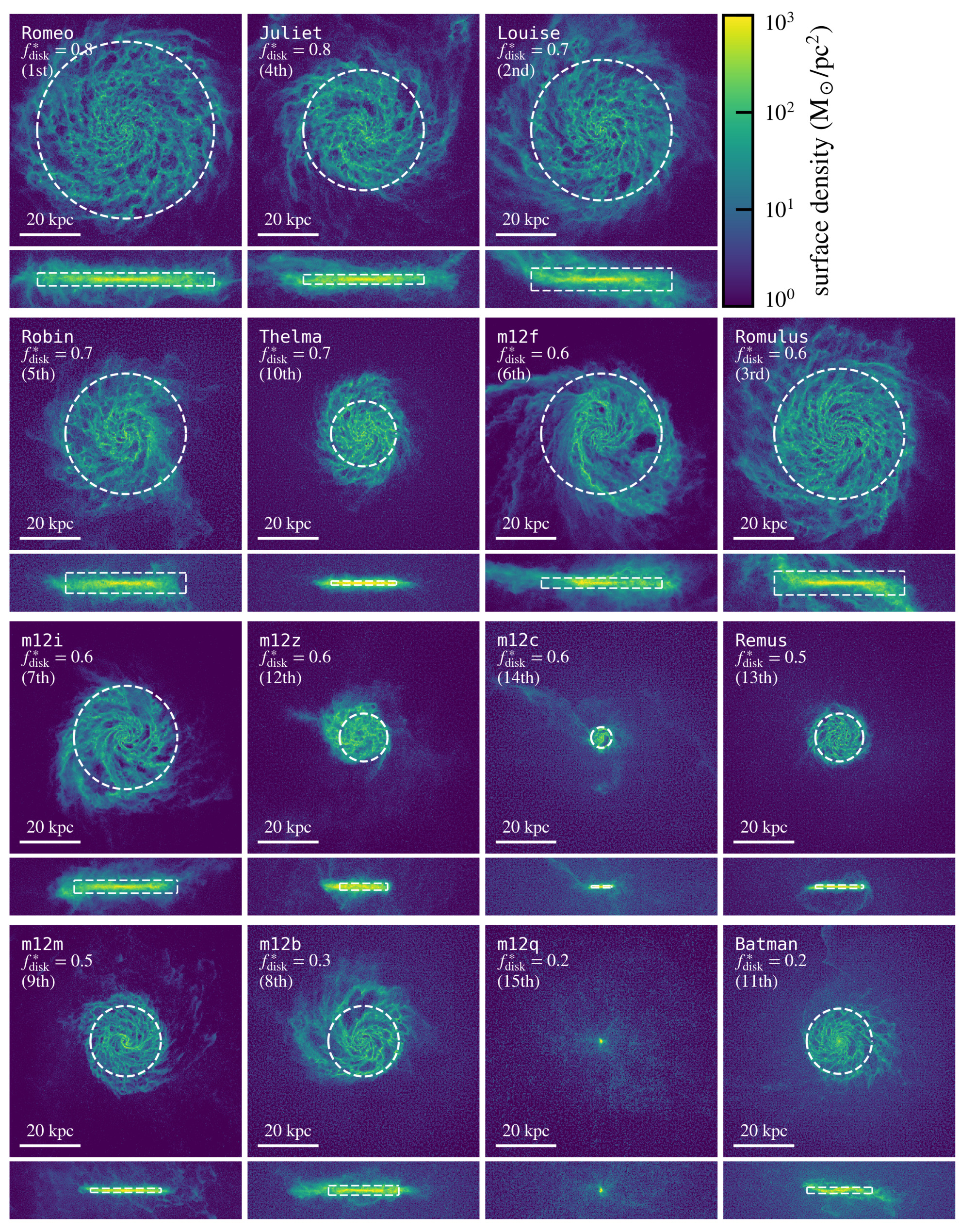}
\caption{Gas column density maps of the galaxies in our sample.  Galaxies 
are again sorted by decreasing $\fsdisk$, but the number in parentheses 
indicates the rank (from largest to smallest) that each galaxy has when 
sorted by $\Rgas$.  \run{Batman} has a significant $\sim20\degree$ warp, 
likely induced because the $z=0$ gas disk is built from an ongoing merger 
that is misaligned with the stars, and \run{m12q} exhausted most of its 
gas supply at $z\sim0.1$; the remainder of the MW-mass FIRE-2 galaxies have 
thin, rotation supported disks ($\fgdisk \geq 0.9$) at $z=0$.  We therefore 
focus on the size of the gas disk, rather than the co-rotation fraction.  
The dashed lines indicate the adopted radial and vertical extents of the 
gas disks; they are not shown for \run{m12q} where they are $<1~\kpc$.}
\label{fig:gas_viz}
\end{figure*}

However, even bulge-dominated galaxies tended to have relatively
strong disks at earlier times before having them destroyed by 
mergers and diluted by misaligned star formation.  At $z\sim1$,
all of \run{Batman}, \run{m12q}, and \run{m12b} had disks that 
comprised 50--70\% of their stellar mass at that time.  That is, 
while bulge-dominated systems arise from a combination of both 
nature and nurture, those in our sample were primarily differentiated 
from disky systems by the latter.  Even though they do have 
more than twice as much retrograde star formation (relative to 
their total stellar mass) as any of the diskier galaxies in our 
sample, the difference between their maximum $\fsdisk$ and their
$z=0$ $\fsdisk$ is much larger, indicating that they were generally 
subject to more disk scrambling than the $z=0$ disk-dominated 
galaxies.

\begin{figure*}
\centering
\includegraphics[width=\textwidth]{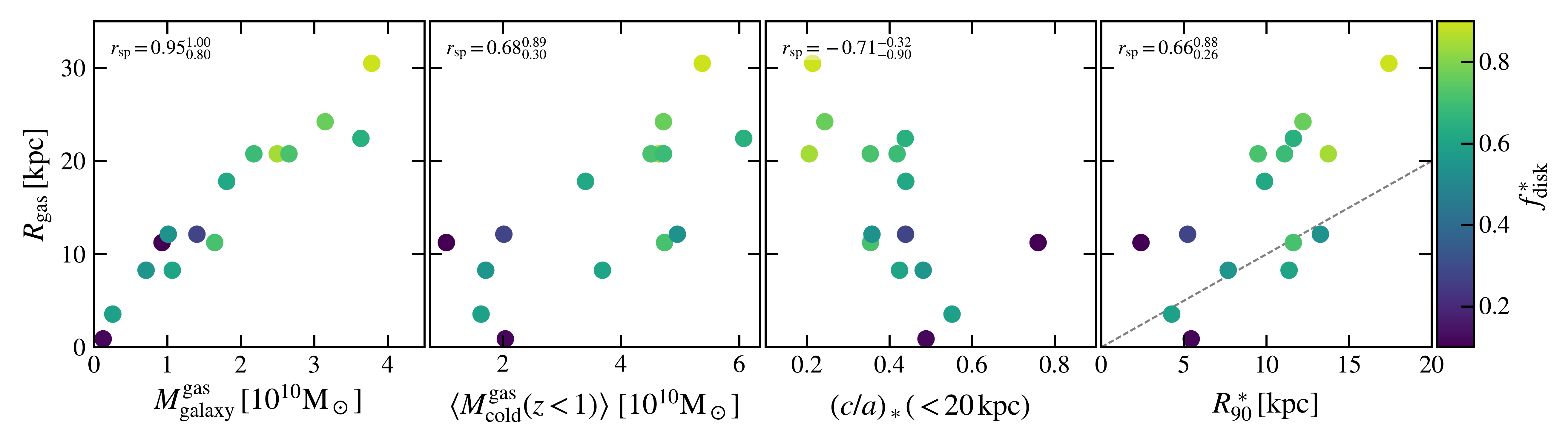}
\caption{The size of the gas disks $\Rgas$ vs the gas mass within
$\Rgas$ and $\Zgas$, $\mgas$; 
the average mass in $T<10^5$~K gas within the virial radius after $z=1$, $\left\langle\mcoldgas(z<1)\right\rangle$;
the ratio
of the shortest to longest principal axes of the stars within $20~\kpc$,
$(c/a)_\mystar(<20~\kpc)$; and the radial extent of the stars, $\Rstar$.  
The dashed gray line in the final column is one-to-one, indicating that 
the majority of the gas disks extend well beyond the stellar disks.  
The points are colored by $\fsdisk$, which also correlates reasonably 
strongly with $\Rgas$.  
}
\label{fig:gas_drivers}
\end{figure*}

Stars may also be shifted to lower circularities, while remaining 
on disk-like orbits, if the overall angular momentum axis of the
galaxy shifts over time.  Most of our sample is only marginally 
impacted by this effect:  the angular momentum axis of our 
galaxies changes by $\lesssim35\degree$ after $z=0.5$ in
all but one of our galaxies.  The exception, moreover, is 
\run{m12b}, which, as described above, builds a large enough young 
disk around the existing compact, bulgy core at late times
to flip the overall angular momentum axis of the galaxy.


As discussed above, however, the $z=0$ morphology is a very weak 
function of mean stellar age:  the most bulge-dominated systems 
do tend to be the oldest, but the youngest systems are not necessarily 
disk-dominated.  Thus, the length of time over which stellar orbits 
can be perturbed is not a primary driver of $z=0$ morphology.  In 
other words, while the bulge-dominated systems in our sample do have 
their pre-existing stellar disks destroyed by mergers or counter-rotating 
star formation, this can happen early or late in cosmic time.


\section{Gas morphologies in MW-mass FIRE-2 galaxies}
\label{sec:gas_morph}

As reflected in \S\ref{sec:evolution} via the stellar circularities
at birth, the star-forming gas in MW-mass galaxies is typically
in a well-ordered disk with the majority of star formation 
occurring with $\abs{\epsilon} \gtrsim 0.75$.  That is, regardless 
of the instantaneous state of the stars in the galaxy, the short 
dynamical memory of the gas leads to thin disks at nearly all times
in massive galaxies.  For a detailed investigation of gas morphologies 
across a larger range of host masses in the FIRE-2 simulations, we 
refer the reader to \citet{KEB2017}, who examined gas angular momentum
and HI morphology as a function of galaxy mass from the dwarf scale 
to the MW scale, and showed that in dwarfs the degree of rotation
support can be much lower.

\subsection{MW-mass galaxies have rotationally supported gas disks}

As demonstrated in Figure~\ref{fig:joverjz}, almost all of the gas 
in the galaxies is on circular, disk-like orbits at $z=0$ (with the 
exception of \run{m12q}).  Even without accounting for elliptical orbits, 
gas rotation curves are consistent with almost complete rotation support 
out to roughly $\Rgas$ in all the galaxies except \run{Batman} and 
\run{m12q}.  In fact, with the exception of \run{m12q}, all of the galaxies 
have $\fgdisk > 0.9$, with nine of the fifteen exceeding $0.98$.  This is 
not particularly surprising at these masses, where pressure support for 
$T\lesssim10^4$~K gas is very weak.


\subsection{Visual morphologies of gas disks}
Figure~\ref{fig:gas_viz} shows face-on and edge-on projections of 
the gas in the FIRE-2 galaxies, again sorted by increasing $\fsdisk$.  
As in Figure~\ref{fig:stellarviz}, the circles and rectangles indicate 
$\Rgas$ and $\Zgas$.  Even with (almost) all of the galaxies having 
$\fgdisk > 0.9$, there exists some diversity in the shape of the 
disks, and even more diversity in the radial extent.  For example, 
\run{m12f} very recently interacted with a gas-rich subhalo, leaving 
a marginally disturbed gas disk at $z=0$.
\run{m12q} has effectively no gas remaining at $z=0$, having consumed 
the last of its gas disk at $z\sim0.1$.  \run{Batman} has a clear 
warp near the center of the disk, likely created because the $z=0$ 
gas disk is formed out of an ongoing accretion event.
\run{Romulus} and \run{Louise} display similar warps near the edges
of their disks.  \run{Batman} is also the only galaxy with a gas disk 
misaligned from the stellar disk by more than $4\degree$.  This 
misalignment presumably survives because the gas is being continually 
replenished at $z\sim0$ \citep{vandeVoort2015}, and likely also because 
\run{Batman} has a relatively spherical stellar distribution:  the ratio 
of the shortest to longest principal axes, $(c/a)$, of the stars within 
$10~\kpc$ is $0.72$.\footnote{The remainder of the sample all have 
$(c/a)^\mystar(<10~\kpc)\lesssim0.5$.}

\subsection{Sizes of gas disks}

Figure~\ref{fig:gas_drivers} explores the radial extent of the gas disks. 
The radius of the gas disk is closely tied to the amount of gas in the 
galaxy, in broad agreement with the observed relationship between the size 
of gas disks and the amount of gas in those disks \citep[e.g.][]{Wang2016}, 
and potentially in agreement with arguments based on the \citet{Toomre1964} 
stability parameter (Schmitz \etal, in prep.).

$\Rgas$ is also correlated with the morphology of the 
stellar component:  the points in Figure~\ref{fig:gas_drivers} 
are colored by $\fsdisk$ and are generally correlated with 
$\Rstar$, though the gas disks are typically more extended.  
To the extent that this is causal, it appears primarily to 
owe to the fact that higher late-time gas masses are associated 
both with larger $\Rgas$ and diskier galaxies.

Though our sample size is small, and we have yet to identify any 
underlying physical drivers, we note galaxies in our paired sample 
tend to have higher $\Rgas$ overall.  This is apparent even by eye in 
Figure~\ref{fig:gas_viz}:  the numbers in parentheses, which indicate 
the rank of $\Rgas$ for each galaxy, show that the five largest gas 
disks are all in halos that reside in a Local Group-like environment.  
However, with such a small sample size, it is impossible to reject the 
null hypothesis that they are drawn from the same distribution and, 
without tying $\Rgas$ to a property of the DMO halo, we cannot directly 
test this hypothesis with a larger sample.  Neither $\Rgas$ nor the 
residuals about a power-law fit of $\Rgas(\mgas)$ strongly correlate 
with any of the DMO halo properties that we have checked, including 
the $z=0$ spin of the DMO halo ($\rspear = -0.21$--$0.79$).  We have also 
tested whether the \citetalias{MoMaoWhite} accurately predicts the 
sizes of \emph{gas} disks at $z=0$ based on the DMO halo, and find 
relatively poor correlations between the model predictions and the 
actual radial extents ($\rspear = -0.05$--$0.79$ for the isothermal 
potential and $\rspear = -0.26$--$0.69$ for the NFW model).

\section{Conclusions}
\label{sec:conclusions}

In this work, we examined the kinematics and morphologies 
of MW-mass galaxies ($10^{10.5}\lesssim\mstar/\msun\lesssim10^{11.5}$)
simulated with the FIRE-2 physics.  Our sample includes fifteen galaxies 
with effective stellar radii ranging from $\sim1~\kpc$~--~$17~\kpc$, and 
kinematic disk fractions varying from $\sim0.2$~--~$0.8$.  We first 
demonstrated that these morphological measures broadly correlate with each 
other (though there is appreciable scatter), and that both also correlate 
with a variety of other morphological measures.  In particular, $\Sigma_1$ 
is a reasonably accurate descriptor of the overall morphology 
over this narrow mass range at $z=0$.

We then showed that the \citet{MoMaoWhite} model, wherein the baryons
that form the disk are assumed to have the same specific angular momentum
of their host DM halos, produces an estimate for galaxy sizes (and how they 
correlate with mass over a large dynamic range) that is accurate that the 
order-of-magnitude level, but fails to recover the actual half-mass radii 
of our galaxies.  This is due primarily to the scatter in the amount of 
specific angular momentum that each galaxy acquires:  $\jd/\md$ has nearly an 
order of magnitude spread overall.  Moreover, there are no obvious trends 
between the morphology of a galaxy and either the mass accretion history or 
the merger history of the host halo in the DMO simulation:  both our most 
bulge-dominated and our most disk-dominated galaxies experience their last 
major merger at $z\sim2$--$3$.  It therefore remains difficult to predict 
the morphology of a galaxy that would form in a given halo based purely 
on the information available from a DMO simulation.


Instead, accurate predictors of morphology within this narrow mass range 
are related to the gas accretion and galaxy merger histories.  Systems 
that maintain high gas fractions to late times tend to be disk-dominated 
at $z=0$ (generally growing inside-out, with $\Rstar$ increasing over time), 
while those that maximize their star forming reservoir early tend to be 
bulge-dominated at $z = 0$.  Based primarily on visual inspection of the 
movies, the amount of gas in the galaxy over time appears to be driven by 
a combination of the impact parameter and timing of galactic mergers, along 
with the amount of gas in the halo that cools and accretes onto the galaxy 
at late times.  We reiterate that our results apply only at this specific 
mass, however:  lower mass galaxies that cannot maintain an ordered gas 
disk at late times \citep[e.g.][]{KEB2017} would not necessarily follow 
the trends identified here.

We find good correlations between morphology and the spin of the gas in the 
halo when the galaxy had formed half of its $z=0$ stellar mass, along with 
the average amount of gas available to form stars at late times.  These 
quantities also correlate (though less strongly) with one another:  gas that 
infalls at later times tends to have more angular momentum.  However, we 
find no clear route from the host properties available only from a DMO 
simulation to galactic morphology:  neither the DMO mass accretion history, 
the half-mass formation time of the halo nor the spin of the dark matter at 
the halo half-mass formation time (either in the hydrodynamic or DMO 
simulation) correlate significantly with morphology, emphasizing the difficulty 
of using DMO simulations to predict morphology.  Nonetheless, our analysis 
does not preclude a multivariate relationship between morphology and DMO 
properties.  In fact, given the correlation between $\lambda_{\rm DMO}$
at $a_\mystar^{50}$ and the morphology at $z=0$, there are hints that such
a relationship may exist, but we lack the sample size to test for those.

The prediction that the spin of the gas in the halo at the stellar 
half-mass formation time (i.e. the angular momentum support of the gas that 
contributes to building the galaxy) drives the late-time morphologies 
of MW-mass galaxies may eventually be observationally falsifiable.  
Wide-field observations with integral field spectroscopy (e.g. with 
instruments similar to the Keck Cosmic Web Imager; \citealp{KCWI}) 
could potentially map out the angular momentum in the cold CGM gas 
and ultimately measure the distribution of that angular momentum 
across halos.  If the picture laid out here is correct, then the 
shapes of that distribution and the distribution of the morphologies
of $z=0$ MW-mass galaxies should broadly agree.

The $z=0$ morphologies can also be viewed as the summation of a Hubble 
time of star formation and subsequent heating of those stars (either 
from mergers, e.g. \citealp{Toomre1972,Hernquist1992,Quinn1993,Sanders1996}, 
or from internal interactions, e.g. \citealp{Minchev2006,Saha2010}).  
We showed that most stars in MW-mass galaxies formed from gas that was 
disky at the time of star formation (i.e.\ with circularity 
$\abs{\epsilon} \geq 0.5$).  The most bulge-dominated galaxies at $z=0$
tend to have the lowest fraction of stars born in a prograde disk 
(and the highest fraction born in a retrograde one), but they also 
show the largest differences between their birth and $z=0$ kinematics.
Therefore, while dispersion supported galaxies arise from a combination 
of birth stellar kinematics and subsequent stellar heating that destroys
ordered rotation, our results suggest the latter effect is far more
important.  

At late times ($z\lesssim1-2$), nearly all of the stars born in MW-mass 
galaxies have disk-like kinematics, such that $\fsdisk$ always grows 
after $z\sim1$.  We do see two exceptions, which actually lower their 
disk fractions (slightly) at $z\lesssim0.5$ by forming stars in a 
counter-rotating disk, but this is only possible because those galaxies 
are already dispersion supported when cold gas is added to the 
central galaxy at low redshift.  Moreover, we emphasize that the 
counter-rotating disks do not \emph{determine} the bulginess of the 
galaxy.  We do not expect ``clump sinking'' to play a significant role 
in bulge formation for systems that are MW-mass at $z=0$ (whose 
progenitors were dwarfs at high redshift). 

The gas in the MW-mass FIRE-2 galaxies, meanwhile, always settles into a 
largely rotation-supported disk at late times.  All but one of our galaxies
maintain that disk to $z=0$, either though fresh accretion from merging 
satellites or condensation out of the CGM.  The size of the gas disk is
primarily driven by its mass.

Our results generally agree with the results of some previous work on the 
formation of galactic disks in hydrodynamic simulations of MW-mass galaxies, 
which have found that star formation is chaotic and bursty at high redshift, 
with well-ordered gas disks only appearing after $z\sim1$ for galaxies with 
MW-masses at $z=0$ (for more/less massive galaxies, the transition occurs 
earlier/later; see \citealp{Muratov2015,Feldmann2015a,Simons2017,Hayward2017,Sparre2017,CAFG2017}.)  
The supply of the gas available to form that disk, therefore, determines the 
amount of stars that form with tangential orbits relative to radial orbits.  
In agreement with several authors \citep[e.g.][]{Scannapieco2009,Sales2012,
Rodriguez-Gomez2017}, we find no strong morphological trends with the $z=0$ 
spin of the DM halo at the MW-mass scale.

While our qualitative results are robust to the mass resolution of the 
simulations (Appendix~\ref{sec:resolution}), the quantitative morphology 
of a given galaxy does change slightly with resolution.  However, these
changes can typically be understood in terms of the trends that we 
identify here (e.g. because a slightly different merger history arises
at different resolutions).  We also caution that half of our sample is
in Local Group-like pairs.  While these are more directly comparable to
the MW and Andromeda galaxies than simulations of isolated MW-mass 
hosts, there may be environmental effects that bias our results.  However,
because these effects should enter via properties of the halo or galaxy, 
such as the halo spin or mass accretion history, our analysis will automatically 
include any changes caused by the 1~Mpc environment.



This work has investigated potential relationships between theoretical measures 
of morphology and (potentially unobservable) physical properties of the simulated 
galaxies in an attempt to understand the physical driver(s) of morphology in the 
simulations, not to compare with observations.  We plan to probe the relationship 
between theoretical morphological measures of the FIRE-2 galaxies, like those 
presented here, and estimates extracted from mock observations, including the 
kinematic distributions \citep[e.g.][]{Zhu2018}, in future work.

\section*{Acknowledgments}

The authors thank Astrid Lamberts, Coral Wheeler, Evan Kirby, 
Laura Sales, and Virginia Kilborn for valuable discussions.  
We also thank the Santa Cruz Galaxy Formation Workshop, the 
Galaxy Formation and Evolution in Southern California 
(GalFRESCA) workshop, and the Swinburne-Caltech workshop for 
spawning useful discussions that significantly improved the 
quality of the manuscript, and we thank Alexander Knebe, Peter 
Behroozi, and Oliver Hahn, respectively, for making \texttt{AHF}, 
\texttt{rockstar} and \texttt{consistent-trees}, and \texttt{MUSIC} 
publicly available.

Support for SGK was provided by NASA through Einstein Postdoctoral Fellowship grant number PF5-160136 awarded by the Chandra X-ray Center, which is operated by the Smithsonian Astrophysical Observatory for NASA under contract NAS8-03060.
Support for PFH was provided by an Alfred P. Sloan Research Fellowship, NSF Collaborative Research Grant \#1715847 and CAREER grant \#1455342. 
AW was supported by a Caltech-Carnegie Fellowship, in part through the Moore Center for Theoretical Cosmology and Physics at Caltech, and by NASA through grants HST-GO-14734 and HST-AR-15057 from STScI.  
RES is supported by an NSF Astronomy \& Astrophysics Postdoctoral Fellowship under grant AST-1400989. 
KEB was supported by a Berkeley graduate fellowship, a Hellman award for graduate study, and an NSF Graduate Research Fellowship.
EQ was supported in part by NSF grant AST-1715070 and a Simons Investigator Award from the Simons Foundation.
JSB was supported by NSF grant AST-1518291 and by NASA through HST theory grants (programs AR-13921, AR-13888, and AR-14282.001) awarded by STScI, which is operated by the Association of Universities for Research in Astronomy (AURA), Inc., under NASA contract NAS5-26555.  
ZH and CAFG were supported by NSF through grants AST-1412836, AST-1517491, AST-1715216, and CAREER award AST-1652522, and by NASA through grant NNX15AB22G,
and ZH additionally acknowledges support from support from Northwestern University through the ``Reach for the Stars'' program.
FvdV acknowledges support from the Klaus Tschira Foundation.
The Flatiron Institute is supported by the Simons Foundation.
DK acknowledges support from NSF grant AST-1412153, NSF grant AST-1715101 and the Cottrell Scholar Award from the Research Corporation for Science Advancement.  
MBK acknowledges support from NSF grant AST-1517226 and from NASA grants NNX17AG29G and HST-AR-13896, HST-AR-14282, HST-AR-14554, HST-GO-12914, and HST-GO-14191 from STScI.

Numerical calculations were run on the Caltech compute cluster ``Wheeler,'' 
allocations from XSEDE TG-AST130039 and PRAC NSF.1713353 supported by the 
NSF, NASA HEC SMD-16-7223 and SMD-16-7592, and High Performance Computing 
at Los Alamos National Labs.  This work also made use of \texttt{Astropy}, 
a community-developed core Python package for Astronomy \citep{Astropy}, 
\texttt{matplotlib} \citep{Matplotlib}, \texttt{numpy} \citep{numpy}, 
\texttt{scipy} \citep{scipy}, \texttt{ipython} \citep{ipython}, and 
NASA's Astrophysics Data System.

\bibliographystyle{mnras}
\bibliography{fire_morphology}

\begin{thebibliography}{}
\makeatletter
\relax
\def\mn@urlcharsother{\let\do\@makeother \do\$\do\&\do\#\do\^\do\_\do\%\do\~}
\def\mn@doi{\begingroup\mn@urlcharsother \@ifnextchar [ {\mn@doi@}
  {\mn@doi@[]}}
\def\mn@doi@[#1]#2{\def\@tempa{#1}\ifx\@tempa\@empty \href
  {http://dx.doi.org/#2} {doi:#2}\else \href {http://dx.doi.org/#2} {#1}\fi
  \endgroup}
\def\mn@eprint#1#2{\mn@eprint@#1:#2::\@nil}
\def\mn@eprint@arXiv#1{\href {http://arxiv.org/abs/#1} {{\tt arXiv:#1}}}
\def\mn@eprint@dblp#1{\href {http://dblp.uni-trier.de/rec/bibtex/#1.xml}
  {dblp:#1}}
\def\mn@eprint@#1:#2:#3:#4\@nil{\def\@tempa {#1}\def\@tempb {#2}\def\@tempc
  {#3}\ifx \@tempc \@empty \let \@tempc \@tempb \let \@tempb \@tempa \fi \ifx
  \@tempb \@empty \def\@tempb {arXiv}\fi \@ifundefined
  {mn@eprint@\@tempb}{\@tempb:\@tempc}{\expandafter \expandafter \csname
  mn@eprint@\@tempb\endcsname \expandafter{\@tempc}}}

\bibitem[\protect\citeauthoryear{{Abadi}, {Navarro}, {Steinmetz}  \&
  {Eke}}{{Abadi} et~al.}{2003}]{Abadi2003}
{Abadi} M.~G.,  {Navarro} J.~F.,  {Steinmetz} M.,   {Eke} V.~R.,  2003, \mn@doi
  [\apj] {10.1086/378316}, \href
  {http://adsabs.harvard.edu/abs/2003ApJ...597...21A} {597, 21}

\bibitem[\protect\citeauthoryear{{Agertz} \& {Kravtsov}}{{Agertz} \&
  {Kravtsov}}{2016}]{Agertz2016}
{Agertz} O.,  {Kravtsov} A.~V.,  2016, \mn@doi [\apj]
  {10.3847/0004-637X/824/2/79}, \href
  {http://adsabs.harvard.edu/abs/2016ApJ...824...79A} {824, 79}

\bibitem[\protect\citeauthoryear{{Agertz}, {Teyssier}  \& {Moore}}{{Agertz}
  et~al.}{2011}]{Agertz2011}
{Agertz} O.,  {Teyssier} R.,   {Moore} B.,  2011, \mn@doi [\mnras]
  {10.1111/j.1365-2966.2010.17530.x}, \href
  {http://adsabs.harvard.edu/abs/2011MNRAS.410.1391A} {410, 1391}

\bibitem[\protect\citeauthoryear{{Angl{\'e}s-Alc{\'a}zar}, {Dav{\'e}},
  {{\"O}zel}  \& {Oppenheimer}}{{Angl{\'e}s-Alc{\'a}zar}
  et~al.}{2014}]{AnglesAlcazar2014}
{Angl{\'e}s-Alc{\'a}zar} D.,  {Dav{\'e}} R.,  {{\"O}zel} F.,   {Oppenheimer}
  B.~D.,  2014, \mn@doi [\apj] {10.1088/0004-637X/782/2/84}, \href
  {http://adsabs.harvard.edu/abs/2014ApJ...782...84A} {782, 84}

\bibitem[\protect\citeauthoryear{{Angl{\'e}s-Alc{\'a}zar},
  {Faucher-Gigu{\`e}re}, {Kere{\v s}}, {Hopkins}, {Quataert}  \&
  {Murray}}{{Angl{\'e}s-Alc{\'a}zar} et~al.}{2017}]{AnglesAcazar2017}
{Angl{\'e}s-Alc{\'a}zar} D.,  {Faucher-Gigu{\`e}re} C.-A.,  {Kere{\v s}} D.,
  {Hopkins} P.~F.,  {Quataert} E.,   {Murray} N.,  2017, \mn@doi [\mnras]
  {10.1093/mnras/stx1517}, \href
  {http://adsabs.harvard.edu/abs/2017MNRAS.470.4698A} {470, 4698}

\bibitem[\protect\citeauthoryear{{Astropy Collaboration} et~al.,}{{Astropy
  Collaboration} et~al.}{2013}]{Astropy}
{Astropy Collaboration} et~al., 2013, \mn@doi [\aap]
  {10.1051/0004-6361/201322068}, \href
  {http://adsabs.harvard.edu/abs/2013A%26A...558A..33A} {558, A33}

\bibitem[\protect\citeauthoryear{{Aumer}, {White}, {Naab}  \&
  {Scannapieco}}{{Aumer} et~al.}{2013}]{Aumer2013}
{Aumer} M.,  {White} S.~D.~M.,  {Naab} T.,   {Scannapieco} C.,  2013, \mn@doi
  [\mnras] {10.1093/mnras/stt1230}, \href
  {http://adsabs.harvard.edu/abs/2013MNRAS.434.3142A} {434, 3142}

\bibitem[\protect\citeauthoryear{{Bamford} et~al.,}{{Bamford}
  et~al.}{2009}]{Bamford2009}
{Bamford} S.~P.,  et~al., 2009, \mn@doi [\mnras]
  {10.1111/j.1365-2966.2008.14252.x}, \href
  {http://adsabs.harvard.edu/abs/2009MNRAS.393.1324B} {393, 1324}

\bibitem[\protect\citeauthoryear{{Barnes} \& {Hernquist}}{{Barnes} \&
  {Hernquist}}{1991}]{Barnes1991}
{Barnes} J.~E.,  {Hernquist} L.~E.,  1991, \mn@doi [\apjl] {10.1086/185978},
  \href {http://adsabs.harvard.edu/abs/1991ApJ...370L..65B} {370, L65}

\bibitem[\protect\citeauthoryear{{Behroozi}, {Wechsler}  \& {Wu}}{{Behroozi}
  et~al.}{2013a}]{rockstar}
{Behroozi} P.~S.,  {Wechsler} R.~H.,   {Wu} H.-Y.,  2013a, \mn@doi [\apj]
  {10.1088/0004-637X/762/2/109}, \href
  {http://adsabs.harvard.edu/abs/2013ApJ...762..109B} {762, 109}

\bibitem[\protect\citeauthoryear{{Behroozi}, {Wechsler}, {Wu}, {Busha},
  {Klypin}  \& {Primack}}{{Behroozi} et~al.}{2013b}]{ctrees}
{Behroozi} P.~S.,  {Wechsler} R.~H.,  {Wu} H.-Y.,  {Busha} M.~T.,  {Klypin}
  A.~A.,   {Primack} J.~R.,  2013b, \mn@doi [\apj]
  {10.1088/0004-637X/763/1/18}, \href
  {http://adsabs.harvard.edu/abs/2013ApJ...763...18B} {763, 18}

\bibitem[\protect\citeauthoryear{{Behroozi}, {Wechsler}  \&
  {Conroy}}{{Behroozi} et~al.}{2013c}]{Behroozi2013}
{Behroozi} P.~S.,  {Wechsler} R.~H.,   {Conroy} C.,  2013c, \mn@doi [\apj]
  {10.1088/0004-637X/770/1/57}, \href
  {http://adsabs.harvard.edu/abs/2013ApJ...770...57B} {770, 57}

\bibitem[\protect\citeauthoryear{{Bell} et~al.,}{{Bell}
  et~al.}{2012}]{Bell2012}
{Bell} E.~F.,  et~al., 2012, \mn@doi [\apj] {10.1088/0004-637X/753/2/167},
  \href {http://adsabs.harvard.edu/abs/2012ApJ...753..167B} {753, 167}

\bibitem[\protect\citeauthoryear{{Bournaud} et~al.,}{{Bournaud}
  et~al.}{2011}]{Bournaud2011}
{Bournaud} F.,  et~al., 2011, \mn@doi [\apj] {10.1088/0004-637X/730/1/4}, \href
  {http://adsabs.harvard.edu/abs/2011ApJ...730....4B} {730, 4}

\bibitem[\protect\citeauthoryear{{Bower}, {Benson}, {Malbon}, {Helly}, {Frenk},
  {Baugh}, {Cole}  \& {Lacey}}{{Bower} et~al.}{2006}]{Bower2006}
{Bower} R.~G.,  {Benson} A.~J.,  {Malbon} R.,  {Helly} J.~C.,  {Frenk} C.~S.,
  {Baugh} C.~M.,  {Cole} S.,   {Lacey} C.~G.,  2006, \mn@doi [\mnras]
  {10.1111/j.1365-2966.2006.10519.x}, \href
  {http://adsabs.harvard.edu/abs/2006MNRAS.370..645B} {370, 645}

\bibitem[\protect\citeauthoryear{{Brooks} \& {Christensen}}{{Brooks} \&
  {Christensen}}{2016}]{Brooks2016}
{Brooks} A.,  {Christensen} C.,  2016, \mn@doi [Galactic Bulges]
  {10.1007/978-3-319-19378-6_12}, \href
  {http://adsabs.harvard.edu/abs/2016ASSL..418..317B} {418, 317}

\bibitem[\protect\citeauthoryear{{Bryan} \& {Norman}}{{Bryan} \&
  {Norman}}{1998}]{Bryan1998}
{Bryan} G.~L.,  {Norman} M.~L.,  1998, \mn@doi [\apj] {10.1086/305262}, \href
  {http://adsabs.harvard.edu/abs/1998ApJ...495...80B} {495, 80}

\bibitem[\protect\citeauthoryear{{Bullock}, {Kravtsov}  \&
  {Weinberg}}{{Bullock} et~al.}{2001}]{Bullock2001}
{Bullock} J.~S.,  {Kravtsov} A.~V.,   {Weinberg} D.~H.,  2001, \mn@doi [\apj]
  {10.1086/318681}, \href {http://adsabs.harvard.edu/abs/2001ApJ...548...33B}
  {548, 33}

\bibitem[\protect\citeauthoryear{{Cattaneo}, {Dekel}, {Devriendt}, {Guiderdoni}
   \& {Blaizot}}{{Cattaneo} et~al.}{2006}]{Cattaneo2006}
{Cattaneo} A.,  {Dekel} A.,  {Devriendt} J.,  {Guiderdoni} B.,   {Blaizot} J.,
  2006, \mn@doi [\mnras] {10.1111/j.1365-2966.2006.10608.x}, \href
  {http://adsabs.harvard.edu/abs/2006MNRAS.370.1651C} {370, 1651}

\bibitem[\protect\citeauthoryear{{Ceverino}, {Primack}, {Dekel}  \&
  {Kassin}}{{Ceverino} et~al.}{2017}]{Ceverino2017}
{Ceverino} D.,  {Primack} J.,  {Dekel} A.,   {Kassin} S.~A.,  2017, \mn@doi
  [\mnras] {10.1093/mnras/stx289}, \href
  {http://adsabs.harvard.edu/abs/2017MNRAS.467.2664C} {467, 2664}

\bibitem[\protect\citeauthoryear{{Chan}, {Kere{\v s}}, {O{\~n}orbe}, {Hopkins},
  {Muratov}, {Faucher-Gigu{\`e}re}  \& {Quataert}}{{Chan}
  et~al.}{2015}]{Chan2015}
{Chan} T.~K.,  {Kere{\v s}} D.,  {O{\~n}orbe} J.,  {Hopkins} P.~F.,  {Muratov}
  A.~L.,  {Faucher-Gigu{\`e}re} C.-A.,   {Quataert} E.,  2015, \mn@doi [\mnras]
  {10.1093/mnras/stv2165}, \href
  {http://adsabs.harvard.edu/abs/2015MNRAS.454.2981C} {454, 2981}

\bibitem[\protect\citeauthoryear{{Cheung} et~al.,}{{Cheung}
  et~al.}{2012}]{Cheung2012}
{Cheung} E.,  et~al., 2012, \mn@doi [\apj] {10.1088/0004-637X/760/2/131}, \href
  {http://adsabs.harvard.edu/abs/2012ApJ...760..131C} {760, 131}

\bibitem[\protect\citeauthoryear{{Col{\'{\i}}n}, {Avila-Reese},
  {Roca-F{\`a}brega}  \& {Valenzuela}}{{Col{\'{\i}}n} et~al.}{2016}]{Colin2016}
{Col{\'{\i}}n} P.,  {Avila-Reese} V.,  {Roca-F{\`a}brega} S.,   {Valenzuela}
  O.,  2016, \mn@doi [\apj] {10.3847/0004-637X/829/2/98}, \href
  {http://adsabs.harvard.edu/abs/2016ApJ...829...98C} {829, 98}

\bibitem[\protect\citeauthoryear{{Crain} et~al.,}{{Crain}
  et~al.}{2009}]{Crain2009}
{Crain} R.~A.,  et~al., 2009, \mn@doi [\mnras]
  {10.1111/j.1365-2966.2009.15402.x}, \href
  {http://adsabs.harvard.edu/abs/2009MNRAS.399.1773C} {399, 1773}

\bibitem[\protect\citeauthoryear{{Croton} et~al.,}{{Croton}
  et~al.}{2006}]{Croton2006}
{Croton} D.~J.,  et~al., 2006, \mn@doi [\mnras]
  {10.1111/j.1365-2966.2005.09675.x}, \href
  {http://adsabs.harvard.edu/abs/2006MNRAS.365...11C} {365, 11}

\bibitem[\protect\citeauthoryear{{Di Cintio}, {Brook}, {Macci{\`o}}, {Stinson},
  {Knebe}, {Dutton}  \& {Wadsley}}{{Di Cintio} et~al.}{2014a}]{diCintio2014a}
{Di Cintio} A.,  {Brook} C.~B.,  {Macci{\`o}} A.~V.,  {Stinson} G.~S.,  {Knebe}
  A.,  {Dutton} A.~A.,   {Wadsley} J.,  2014a, \mn@doi [\mnras]
  {10.1093/mnras/stt1891}, \href
  {http://adsabs.harvard.edu/abs/2014MNRAS.437..415D} {437, 415}

\bibitem[\protect\citeauthoryear{{Di Cintio}, {Brook}, {Dutton}, {Macci{\`o}},
  {Stinson}  \& {Knebe}}{{Di Cintio} et~al.}{2014b}]{DiCintio2014b}
{Di Cintio} A.,  {Brook} C.~B.,  {Dutton} A.~A.,  {Macci{\`o}} A.~V.,
  {Stinson} G.~S.,   {Knebe} A.,  2014b, \mn@doi [\mnras]
  {10.1093/mnras/stu729}, \href
  {http://adsabs.harvard.edu/abs/2014MNRAS.441.2986D} {441, 2986}

\bibitem[\protect\citeauthoryear{{El-Badry}, {Wetzel}, {Geha}, {Hopkins},
  {Kere{\v s}}, {Chan}  \& {Faucher-Gigu{\`e}re}}{{El-Badry}
  et~al.}{2016}]{ElBadry2016}
{El-Badry} K.,  {Wetzel} A.,  {Geha} M.,  {Hopkins} P.~F.,  {Kere{\v s}} D.,
  {Chan} T.~K.,   {Faucher-Gigu{\`e}re} C.-A.,  2016, \mn@doi [\apj]
  {10.3847/0004-637X/820/2/131}, \href
  {http://adsabs.harvard.edu/abs/2016ApJ...820..131E} {820, 131}

\bibitem[\protect\citeauthoryear{{El-Badry} et~al.,}{{El-Badry}
  et~al.}{2017}]{KEB2017}
{El-Badry} K.,  et~al., 2017, preprint, \href
  {http://adsabs.harvard.edu/abs/2017arXiv170510321E} {} (\mn@eprint {arXiv}
  {1705.10321})

\bibitem[\protect\citeauthoryear{{Escala} et~al.,}{{Escala}
  et~al.}{2017}]{Escala2017}
{Escala} I.,  et~al., 2017, preprint, \href
  {http://adsabs.harvard.edu/abs/2017arXiv171006533E} {} (\mn@eprint {arXiv}
  {1710.06533})

\bibitem[\protect\citeauthoryear{{Fall}}{{Fall}}{1983}]{Fall1983}
{Fall} S.~M.,  1983, in {Athanassoula} E.,  ed.,  IAU Symposium Vol. 100,
  Internal Kinematics and Dynamics of Galaxies. pp 391--398

\bibitem[\protect\citeauthoryear{{Fall} \& {Efstathiou}}{{Fall} \&
  {Efstathiou}}{1980}]{Fall1980}
{Fall} S.~M.,  {Efstathiou} G.,  1980, \mn@doi [\mnras]
  {10.1093/mnras/193.2.189}, \href
  {http://adsabs.harvard.edu/abs/1980MNRAS.193..189F} {193, 189}

\bibitem[\protect\citeauthoryear{{Fall} \& {Romanowsky}}{{Fall} \&
  {Romanowsky}}{2013}]{Fall2013}
{Fall} S.~M.,  {Romanowsky} A.~J.,  2013, \mn@doi [\apjl]
  {10.1088/2041-8205/769/2/L26}, \href
  {http://adsabs.harvard.edu/abs/2013ApJ...769L..26F} {769, L26}

\bibitem[\protect\citeauthoryear{{Fang}, {Faber}, {Koo}  \& {Dekel}}{{Fang}
  et~al.}{2013}]{Fang2013}
{Fang} J.~J.,  {Faber} S.~M.,  {Koo} D.~C.,   {Dekel} A.,  2013, \mn@doi [\apj]
  {10.1088/0004-637X/776/1/63}, \href
  {http://adsabs.harvard.edu/abs/2013ApJ...776...63F} {776, 63}

\bibitem[\protect\citeauthoryear{{Faucher-Gigu{\`e}re}}{{Faucher-Gigu{\`e}re}}{2018}]{CAFG2017}
{Faucher-Gigu{\`e}re} C.-A.,  2018, \mn@doi [\mnras] {10.1093/mnras/stx2595},
  \href {http://adsabs.harvard.edu/abs/2018MNRAS.473.3717F} {473, 3717}

\bibitem[\protect\citeauthoryear{{Faucher-Gigu{\`e}re}, {Lidz}, {Zaldarriaga}
  \& {Hernquist}}{{Faucher-Gigu{\`e}re} et~al.}{2009}]{FaucherGiguere2009}
{Faucher-Gigu{\`e}re} C.-A.,  {Lidz} A.,  {Zaldarriaga} M.,   {Hernquist} L.,
  2009, \mn@doi [\apj] {10.1088/0004-637X/703/2/1416}, \href
  {http://adsabs.harvard.edu/abs/2009ApJ...703.1416F} {703, 1416}

\bibitem[\protect\citeauthoryear{{Feldmann} \& {Mayer}}{{Feldmann} \&
  {Mayer}}{2015}]{Feldmann2015a}
{Feldmann} R.,  {Mayer} L.,  2015, \mn@doi [\mnras] {10.1093/mnras/stu2207},
  \href {http://adsabs.harvard.edu/abs/2015MNRAS.446.1939F} {446, 1939}

\bibitem[\protect\citeauthoryear{{Fiacconi}, {Feldmann}  \& {Mayer}}{{Fiacconi}
  et~al.}{2015}]{Fiacconi2015}
{Fiacconi} D.,  {Feldmann} R.,   {Mayer} L.,  2015, \mn@doi [\mnras]
  {10.1093/mnras/stu2228}, \href
  {http://adsabs.harvard.edu/abs/2015MNRAS.446.1957F} {446, 1957}

\bibitem[\protect\citeauthoryear{{F{\"o}rster Schreiber} et~al.,}{{F{\"o}rster
  Schreiber} et~al.}{2011}]{ForsterSchreiber2011}
{F{\"o}rster Schreiber} N.~M.,  et~al., 2011, \mn@doi [\apj]
  {10.1088/0004-637X/739/1/45}, \href
  {http://adsabs.harvard.edu/abs/2011ApJ...739...45F} {739, 45}

\bibitem[\protect\citeauthoryear{{Garrison-Kimmel}, {Boylan-Kolchin}, {Bullock}
   \& {Lee}}{{Garrison-Kimmel} et~al.}{2014}]{ELVIS}
{Garrison-Kimmel} S.,  {Boylan-Kolchin} M.,  {Bullock} J.~S.,   {Lee} K.,
  2014, \mn@doi [\mnras] {10.1093/mnras/stt2377}, \href
  {http://adsabs.harvard.edu/abs/2014MNRAS.438.2578G} {438, 2578}

\bibitem[\protect\citeauthoryear{{Garrison-Kimmel} et~al.,}{{Garrison-Kimmel}
  et~al.}{2017}]{GKDisk}
{Garrison-Kimmel} S.,  et~al., 2017, preprint, \href
  {http://adsabs.harvard.edu/abs/2017arXiv170103792G} {} (\mn@eprint {arXiv}
  {1701.03792})

\bibitem[\protect\citeauthoryear{{Genel}, {Fall}, {Hernquist}, {Vogelsberger},
  {Snyder}, {Rodriguez-Gomez}, {Sijacki}  \& {Springel}}{{Genel}
  et~al.}{2015}]{Genel2015}
{Genel} S.,  {Fall} S.~M.,  {Hernquist} L.,  {Vogelsberger} M.,  {Snyder}
  G.~F.,  {Rodriguez-Gomez} V.,  {Sijacki} D.,   {Springel} V.,  2015, \mn@doi
  [\apjl] {10.1088/2041-8205/804/2/L40}, \href
  {http://adsabs.harvard.edu/abs/2015ApJ...804L..40G} {804, L40}

\bibitem[\protect\citeauthoryear{{Genel} et~al.,}{{Genel}
  et~al.}{2017}]{Genel2017}
{Genel} S.,  et~al., 2017, preprint, \href
  {http://adsabs.harvard.edu/abs/2017arXiv170705327G} {} (\mn@eprint {arXiv}
  {1707.05327})

\bibitem[\protect\citeauthoryear{{Governato}, {Willman}, {Mayer}, {Brooks},
  {Stinson}, {Valenzuela}, {Wadsley}  \& {Quinn}}{{Governato}
  et~al.}{2007}]{Governato2007}
{Governato} F.,  {Willman} B.,  {Mayer} L.,  {Brooks} A.,  {Stinson} G.,
  {Valenzuela} O.,  {Wadsley} J.,   {Quinn} T.,  2007, \mn@doi [\mnras]
  {10.1111/j.1365-2966.2006.11266.x}, \href
  {http://adsabs.harvard.edu/abs/2007MNRAS.374.1479G} {374, 1479}

\bibitem[\protect\citeauthoryear{{Governato} et~al.,}{{Governato}
  et~al.}{2009}]{Governato2009}
{Governato} F.,  et~al., 2009, \mn@doi [\mnras]
  {10.1111/j.1365-2966.2009.15143.x}, \href
  {http://adsabs.harvard.edu/abs/2009MNRAS.398..312G} {398, 312}

\bibitem[\protect\citeauthoryear{{Governato} et~al.,}{{Governato}
  et~al.}{2012}]{Governato2012}
{Governato} F.,  et~al., 2012, \mn@doi [\mnras]
  {10.1111/j.1365-2966.2012.20696.x}, \href
  {http://adsabs.harvard.edu/abs/2012MNRAS.422.1231G} {422, 1231}

\bibitem[\protect\citeauthoryear{{Grand} et~al.,}{{Grand}
  et~al.}{2017}]{Grand2017}
{Grand} R.~J.~J.,  et~al., 2017, \mn@doi [\mnras] {10.1093/mnras/stx071}, \href
  {http://adsabs.harvard.edu/abs/2017MNRAS.467..179G} {467, 179}

\bibitem[\protect\citeauthoryear{{Guedes}, {Callegari}, {Madau}  \&
  {Mayer}}{{Guedes} et~al.}{2011}]{Guedes2011}
{Guedes} J.,  {Callegari} S.,  {Madau} P.,   {Mayer} L.,  2011, \mn@doi [\apj]
  {10.1088/0004-637X/742/2/76}, \href
  {http://adsabs.harvard.edu/abs/2011ApJ...742...76G} {742, 76}

\bibitem[\protect\citeauthoryear{{Hafen} et~al.,}{{Hafen}
  et~al.}{2017}]{Hafen2016}
{Hafen} Z.,  et~al., 2017, \mn@doi [\mnras] {10.1093/mnras/stx952}, \href
  {http://adsabs.harvard.edu/abs/2017MNRAS.469.2292H} {469, 2292}

\bibitem[\protect\citeauthoryear{{Hayward} \& {Hopkins}}{{Hayward} \&
  {Hopkins}}{2017}]{Hayward2017}
{Hayward} C.~C.,  {Hopkins} P.~F.,  2017, \mn@doi [\mnras]
  {10.1093/mnras/stw2888}, \href
  {http://adsabs.harvard.edu/abs/2017MNRAS.465.1682H} {465, 1682}

\bibitem[\protect\citeauthoryear{{Hernquist}}{{Hernquist}}{1992}]{Hernquist1992}
{Hernquist} L.,  1992, \mn@doi [\apj] {10.1086/172009}, \href
  {http://adsabs.harvard.edu/abs/1992ApJ...400..460H} {400, 460}

\bibitem[\protect\citeauthoryear{{Hopkins}}{{Hopkins}}{2015}]{GIZMO}
{Hopkins} P.~F.,  2015, \mn@doi [\mnras] {10.1093/mnras/stv195}, \href
  {http://adsabs.harvard.edu/abs/2015MNRAS.450...53H} {450, 53}

\bibitem[\protect\citeauthoryear{{Hopkins}}{{Hopkins}}{2017}]{Hopkinsmetaldiff}
{Hopkins} P.~F.,  2017, \mn@doi [\mnras] {10.1093/mnras/stw3306}, \href
  {http://adsabs.harvard.edu/abs/2017MNRAS.466.3387H} {466, 3387}

\bibitem[\protect\citeauthoryear{{Hopkins} et~al.,}{{Hopkins}
  et~al.}{2009a}]{Hopkins2009b}
{Hopkins} P.~F.,  et~al., 2009a, \mn@doi [\mnras]
  {10.1111/j.1365-2966.2009.14983.x}, \href
  {http://adsabs.harvard.edu/abs/2009MNRAS.397..802H} {397, 802}

\bibitem[\protect\citeauthoryear{{Hopkins}, {Cox}, {Younger}  \&
  {Hernquist}}{{Hopkins} et~al.}{2009b}]{Hopkins2009a}
{Hopkins} P.~F.,  {Cox} T.~J.,  {Younger} J.~D.,   {Hernquist} L.,  2009b,
  \mn@doi [\apj] {10.1088/0004-637X/691/2/1168}, \href
  {http://adsabs.harvard.edu/abs/2009ApJ...691.1168H} {691, 1168}

\bibitem[\protect\citeauthoryear{{Hopkins} et~al.,}{{Hopkins}
  et~al.}{2010}]{Hopkins2010}
{Hopkins} P.~F.,  et~al., 2010, \mn@doi [\apj] {10.1088/0004-637X/715/1/202},
  \href {http://adsabs.harvard.edu/abs/2010ApJ...715..202H} {715, 202}

\bibitem[\protect\citeauthoryear{{Hopkins}, {Narayanan}  \& {Murray}}{{Hopkins}
  et~al.}{2013}]{Hopkins2013sf_criteria}
{Hopkins} P.~F.,  {Narayanan} D.,   {Murray} N.,  2013, \mn@doi [\mnras]
  {10.1093/mnras/stt723}, \href
  {http://adsabs.harvard.edu/abs/2013MNRAS.432.2647H} {432, 2647}

\bibitem[\protect\citeauthoryear{{Hopkins}, {Kere{\v s}}, {O{\~n}orbe},
  {Faucher-Gigu{\`e}re}, {Quataert}, {Murray}  \& {Bullock}}{{Hopkins}
  et~al.}{2014}]{FIRE}
{Hopkins} P.~F.,  {Kere{\v s}} D.,  {O{\~n}orbe} J.,  {Faucher-Gigu{\`e}re}
  C.-A.,  {Quataert} E.,  {Murray} N.,   {Bullock} J.~S.,  2014, \mn@doi
  [\mnras] {10.1093/mnras/stu1738}, \href
  {http://adsabs.harvard.edu/abs/2014MNRAS.445..581H} {445, 581}

\bibitem[\protect\citeauthoryear{{Hopkins} et~al.,}{{Hopkins}
  et~al.}{2017}]{FIRE2}
{Hopkins} P.~F.,  et~al., 2017, preprint, \href
  {http://adsabs.harvard.edu/abs/2017arXiv170206148H} {} (\mn@eprint {arXiv}
  {1702.06148})

\bibitem[\protect\citeauthoryear{{Howes} et~al.,}{{Howes}
  et~al.}{2015}]{Howes2015}
{Howes} L.~M.,  et~al., 2015, \mn@doi [\nat] {10.1038/nature15747}, \href
  {http://adsabs.harvard.edu/abs/2015Natur.527..484H} {527, 484}

\bibitem[\protect\citeauthoryear{{Hubble}}{{Hubble}}{1926}]{Hubble1926}
{Hubble} E.~P.,  1926, \mn@doi [\apj] {10.1086/143018}, \href
  {http://adsabs.harvard.edu/abs/1926ApJ....64..321H} {64}

\bibitem[\protect\citeauthoryear{{Huertas-Company}, {Aguerri}, {Bernardi},
  {Mei}  \& {S{\'a}nchez Almeida}}{{Huertas-Company}
  et~al.}{2011}]{Huertas-Company2011}
{Huertas-Company} M.,  {Aguerri} J.~A.~L.,  {Bernardi} M.,  {Mei} S.,
  {S{\'a}nchez Almeida} J.,  2011, \mn@doi [\aap]
  {10.1051/0004-6361/201015735}, \href
  {http://adsabs.harvard.edu/abs/2011A%26A...525A.157H} {525, A157}

\bibitem[\protect\citeauthoryear{Hunter}{Hunter}{2007}]{Matplotlib}
Hunter J.~D.,  2007, Computing In Science \& Engineering, 9, 90

\bibitem[\protect\citeauthoryear{Jones, Oliphant, Peterson  et~al.}{Jones
  et~al.}{01  }]{scipy}
Jones E.,  Oliphant T.,  Peterson P.,   et~al., 2001--, {SciPy}: Open source
  scientific tools for {Python}, \url {http://www.scipy.org/}

\bibitem[\protect\citeauthoryear{{Katz} \& {White}}{{Katz} \&
  {White}}{1993}]{Katz1993}
{Katz} N.,  {White} S.~D.~M.,  1993, \mn@doi [\apj] {10.1086/172935}, \href
  {http://adsabs.harvard.edu/abs/1993ApJ...412..455K} {412, 455}

\bibitem[\protect\citeauthoryear{{Kaufmann}, {Wheeler}  \&
  {Bullock}}{{Kaufmann} et~al.}{2007}]{Kaufmann2007}
{Kaufmann} T.,  {Wheeler} C.,   {Bullock} J.~S.,  2007, \mn@doi [\mnras]
  {10.1111/j.1365-2966.2007.12436.x}, \href
  {http://adsabs.harvard.edu/abs/2007MNRAS.382.1187K} {382, 1187}

\bibitem[\protect\citeauthoryear{{Kepner}}{{Kepner}}{1999}]{Kepner1999}
{Kepner} J.~V.,  1999, \mn@doi [\apj] {10.1086/307419}, \href
  {http://adsabs.harvard.edu/abs/1999ApJ...520...59K} {520, 59}

\bibitem[\protect\citeauthoryear{{Kere{\v s}} \& {Hernquist}}{{Kere{\v s}} \&
  {Hernquist}}{2009}]{Keres2009}
{Kere{\v s}} D.,  {Hernquist} L.,  2009, \mn@doi [\apjl]
  {10.1088/0004-637X/700/1/L1}, \href
  {http://adsabs.harvard.edu/abs/2009ApJ...700L...1K} {700, L1}

\bibitem[\protect\citeauthoryear{{Kere{\v s}}, {Katz}, {Weinberg}  \&
  {Dav{\'e}}}{{Kere{\v s}} et~al.}{2005}]{Keres2005}
{Kere{\v s}} D.,  {Katz} N.,  {Weinberg} D.~H.,   {Dav{\'e}} R.,  2005, \mn@doi
  [\mnras] {10.1111/j.1365-2966.2005.09451.x}, \href
  {http://adsabs.harvard.edu/abs/2005MNRAS.363....2K} {363, 2}

\bibitem[\protect\citeauthoryear{{Kravtsov}}{{Kravtsov}}{2013}]{Kravtsov2013}
{Kravtsov} A.~V.,  2013, \mn@doi [\apjl] {10.1088/2041-8205/764/2/L31}, \href
  {http://adsabs.harvard.edu/abs/2013ApJ...764L..31K} {764, L31}

\bibitem[\protect\citeauthoryear{{Kroupa}}{{Kroupa}}{2001}]{Kroupa2001}
{Kroupa} P.,  2001, \mn@doi [\mnras] {10.1046/j.1365-8711.2001.04022.x}, \href
  {http://adsabs.harvard.edu/abs/2001MNRAS.322..231K} {322, 231}

\bibitem[\protect\citeauthoryear{{Krumholz} \& {Gnedin}}{{Krumholz} \&
  {Gnedin}}{2011}]{Krumholz2011}
{Krumholz} M.~R.,  {Gnedin} N.~Y.,  2011, \mn@doi [\apj]
  {10.1088/0004-637X/729/1/36}, \href
  {http://adsabs.harvard.edu/abs/2011ApJ...729...36K} {729, 36}

\bibitem[\protect\citeauthoryear{{Lagos}, {Theuns}, {Stevens}, {Cortese},
  {Padilla}, {Davis}, {Contreras}  \& {Croton}}{{Lagos}
  et~al.}{2017}]{Lagos2017}
{Lagos} C.~d.~P.,  {Theuns} T.,  {Stevens} A.~R.~H.,  {Cortese} L.,  {Padilla}
  N.~D.,  {Davis} T.~A.,  {Contreras} S.,   {Croton} D.,  2017, \mn@doi
  [\mnras] {10.1093/mnras/stw2610}, \href
  {http://adsabs.harvard.edu/abs/2017MNRAS.464.3850L} {464, 3850}

\bibitem[\protect\citeauthoryear{{Larson} et~al.,}{{Larson}
  et~al.}{2011}]{Larson2011}
{Larson} D.,  et~al., 2011, \mn@doi [\apjs] {10.1088/0067-0049/192/2/16}, \href
  {http://adsabs.harvard.edu/abs/2011ApJS..192...16L} {192, 16}

\bibitem[\protect\citeauthoryear{{Leitherer} et~al.,}{{Leitherer}
  et~al.}{1999}]{Leitherer1999}
{Leitherer} C.,  et~al., 1999, \mn@doi [\apjs] {10.1086/313233}, \href
  {http://adsabs.harvard.edu/abs/1999ApJS..123....3L} {123, 3}

\bibitem[\protect\citeauthoryear{{Ludlow}, {Navarro}, {Angulo},
  {Boylan-Kolchin}, {Springel}, {Frenk}  \& {White}}{{Ludlow}
  et~al.}{2014}]{Ludlow2014}
{Ludlow} A.~D.,  {Navarro} J.~F.,  {Angulo} R.~E.,  {Boylan-Kolchin} M.,
  {Springel} V.,  {Frenk} C.,   {White} S.~D.~M.,  2014, \mn@doi [\mnras]
  {10.1093/mnras/stu483}, \href
  {http://adsabs.harvard.edu/abs/2014MNRAS.441..378L} {441, 378}

\bibitem[\protect\citeauthoryear{{Ma}, {Hopkins}, {Faucher-Gigu{\`e}re},
  {Zolman}, {Muratov}, {Kere{\v s}}  \& {Quataert}}{{Ma}
  et~al.}{2016}]{MaMassMetallicity}
{Ma} X.,  {Hopkins} P.~F.,  {Faucher-Gigu{\`e}re} C.-A.,  {Zolman} N.,
  {Muratov} A.~L.,  {Kere{\v s}} D.,   {Quataert} E.,  2016, \mn@doi [\mnras]
  {10.1093/mnras/stv2659}, \href
  {http://adsabs.harvard.edu/abs/2016MNRAS.456.2140M} {456, 2140}

\bibitem[\protect\citeauthoryear{{Ma}, {Hopkins}, {Feldmann}, {Torrey},
  {Faucher-Gigu{\`e}re}  \& {Kere{\v s}}}{{Ma} et~al.}{2017a}]{Mametalgrad}
{Ma} X.,  {Hopkins} P.~F.,  {Feldmann} R.,  {Torrey} P.,  {Faucher-Gigu{\`e}re}
  C.-A.,   {Kere{\v s}} D.,  2017a, \mn@doi [\mnras] {10.1093/mnras/stw034},
  \href {http://adsabs.harvard.edu/abs/2017MNRAS.466.4780M} {466, 4780}

\bibitem[\protect\citeauthoryear{{Ma}, {Hopkins}, {Wetzel}, {Kirby},
  {Angl{\'e}s-Alc{\'a}zar}, {Faucher-Gigu{\`e}re}, {Kere{\v s}}  \&
  {Quataert}}{{Ma} et~al.}{2017b}]{Ma2016b}
{Ma} X.,  {Hopkins} P.~F.,  {Wetzel} A.~R.,  {Kirby} E.~N.,
  {Angl{\'e}s-Alc{\'a}zar} D.,  {Faucher-Gigu{\`e}re} C.-A.,  {Kere{\v s}} D.,
   {Quataert} E.,  2017b, \mn@doi [\mnras] {10.1093/mnras/stx273}, \href
  {http://adsabs.harvard.edu/abs/2017MNRAS.467.2430M} {467, 2430}

\bibitem[\protect\citeauthoryear{{Mandelker}, {Dekel}, {Ceverino}, {DeGraf},
  {Guo}  \& {Primack}}{{Mandelker} et~al.}{2017}]{Mandelker2017}
{Mandelker} N.,  {Dekel} A.,  {Ceverino} D.,  {DeGraf} C.,  {Guo} Y.,
  {Primack} J.,  2017, \mn@doi [\mnras] {10.1093/mnras/stw2358}, \href
  {http://adsabs.harvard.edu/abs/2017MNRAS.464..635M} {464, 635}

\bibitem[\protect\citeauthoryear{{Marinacci}, {Pakmor}  \&
  {Springel}}{{Marinacci} et~al.}{2014}]{Marinacci2014}
{Marinacci} F.,  {Pakmor} R.,   {Springel} V.,  2014, \mn@doi [\mnras]
  {10.1093/mnras/stt2003}, \href
  {http://adsabs.harvard.edu/abs/2014MNRAS.437.1750M} {437, 1750}

\bibitem[\protect\citeauthoryear{{Merritt}, {van Dokkum}, {Abraham}  \&
  {Zhang}}{{Merritt} et~al.}{2016}]{Merritt2016}
{Merritt} A.,  {van Dokkum} P.,  {Abraham} R.,   {Zhang} J.,  2016, \mn@doi
  [\apj] {10.3847/0004-637X/830/2/62}, \href
  {http://adsabs.harvard.edu/abs/2016ApJ...830...62M} {830, 62}

\bibitem[\protect\citeauthoryear{{Minchev} \& {Quillen}}{{Minchev} \&
  {Quillen}}{2006}]{Minchev2006}
{Minchev} I.,  {Quillen} A.~C.,  2006, \mn@doi [\mnras]
  {10.1111/j.1365-2966.2006.10129.x}, \href
  {http://adsabs.harvard.edu/abs/2006MNRAS.368..623M} {368, 623}

\bibitem[\protect\citeauthoryear{{Mo}, {Mao}  \& {White}}{{Mo}
  et~al.}{1998}]{MoMaoWhite}
{Mo} H.~J.,  {Mao} S.,   {White} S.~D.~M.,  1998, \mn@doi [\mnras]
  {10.1046/j.1365-8711.1998.01227.x}, \href
  {http://adsabs.harvard.edu/abs/1998MNRAS.295..319M} {295, 319}

\bibitem[\protect\citeauthoryear{{Morrissey} \& {KCWI Team}}{{Morrissey} \&
  {KCWI Team}}{2013}]{KCWI}
{Morrissey} P.,  {KCWI Team} 2013, in American Astronomical Society Meeting
  Abstracts \#221. p. 345.05

\bibitem[\protect\citeauthoryear{{Murante}, {Monaco}, {Borgani}, {Tornatore},
  {Dolag}  \& {Goz}}{{Murante} et~al.}{2015}]{Murante2015}
{Murante} G.,  {Monaco} P.,  {Borgani} S.,  {Tornatore} L.,  {Dolag} K.,
  {Goz} D.,  2015, \mn@doi [\mnras] {10.1093/mnras/stu2400}, \href
  {http://adsabs.harvard.edu/abs/2015MNRAS.447..178M} {447, 178}

\bibitem[\protect\citeauthoryear{{Muratov}, {Kere{\v s}},
  {Faucher-Gigu{\`e}re}, {Hopkins}, {Quataert}  \& {Murray}}{{Muratov}
  et~al.}{2015}]{Muratov2015}
{Muratov} A.~L.,  {Kere{\v s}} D.,  {Faucher-Gigu{\`e}re} C.-A.,  {Hopkins}
  P.~F.,  {Quataert} E.,   {Murray} N.,  2015, \mn@doi [\mnras]
  {10.1093/mnras/stv2126}, \href
  {http://adsabs.harvard.edu/abs/2015MNRAS.454.2691M} {454, 2691}

\bibitem[\protect\citeauthoryear{{Navarro}, {Frenk}  \& {White}}{{Navarro}
  et~al.}{1996}]{Navarro1996}
{Navarro} J.~F.,  {Frenk} C.~S.,   {White} S.~D.~M.,  1996, \mn@doi [\apj]
  {10.1086/177173}, \href {http://adsabs.harvard.edu/abs/1996ApJ...462..563N}
  {462, 563}

\bibitem[\protect\citeauthoryear{{O{\~n}orbe}, {Garrison-Kimmel}, {Maller},
  {Bullock}, {Rocha}  \& {Hahn}}{{O{\~n}orbe} et~al.}{2014}]{Onorbe2014}
{O{\~n}orbe} J.,  {Garrison-Kimmel} S.,  {Maller} A.~H.,  {Bullock} J.~S.,
  {Rocha} M.,   {Hahn} O.,  2014, \mn@doi [\mnras] {10.1093/mnras/stt2020},
  \href {http://adsabs.harvard.edu/abs/2014MNRAS.437.1894O} {437, 1894}

\bibitem[\protect\citeauthoryear{{O{\~n}orbe}, {Boylan-Kolchin}, {Bullock},
  {Hopkins}, {Kere{\v s}}, {Faucher-Gigu{\`e}re}, {Quataert}  \&
  {Murray}}{{O{\~n}orbe} et~al.}{2015}]{Onorbe2015}
{O{\~n}orbe} J.,  {Boylan-Kolchin} M.,  {Bullock} J.~S.,  {Hopkins} P.~F.,
  {Kere{\v s}} D.,  {Faucher-Gigu{\`e}re} C.-A.,  {Quataert} E.,   {Murray} N.,
   2015, \mn@doi [\mnras] {10.1093/mnras/stv2072}, \href
  {http://adsabs.harvard.edu/abs/2015MNRAS.454.2092O} {454, 2092}

\bibitem[\protect\citeauthoryear{{Okamoto}, {Eke}, {Frenk}  \&
  {Jenkins}}{{Okamoto} et~al.}{2005}]{Okamoto2005}
{Okamoto} T.,  {Eke} V.~R.,  {Frenk} C.~S.,   {Jenkins} A.,  2005, \mn@doi
  [\mnras] {10.1111/j.1365-2966.2005.09525.x}, \href
  {http://adsabs.harvard.edu/abs/2005MNRAS.363.1299O} {363, 1299}

\bibitem[\protect\citeauthoryear{{Oklop{\v c}i{\'c}}, {Hopkins}, {Feldmann},
  {Kere{\v s}}, {Faucher-Gigu{\`e}re}  \& {Murray}}{{Oklop{\v c}i{\'c}}
  et~al.}{2017}]{Oklopcic2017}
{Oklop{\v c}i{\'c}} A.,  {Hopkins} P.~F.,  {Feldmann} R.,  {Kere{\v s}} D.,
  {Faucher-Gigu{\`e}re} C.-A.,   {Murray} N.,  2017, \mn@doi [\mnras]
  {10.1093/mnras/stw2754}, \href
  {http://adsabs.harvard.edu/abs/2017MNRAS.465..952O} {465, 952}

\bibitem[\protect\citeauthoryear{{Orr} et~al.,}{{Orr} et~al.}{2017}]{Orr2017}
{Orr} M.~E.,  et~al., 2017, preprint, \href
  {http://adsabs.harvard.edu/abs/2017arXiv170101788O} {} (\mn@eprint {arXiv}
  {1701.01788})

\bibitem[\protect\citeauthoryear{{Pedrosa} \& {Tissera}}{{Pedrosa} \&
  {Tissera}}{2015}]{Pedrosa2015}
{Pedrosa} S.~E.,  {Tissera} P.~B.,  2015, \mn@doi [\aap]
  {10.1051/0004-6361/201526440}, \href
  {http://adsabs.harvard.edu/abs/2015A%26A...584A..43P} {584, A43}

\bibitem[\protect\citeauthoryear{{Pedrosa}, {Tissera}  \& {De Rossi}}{{Pedrosa}
  et~al.}{2014}]{Pedrosa2014}
{Pedrosa} S.~E.,  {Tissera} P.~B.,   {De Rossi} M.~E.,  2014, \mn@doi [\aap]
  {10.1051/0004-6361/201323079}, \href
  {http://adsabs.harvard.edu/abs/2014A%26A...567A..47P} {567, A47}

\bibitem[\protect\citeauthoryear{{Peebles}}{{Peebles}}{1969}]{PeeblesSpin}
{Peebles} P.~J.~E.,  1969, \mn@doi [\apj] {10.1086/149876}, \href
  {http://adsabs.harvard.edu/abs/1969ApJ...155..393P} {155, 393}

\bibitem[\protect\citeauthoryear{{Peirani}, {Mohayaee}  \& {de Freitas
  Pacheco}}{{Peirani} et~al.}{2004}]{Peirani2004}
{Peirani} S.,  {Mohayaee} R.,   {de Freitas Pacheco} J.~A.,  2004, \mn@doi
  [\mnras] {10.1111/j.1365-2966.2004.07412.x}, \href
  {http://adsabs.harvard.edu/abs/2004MNRAS.348..921P} {348, 921}

\bibitem[\protect\citeauthoryear{Perez \& Granger}{Perez \&
  Granger}{2007}]{ipython}
Perez F.,  Granger B.~E.,  2007, \mn@doi [Computing in Science Engineering]
  {10.1109/MCSE.2007.53}, 9, 21

\bibitem[\protect\citeauthoryear{{Pichon}, {Pogosyan}, {Kimm}, {Slyz},
  {Devriendt}  \& {Dubois}}{{Pichon} et~al.}{2011}]{Pichon2011}
{Pichon} C.,  {Pogosyan} D.,  {Kimm} T.,  {Slyz} A.,  {Devriendt} J.,
  {Dubois} Y.,  2011, \mn@doi [\mnras] {10.1111/j.1365-2966.2011.19640.x},
  \href {http://adsabs.harvard.edu/abs/2011MNRAS.418.2493P} {418, 2493}

\bibitem[\protect\citeauthoryear{{Pilkington} et~al.,}{{Pilkington}
  et~al.}{2012}]{Pilkington2012}
{Pilkington} K.,  et~al., 2012, \mn@doi [\aap] {10.1051/0004-6361/201117466},
  \href {http://adsabs.harvard.edu/abs/2012A%26A...540A..56P} {540, A56}

\bibitem[\protect\citeauthoryear{{Pillepich} et~al.,}{{Pillepich}
  et~al.}{2017}]{IllustrisTNG1}
{Pillepich} A.,  et~al., 2017, preprint, \href
  {http://adsabs.harvard.edu/abs/2017arXiv170302970P} {} (\mn@eprint {arXiv}
  {1703.02970})

\bibitem[\protect\citeauthoryear{{Planck Collaboration} et~al.,}{{Planck
  Collaboration} et~al.}{2016}]{Planck15}
{Planck Collaboration} et~al., 2016, \mn@doi [\aap]
  {10.1051/0004-6361/201525830}, \href
  {http://adsabs.harvard.edu/abs/2016A%26A...594A..13P} {594, A13}

\bibitem[\protect\citeauthoryear{{Pontzen} \& {Governato}}{{Pontzen} \&
  {Governato}}{2012}]{Pontzen2012}
{Pontzen} A.,  {Governato} F.,  2012, \mn@doi [\mnras]
  {10.1111/j.1365-2966.2012.20571.x}, \href
  {http://adsabs.harvard.edu/abs/2012MNRAS.421.3464P} {421, 3464}

\bibitem[\protect\citeauthoryear{{Power}, {Navarro}, {Jenkins}, {Frenk},
  {White}, {Springel}, {Stadel}  \& {Quinn}}{{Power} et~al.}{2003}]{Power2003}
{Power} C.,  {Navarro} J.~F.,  {Jenkins} A.,  {Frenk} C.~S.,  {White} S.~D.~M.,
   {Springel} V.,  {Stadel} J.,   {Quinn} T.,  2003, \mn@doi [\mnras]
  {10.1046/j.1365-8711.2003.05925.x}, \href
  {http://adsabs.harvard.edu/abs/2003MNRAS.338...14P} {338, 14}

\bibitem[\protect\citeauthoryear{{Price} \& {Monaghan}}{{Price} \&
  {Monaghan}}{2007}]{Price2007}
{Price} D.~J.,  {Monaghan} J.~J.,  2007, \mn@doi [\mnras]
  {10.1111/j.1365-2966.2006.11241.x}, \href
  {http://adsabs.harvard.edu/abs/2007MNRAS.374.1347P} {374, 1347}

\bibitem[\protect\citeauthoryear{{Quinn}, {Hernquist}  \& {Fullagar}}{{Quinn}
  et~al.}{1993}]{Quinn1993}
{Quinn} P.~J.,  {Hernquist} L.,   {Fullagar} D.~P.,  1993, \mn@doi [\apj]
  {10.1086/172184}, \href {http://adsabs.harvard.edu/abs/1993ApJ...403...74Q}
  {403, 74}

\bibitem[\protect\citeauthoryear{{Robertson} \& {Bullock}}{{Robertson} \&
  {Bullock}}{2008}]{Robertson2008}
{Robertson} B.~E.,  {Bullock} J.~S.,  2008, \mn@doi [\apjl] {10.1086/592329},
  \href {http://adsabs.harvard.edu/abs/2008ApJ...685L..27R} {685, L27}

\bibitem[\protect\citeauthoryear{{Robertson}, {Bullock}, {Cox}, {Di Matteo},
  {Hernquist}, {Springel}  \& {Yoshida}}{{Robertson}
  et~al.}{2006}]{Robertson2006}
{Robertson} B.,  {Bullock} J.~S.,  {Cox} T.~J.,  {Di Matteo} T.,  {Hernquist}
  L.,  {Springel} V.,   {Yoshida} N.,  2006, \mn@doi [\apj] {10.1086/504412},
  \href {http://adsabs.harvard.edu/abs/2006ApJ...645..986R} {645, 986}

\bibitem[\protect\citeauthoryear{{Rodriguez-Gomez} et~al.,}{{Rodriguez-Gomez}
  et~al.}{2017}]{Rodriguez-Gomez2017}
{Rodriguez-Gomez} V.,  et~al., 2017, \mn@doi [\mnras] {10.1093/mnras/stx305},
  \href {http://adsabs.harvard.edu/abs/2017MNRAS.467.3083R} {467, 3083}

\bibitem[\protect\citeauthoryear{{Rodr{\'{\i}}guez-Puebla}, {Primack},
  {Avila-Reese}  \& {Faber}}{{Rodr{\'{\i}}guez-Puebla}
  et~al.}{2017}]{Rodriguez-Puebla2017}
{Rodr{\'{\i}}guez-Puebla} A.,  {Primack} J.~R.,  {Avila-Reese} V.,   {Faber}
  S.~M.,  2017, \mn@doi [\mnras] {10.1093/mnras/stx1172}, \href
  {http://adsabs.harvard.edu/abs/2017MNRAS.470..651R} {470, 651}

\bibitem[\protect\citeauthoryear{{Romanowsky} \& {Fall}}{{Romanowsky} \&
  {Fall}}{2012}]{Romanowsky2012}
{Romanowsky} A.~J.,  {Fall} S.~M.,  2012, \mn@doi [\apjs]
  {10.1088/0067-0049/203/2/17}, \href
  {http://adsabs.harvard.edu/abs/2012ApJS..203...17R} {203, 17}

\bibitem[\protect\citeauthoryear{{Ro{\v s}kar}, {Teyssier}, {Agertz},
  {Wetzstein}  \& {Moore}}{{Ro{\v s}kar} et~al.}{2014}]{Roskar2014}
{Ro{\v s}kar} R.,  {Teyssier} R.,  {Agertz} O.,  {Wetzstein} M.,   {Moore} B.,
  2014, \mn@doi [\mnras] {10.1093/mnras/stu1548}, \href
  {http://adsabs.harvard.edu/abs/2014MNRAS.444.2837R} {444, 2837}

\bibitem[\protect\citeauthoryear{{Saha}, {Tseng}  \& {Taam}}{{Saha}
  et~al.}{2010}]{Saha2010}
{Saha} K.,  {Tseng} Y.-H.,   {Taam} R.~E.,  2010, \mn@doi [\apj]
  {10.1088/0004-637X/721/2/1878}, \href
  {http://adsabs.harvard.edu/abs/2010ApJ...721.1878S} {721, 1878}

\bibitem[\protect\citeauthoryear{{Sales}, {Navarro}, {Theuns}, {Schaye},
  {White}, {Frenk}, {Crain}  \& {Dalla Vecchia}}{{Sales}
  et~al.}{2012}]{Sales2012}
{Sales} L.~V.,  {Navarro} J.~F.,  {Theuns} T.,  {Schaye} J.,  {White} S.~D.~M.,
   {Frenk} C.~S.,  {Crain} R.~A.,   {Dalla Vecchia} C.,  2012, \mn@doi [\mnras]
  {10.1111/j.1365-2966.2012.20975.x}, \href
  {http://adsabs.harvard.edu/abs/2012MNRAS.423.1544S} {423, 1544}

\bibitem[\protect\citeauthoryear{{Salim} et~al.,}{{Salim}
  et~al.}{2007}]{Salim2007}
{Salim} S.,  et~al., 2007, \mn@doi [\apjs] {10.1086/519218}, \href
  {http://adsabs.harvard.edu/abs/2007ApJS..173..267S} {173, 267}

\bibitem[\protect\citeauthoryear{{Sanders} \& {Mirabel}}{{Sanders} \&
  {Mirabel}}{1996}]{Sanders1996}
{Sanders} D.~B.,  {Mirabel} I.~F.,  1996, \mn@doi [\araa]
  {10.1146/annurev.astro.34.1.749}, \href
  {http://adsabs.harvard.edu/abs/1996ARA%26A..34..749S} {34, 749}

\bibitem[\protect\citeauthoryear{{Scannapieco}, {Tissera}, {White}  \&
  {Springel}}{{Scannapieco} et~al.}{2008}]{Scannapieco2008}
{Scannapieco} C.,  {Tissera} P.~B.,  {White} S.~D.~M.,   {Springel} V.,  2008,
  \mn@doi [\mnras] {10.1111/j.1365-2966.2008.13678.x}, \href
  {http://adsabs.harvard.edu/abs/2008MNRAS.389.1137S} {389, 1137}

\bibitem[\protect\citeauthoryear{{Scannapieco}, {White}, {Springel}  \&
  {Tissera}}{{Scannapieco} et~al.}{2009}]{Scannapieco2009}
{Scannapieco} C.,  {White} S.~D.~M.,  {Springel} V.,   {Tissera} P.~B.,  2009,
  \mn@doi [\mnras] {10.1111/j.1365-2966.2009.14764.x}, \href
  {http://adsabs.harvard.edu/abs/2009MNRAS.396..696S} {396, 696}

\bibitem[\protect\citeauthoryear{{Scannapieco}, {Gadotti}, {Jonsson}  \&
  {White}}{{Scannapieco} et~al.}{2010}]{Scannapieco2010}
{Scannapieco} C.,  {Gadotti} D.~A.,  {Jonsson} P.,   {White} S.~D.~M.,  2010,
  \mn@doi [\mnras] {10.1111/j.1745-3933.2010.00900.x}, \href
  {http://adsabs.harvard.edu/abs/2010MNRAS.407L..41S} {407, L41}

\bibitem[\protect\citeauthoryear{{Schaye} et~al.,}{{Schaye}
  et~al.}{2015}]{EAGLE}
{Schaye} J.,  et~al., 2015, \mn@doi [\mnras] {10.1093/mnras/stu2058}, \href
  {http://adsabs.harvard.edu/abs/2015MNRAS.446..521S} {446, 521}

\bibitem[\protect\citeauthoryear{{S{\'e}rsic}}{{S{\'e}rsic}}{1963}]{Sersic1963}
{S{\'e}rsic} J.~L.,  1963, Boletin de la Asociacion Argentina de Astronomia La
  Plata Argentina, \href {http://adsabs.harvard.edu/abs/1963BAAA....6...41S}
  {6, 41}

\bibitem[\protect\citeauthoryear{{Shen}, {Mo}, {White}, {Blanton}, {Kauffmann},
  {Voges}, {Brinkmann}  \& {Csabai}}{{Shen} et~al.}{2003}]{Shen2003}
{Shen} S.,  {Mo} H.~J.,  {White} S.~D.~M.,  {Blanton} M.~R.,  {Kauffmann} G.,
  {Voges} W.,  {Brinkmann} J.,   {Csabai} I.,  2003, \mn@doi [\mnras]
  {10.1046/j.1365-8711.2003.06740.x}, \href
  {http://adsabs.harvard.edu/abs/2003MNRAS.343..978S} {343, 978}

\bibitem[\protect\citeauthoryear{{Simons}, {Kassin}, {Weiner}, {Heckman},
  {Lee}, {Lotz}, {Peth}  \& {Tchernyshyov}}{{Simons} et~al.}{2015}]{Simons2015}
{Simons} R.~C.,  {Kassin} S.~A.,  {Weiner} B.~J.,  {Heckman} T.~M.,  {Lee}
  J.~C.,  {Lotz} J.~M.,  {Peth} M.,   {Tchernyshyov} K.,  2015, \mn@doi
  [\mnras] {10.1093/mnras/stv1298}, \href
  {http://adsabs.harvard.edu/abs/2015MNRAS.452..986S} {452, 986}

\bibitem[\protect\citeauthoryear{{Simons} et~al.,}{{Simons}
  et~al.}{2017}]{Simons2017}
{Simons} R.~C.,  et~al., 2017, \mn@doi [\apj] {10.3847/1538-4357/aa740c}, \href
  {http://adsabs.harvard.edu/abs/2017ApJ...843...46S} {843, 46}

\bibitem[\protect\citeauthoryear{{Soko{\l}owska}, {Capelo}, {Fall}, {Mayer},
  {Shen}  \& {Bonoli}}{{Soko{\l}owska} et~al.}{2017}]{Sokolowska2017}
{Soko{\l}owska} A.,  {Capelo} P.~R.,  {Fall} S.~M.,  {Mayer} L.,  {Shen} S.,
  {Bonoli} S.,  2017, \mn@doi [\apj] {10.3847/1538-4357/835/2/289}, \href
  {http://adsabs.harvard.edu/abs/2017ApJ...835..289S} {835, 289}

\bibitem[\protect\citeauthoryear{{Somerville}, {Hopkins}, {Cox}, {Robertson}
  \& {Hernquist}}{{Somerville} et~al.}{2008}]{Somerville2008}
{Somerville} R.~S.,  {Hopkins} P.~F.,  {Cox} T.~J.,  {Robertson} B.~E.,
  {Hernquist} L.,  2008, \mn@doi [\mnras] {10.1111/j.1365-2966.2008.13805.x},
  \href {http://adsabs.harvard.edu/abs/2008MNRAS.391..481S} {391, 481}

\bibitem[\protect\citeauthoryear{{Sparre}, {Hayward}, {Feldmann},
  {Faucher-Gigu{\`e}re}, {Muratov}, {Kere{\v s}}  \& {Hopkins}}{{Sparre}
  et~al.}{2017}]{Sparre2017}
{Sparre} M.,  {Hayward} C.~C.,  {Feldmann} R.,  {Faucher-Gigu{\`e}re} C.-A.,
  {Muratov} A.~L.,  {Kere{\v s}} D.,   {Hopkins} P.~F.,  2017, \mn@doi [\mnras]
  {10.1093/mnras/stw3011}, \href
  {http://adsabs.harvard.edu/abs/2017MNRAS.466...88S} {466, 88}

\bibitem[\protect\citeauthoryear{{Springel}}{{Springel}}{2005}]{Springel2005}
{Springel} V.,  2005, \mn@doi [\mnras] {10.1111/j.1365-2966.2005.09655.x},
  \href {http://adsabs.harvard.edu/abs/2005MNRAS.364.1105S} {364, 1105}

\bibitem[\protect\citeauthoryear{{Springel} \& {Hernquist}}{{Springel} \&
  {Hernquist}}{2005}]{SpringelHernquist2005}
{Springel} V.,  {Hernquist} L.,  2005, \mn@doi [\apjl] {10.1086/429486}, \href
  {http://adsabs.harvard.edu/abs/2005ApJ...622L...9S} {622, L9}

\bibitem[\protect\citeauthoryear{{Springel} et~al.,}{{Springel}
  et~al.}{2005}]{millennium}
{Springel} V.,  et~al., 2005, \mn@doi [\nat] {10.1038/nature03597}, \href
  {http://adsabs.harvard.edu/abs/2005Natur.435..629S} {435, 629}

\bibitem[\protect\citeauthoryear{{Stewart}, {Bullock}, {Wechsler}  \&
  {Maller}}{{Stewart} et~al.}{2009}]{Stewart2009}
{Stewart} K.~R.,  {Bullock} J.~S.,  {Wechsler} R.~H.,   {Maller} A.~H.,  2009,
  \mn@doi [\apj] {10.1088/0004-637X/702/1/307}, \href
  {http://adsabs.harvard.edu/abs/2009ApJ...702..307S} {702, 307}

\bibitem[\protect\citeauthoryear{{Stewart}, {Kaufmann}, {Bullock}, {Barton},
  {Maller}, {Diemand}  \& {Wadsley}}{{Stewart} et~al.}{2011}]{Stewart2011}
{Stewart} K.~R.,  {Kaufmann} T.,  {Bullock} J.~S.,  {Barton} E.~J.,  {Maller}
  A.~H.,  {Diemand} J.,   {Wadsley} J.,  2011, \mn@doi [\apj]
  {10.1088/0004-637X/738/1/39}, \href
  {http://adsabs.harvard.edu/abs/2011ApJ...738...39S} {738, 39}

\bibitem[\protect\citeauthoryear{{Stewart}, {Brooks}, {Bullock}, {Maller},
  {Diemand}, {Wadsley}  \& {Moustakas}}{{Stewart} et~al.}{2013}]{Stewart2013}
{Stewart} K.~R.,  {Brooks} A.~M.,  {Bullock} J.~S.,  {Maller} A.~H.,  {Diemand}
  J.,  {Wadsley} J.,   {Moustakas} L.~A.,  2013, \mn@doi [\apj]
  {10.1088/0004-637X/769/1/74}, \href
  {http://adsabs.harvard.edu/abs/2013ApJ...769...74S} {769, 74}

\bibitem[\protect\citeauthoryear{{Stewart} et~al.,}{{Stewart}
  et~al.}{2017}]{Stewart2017}
{Stewart} K.~R.,  et~al., 2017, \mn@doi [\apj] {10.3847/1538-4357/aa6dff},
  \href {http://adsabs.harvard.edu/abs/2017ApJ...843...47S} {843, 47}

\bibitem[\protect\citeauthoryear{{Stinson} et~al.,}{{Stinson}
  et~al.}{2012}]{Stinson2012}
{Stinson} G.~S.,  et~al., 2012, \mn@doi [\mnras]
  {10.1111/j.1365-2966.2012.21522.x}, \href
  {http://adsabs.harvard.edu/abs/2012MNRAS.425.1270S} {425, 1270}

\bibitem[\protect\citeauthoryear{{Su}, {Hopkins}, {Hayward}, {Faucher-Giguere},
  {Keres}, {Ma}  \& {Robles}}{{Su} et~al.}{2016}]{Su2016}
{Su} K.-Y.,  {Hopkins} P.~F.,  {Hayward} C.~C.,  {Faucher-Giguere} C.-A.,
  {Keres} D.,  {Ma} X.,   {Robles} V.~H.,  2016, preprint, \href
  {http://adsabs.harvard.edu/abs/2016arXiv160705274S} {} (\mn@eprint {arXiv}
  {1607.05274})

\bibitem[\protect\citeauthoryear{{Teklu}, {Remus}, {Dolag}, {Beck}, {Burkert},
  {Schmidt}, {Schulze}  \& {Steinborn}}{{Teklu} et~al.}{2015}]{Teklu2015}
{Teklu} A.~F.,  {Remus} R.-S.,  {Dolag} K.,  {Beck} A.~M.,  {Burkert} A.,
  {Schmidt} A.~S.,  {Schulze} F.,   {Steinborn} L.~K.,  2015, \mn@doi [\apj]
  {10.1088/0004-637X/812/1/29}, \href
  {http://adsabs.harvard.edu/abs/2015ApJ...812...29T} {812, 29}

\bibitem[\protect\citeauthoryear{{Toomre}}{{Toomre}}{1964}]{Toomre1964}
{Toomre} A.,  1964, \mn@doi [\apj] {10.1086/147861}, \href
  {http://adsabs.harvard.edu/abs/1964ApJ...139.1217T} {139, 1217}

\bibitem[\protect\citeauthoryear{{Toomre} \& {Toomre}}{{Toomre} \&
  {Toomre}}{1972}]{Toomre1972}
{Toomre} A.,  {Toomre} J.,  1972, \mn@doi [\apj] {10.1086/151823}, \href
  {http://adsabs.harvard.edu/abs/1972ApJ...178..623T} {178, 623}

\bibitem[\protect\citeauthoryear{{Vogelsberger} et~al.,}{{Vogelsberger}
  et~al.}{2014a}]{Illustris2}
{Vogelsberger} M.,  et~al., 2014a, \mn@doi [\mnras] {10.1093/mnras/stu1536},
  \href {http://adsabs.harvard.edu/abs/2014MNRAS.444.1518V} {444, 1518}

\bibitem[\protect\citeauthoryear{{Vogelsberger} et~al.,}{{Vogelsberger}
  et~al.}{2014b}]{Illustris1}
{Vogelsberger} M.,  et~al., 2014b, \mn@doi [\nat] {10.1038/nature13316}, \href
  {http://adsabs.harvard.edu/abs/2014Natur.509..177V} {509, 177}

\bibitem[\protect\citeauthoryear{{Wang}, {Dutton}, {Stinson}, {Macci{\`o}},
  {Penzo}, {Kang}, {Keller}  \& {Wadsley}}{{Wang} et~al.}{2015}]{Wang2015}
{Wang} L.,  {Dutton} A.~A.,  {Stinson} G.~S.,  {Macci{\`o}} A.~V.,  {Penzo} C.,
   {Kang} X.,  {Keller} B.~W.,   {Wadsley} J.,  2015, \mn@doi [\mnras]
  {10.1093/mnras/stv1937}, \href
  {http://adsabs.harvard.edu/abs/2015MNRAS.454...83W} {454, 83}

\bibitem[\protect\citeauthoryear{Wang, Koribalski, Serra, van~der Hulst,
  Roychowdhury, Kamphuis  \& N.~Chengalur}{Wang et~al.}{2016}]{Wang2016}
Wang J.,  Koribalski B.~S.,  Serra P.,  van~der Hulst T.,  Roychowdhury S.,
  Kamphuis P.,   N.~Chengalur J.,  2016, \mn@doi [Monthly Notices of the Royal
  Astronomical Society] {10.1093/mnras/stw1099}, 460, 2143

\bibitem[\protect\citeauthoryear{{Weinberger} et~al.,}{{Weinberger}
  et~al.}{2017}]{IllustrisTNG2}
{Weinberger} R.,  et~al., 2017, \mn@doi [\mnras] {10.1093/mnras/stw2944}, \href
  {http://adsabs.harvard.edu/abs/2017MNRAS.465.3291W} {465, 3291}

\bibitem[\protect\citeauthoryear{{Wetzel}, {Hopkins}, {Kim},
  {Faucher-Gigu{\`e}re}, {Kere{\v s}}  \& {Quataert}}{{Wetzel}
  et~al.}{2016}]{Wetzel2016}
{Wetzel} A.~R.,  {Hopkins} P.~F.,  {Kim} J.-h.,  {Faucher-Gigu{\`e}re} C.-A.,
  {Kere{\v s}} D.,   {Quataert} E.,  2016, \mn@doi [\apjl]
  {10.3847/2041-8205/827/2/L23}, \href
  {http://adsabs.harvard.edu/abs/2016ApJ...827L..23W} {827, L23}

\bibitem[\protect\citeauthoryear{{Wheeler} et~al.,}{{Wheeler}
  et~al.}{2017}]{Wheeler2017}
{Wheeler} C.,  et~al., 2017, \mn@doi [\mnras] {10.1093/mnras/stw2583}, \href
  {http://adsabs.harvard.edu/abs/2017MNRAS.465.2420W} {465, 2420}

\bibitem[\protect\citeauthoryear{{White}}{{White}}{1984}]{White1984}
{White} S.~D.~M.,  1984, \mn@doi [\apj] {10.1086/162573}, \href
  {http://adsabs.harvard.edu/abs/1984ApJ...286...38W} {286, 38}

\bibitem[\protect\citeauthoryear{{White} \& {Rees}}{{White} \&
  {Rees}}{1978}]{White78}
{White} S.~D.~M.,  {Rees} M.~J.,  1978, \mnras, \href
  {http://adsabs.harvard.edu/abs/1978MNRAS.183..341W} {183, 341}

\bibitem[\protect\citeauthoryear{{Woo} et~al.,}{{Woo} et~al.}{2013}]{Woo2013}
{Woo} J.,  et~al., 2013, \mn@doi [\mnras] {10.1093/mnras/sts274}, \href
  {http://adsabs.harvard.edu/abs/2013MNRAS.428.3306W} {428, 3306}

\bibitem[\protect\citeauthoryear{{Woo}, {Dekel}, {Faber}  \& {Koo}}{{Woo}
  et~al.}{2015}]{Woo2015}
{Woo} J.,  {Dekel} A.,  {Faber} S.~M.,   {Koo} D.~C.,  2015, \mn@doi [\mnras]
  {10.1093/mnras/stu2755}, \href
  {http://adsabs.harvard.edu/abs/2015MNRAS.448..237W} {448, 237}

\bibitem[\protect\citeauthoryear{{Woo}, {Carollo}, {Faber}, {Dekel}  \&
  {Tacchella}}{{Woo} et~al.}{2017}]{Woo2017}
{Woo} J.,  {Carollo} C.~M.,  {Faber} S.~M.,  {Dekel} A.,   {Tacchella} S.,
  2017, \mn@doi [\mnras] {10.1093/mnras/stw2403}, \href
  {http://adsabs.harvard.edu/abs/2017MNRAS.464.1077W} {464, 1077}

\bibitem[\protect\citeauthoryear{{Wuyts}, {Cox}, {Hayward}, {Franx},
  {Hernquist}, {Hopkins}, {Jonsson}  \& {van Dokkum}}{{Wuyts}
  et~al.}{2010}]{Wuyts2010}
{Wuyts} S.,  {Cox} T.~J.,  {Hayward} C.~C.,  {Franx} M.,  {Hernquist} L.,
  {Hopkins} P.~F.,  {Jonsson} P.,   {van Dokkum} P.~G.,  2010, \mn@doi [\apj]
  {10.1088/0004-637X/722/2/1666}, \href
  {http://adsabs.harvard.edu/abs/2010ApJ...722.1666W} {722, 1666}

\bibitem[\protect\citeauthoryear{{Zavala} et~al.,}{{Zavala}
  et~al.}{2016}]{Zavala2016}
{Zavala} J.,  et~al., 2016, \mn@doi [\mnras] {10.1093/mnras/stw1286}, \href
  {http://adsabs.harvard.edu/abs/2016MNRAS.460.4466Z} {460, 4466}

\bibitem[\protect\citeauthoryear{{Zhu} et~al.,}{{Zhu} et~al.}{2018}]{Zhu2018}
{Zhu} L.,  et~al., 2018, \mn@doi [\mnras] {10.1093/mnras/stx2409}, \href
  {http://adsabs.harvard.edu/abs/2018MNRAS.473.3000Z} {473, 3000}

\bibitem[\protect\citeauthoryear{{Zolotov} et~al.,}{{Zolotov}
  et~al.}{2015}]{Zolotov2015}
{Zolotov} A.,  et~al., 2015, \mn@doi [\mnras] {10.1093/mnras/stv740}, \href
  {http://adsabs.harvard.edu/abs/2015MNRAS.450.2327Z} {450, 2327}

\bibitem[\protect\citeauthoryear{{van Dokkum}}{{van
  Dokkum}}{2005}]{vanDokkum2005}
{van Dokkum} P.~G.,  2005, \mn@doi [\aj] {10.1086/497593}, \href
  {http://adsabs.harvard.edu/abs/2005AJ....130.2647V} {130, 2647}

\bibitem[\protect\citeauthoryear{{van Dokkum} et~al.,}{{van Dokkum}
  et~al.}{2010}]{vanDokkum2010}
{van Dokkum} P.~G.,  et~al., 2010, \mn@doi [\apj]
  {10.1088/0004-637X/709/2/1018}, \href
  {http://adsabs.harvard.edu/abs/2010ApJ...709.1018V} {709, 1018}

\bibitem[\protect\citeauthoryear{{van de Voort}, {Davis}, {Kere{\v s}},
  {Quataert}, {Faucher-Gigu{\`e}re}  \& {Hopkins}}{{van de Voort}
  et~al.}{2015}]{vandeVoort2015}
{van de Voort} F.,  {Davis} T.~A.,  {Kere{\v s}} D.,  {Quataert} E.,
  {Faucher-Gigu{\`e}re} C.-A.,   {Hopkins} P.~F.,  2015, \mn@doi [\mnras]
  {10.1093/mnras/stv1217}, \href
  {http://adsabs.harvard.edu/abs/2015MNRAS.451.3269V} {451, 3269}

\bibitem[\protect\citeauthoryear{van~der Walt, Colbert  \& Varoquaux}{van~der
  Walt et~al.}{2011}]{numpy}
van~der Walt S.,  Colbert S.~C.,   Varoquaux G.,  2011, \mn@doi [Computing in
  Science Engineering] {10.1109/MCSE.2011.37}, 13, 22

\makeatother
\end{thebibliography}

\appendix
\section{Robustness to resolution}
\label{sec:resolution}
\begin{figure}
\centering
\includegraphics[width=\columnwidth]{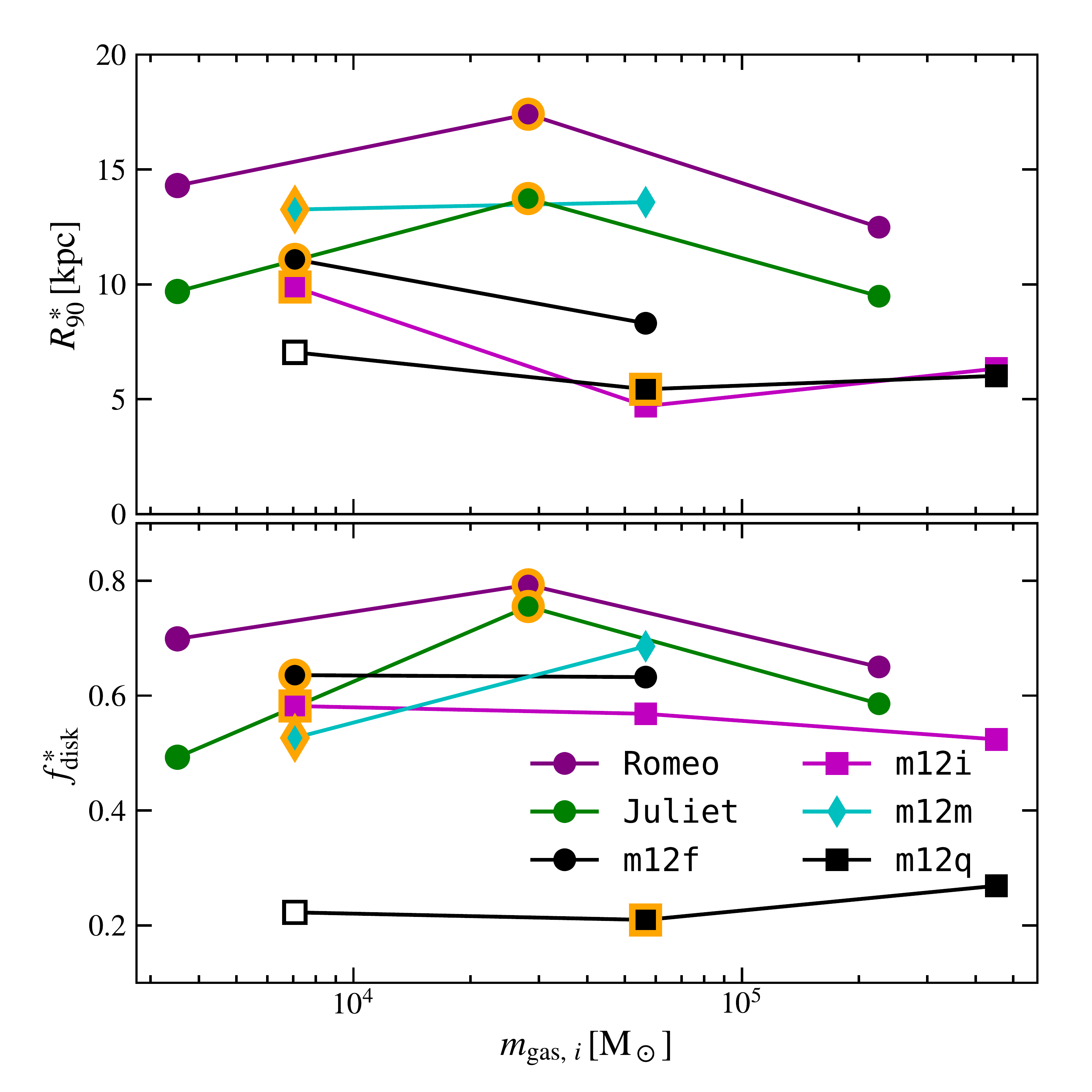}
\vspace{-1.5em}
\caption{$\Rstar$ (top) and $\fsdisk$ (bottom) for several of the 
galaxies as a function of the numerical resolution of the simulation, 
quantified here by the initial gas particle mass.  Though the exact 
values vary somewhat as a function of resolution, the relative 
morphologies of the galaxies remain rather consistent; any variations 
are relatively small (bulgy galaxies remain bulgy, and disky galaxies
remain disky), and would not change our conclusions. The open point of 
\run{m12q} is taken from an incomplete run at $z\sim1.5$ and is primarily 
shown for illustrative purposes.  The orange-rimmed points indicate the 
runs analyzed in the main body.  The higher resolution \run{Romeo} 
\run{\&} \run{Juliet} finished while the manuscript was in the late 
stages of preparation, and a full analysis (particularly as a function of 
time) is still ongoing.}
\label{fig:resolution}
\end{figure}

\renewcommand\thefigure{B\arabic{figure}}    
\setcounter{figure}{0}  
\begin{figure*}
\includegraphics[width=\textwidth]{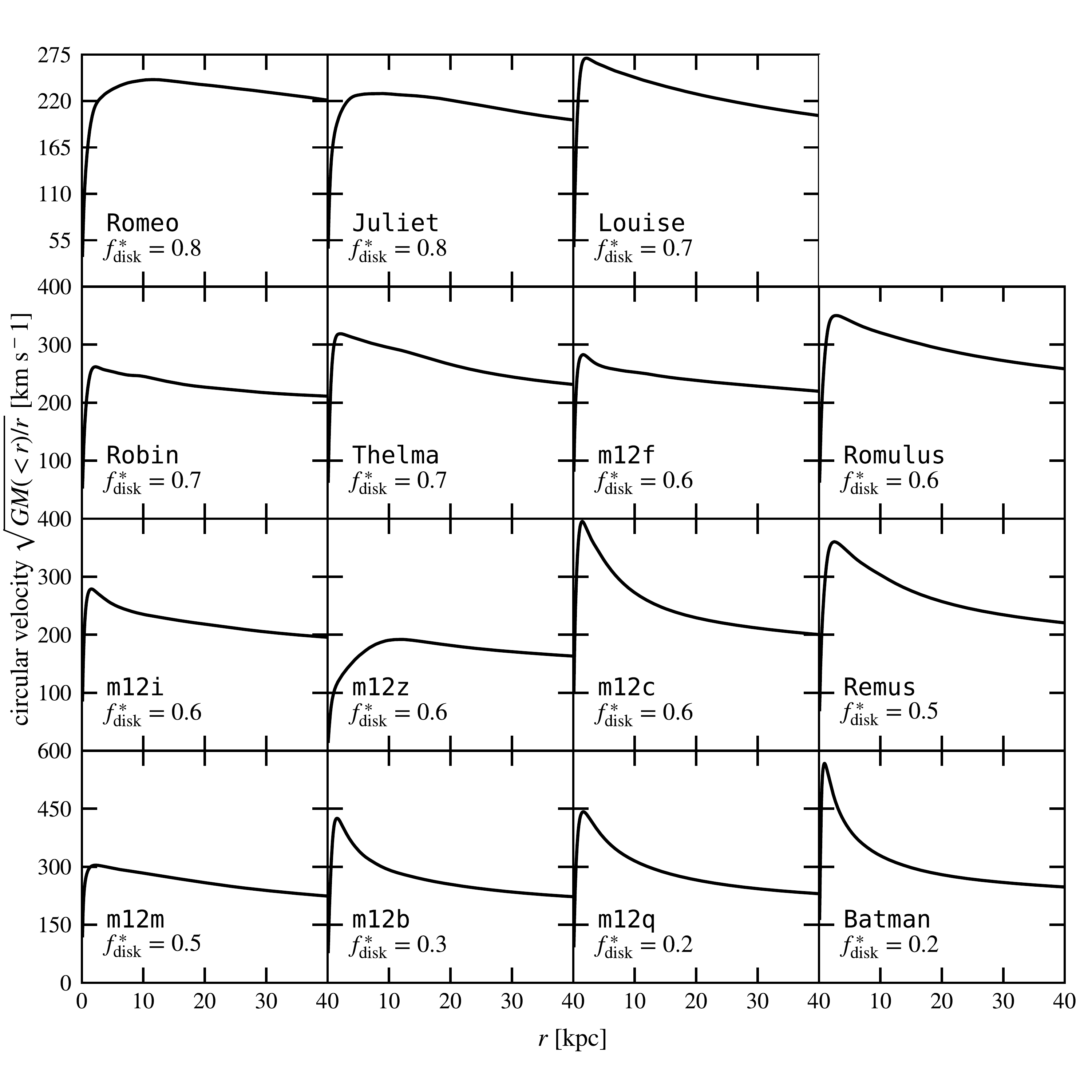}
\vspace{-3em}
\caption{The circular velocity curves of our galaxies, again sorted 
by decreasing $\fsdisk$.  The bulge-dominated galaxies are also 
generally the most concentrated, and therefore tend to be more 
centrally peaked.  The diskiest galaxies, however, have smoothly 
rising rotation curves and peak at $\lesssim250~\kms$ at 
5--10~$\kpc$, as does \run{m12z}, our lowest mass host.  Note that
the scales of the $y$ axes vary from row to row to capture the 
behavior of the more centrally concentrated galaxies.}
\label{fig:vccurves}
\end{figure*}

\renewcommand\thefigure{C\arabic{figure}}    
\setcounter{figure}{0}  
\begin{figure*}
\includegraphics[width=\textwidth]{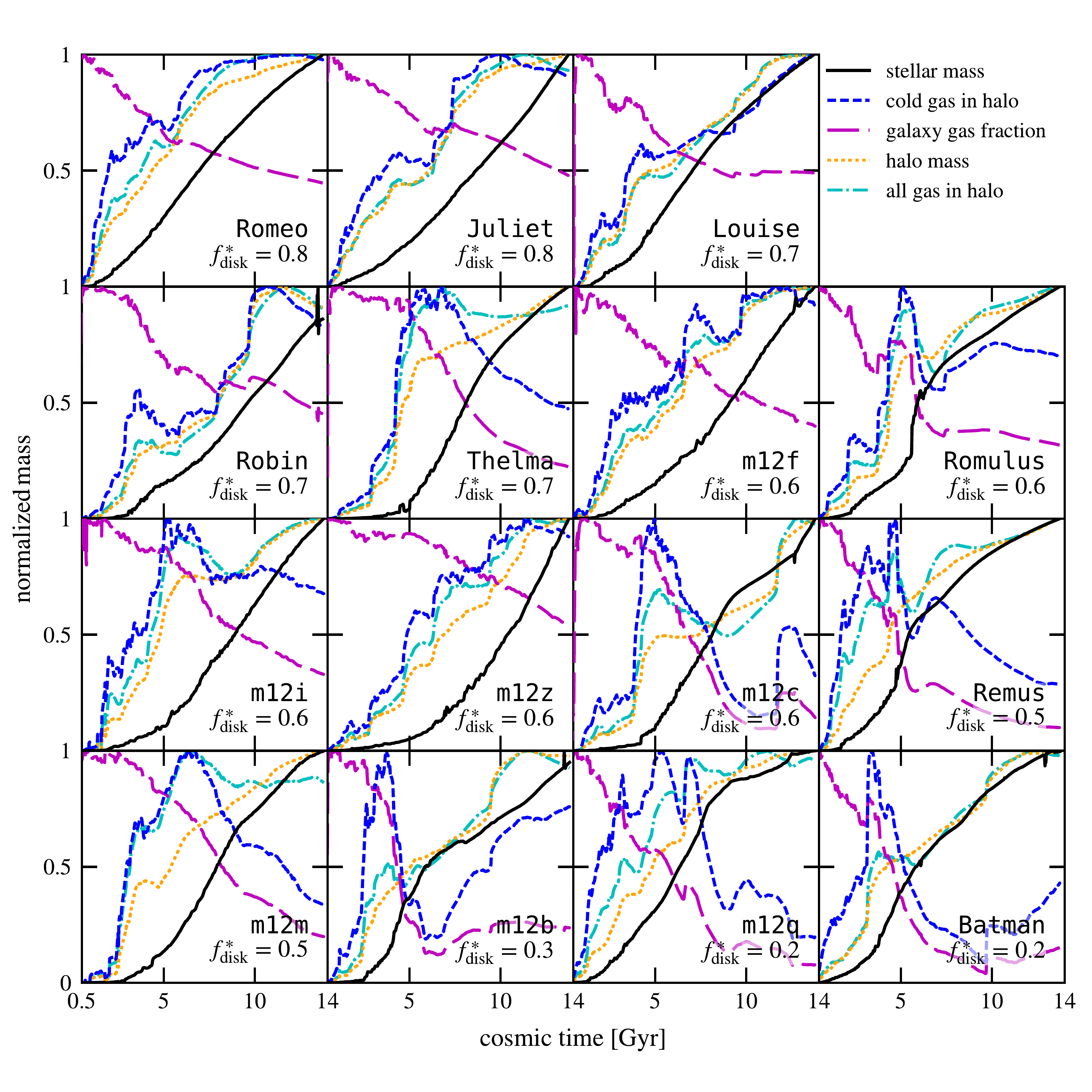}
\vspace{-3em}
\caption{Identical to Figure~\ref{fig:growth_hist}, but showing our
full sample.}
\label{fig:growth_hist_all}
\label{lastpage}
\end{figure*}

As with any numerical work, it is important to establish that our results 
are not driven by the resolution of the simulations, which we quantify 
here by the initial baryonic particle mass $m_{\mathrm{gas},\,i}$.  
As we demonstrate here, the morphological parameters of a single galaxy 
are relatively stable to changes in the resolution, indicating that the 
morphology is driven by fundamental properties of either the galaxy or 
halo, not the resolution at which that galaxy is simulated.  

Figure~\ref{fig:resolution} plots $\Rstar$ and $\fsdisk$ explicitly as 
a function of the resolution of the simulation for several of our galaxies 
that have been simulated at a variety of resolutions.  Lines connect runs 
of the same galaxy at different resolutions; the runs analyzed in the
main text are highlighted in orange.  Though there are changes with 
resolution (in particular, galaxies typically tend to be marginally more 
radially extended and diskier at higher resolution), the relative ordering 
of the simulations stays generally constant and the morphological 
parameters are generally resolved.  Moreover, our results indicate that
the FIRE-2 physics can yield roughly flat stellar rotation curves even
with initial gas particle masses $m_{i,\,\mathrm{gas}} = 57,000\msun$
(\run{Robin}).

The largest outlier from this trend is \run{Juliet}.  At the resolution we 
analyze here, \run{Juliet} experiences an extended prograde encounter with 
a gas-rich satellite that both torques up the existing disk and deposits
substantial fuel for additional disky star formation.  Though analysis is
still ongoing, preliminary investigations of the high resolution version 
of this simulation suggest that the satellite punches directly through the
main galaxy at $z\sim0.2$, partially disrupting the galaxy and in line with
the trends that we discuss above. 
We plot an incomplete run of \run{m12q} from $z\sim1.5$ as the open point; 
because the disk that \run{m12q} does build (Figure~\ref{fig:birth_v_now})
does not appear until $z\sim1$ (after which is it quickly destroyed), the 
higher resolution version of this galaxy is remarkably similar to the 
fiducial version, even at this time.

Therefore, we assert that the relative morphologies of several systems may 
be fairly compared, even if the exact values vary slightly as a function of 
resolution.  That is, there are fundamental, underlying properties of a given 
host that determine the morphology of that galaxy.  We demonstrated above that 
these properties are most strongly related to the formation histories of 
galaxies and their spin at high redshift.

\section{Rotation curves}
\label{sec:rotcurves}
Figure~\ref{fig:vccurves} plots the circular velocity curves of our 
galaxies, defined as $\sqrt{GM(<r)/r}$.  Generally speaking, galaxies 
with higher kinematic disk fractions also have flatter, less 
centrally-peaked rotation curves.  The exceptions are \run{m12z}, 
the least massive galaxy in our sample, and \run{m12m}, which recently 
underwent a bar-buckling event as noted in the main text. 


\section{Growth histories}
\label{sec:allgrowth}
Figure~\ref{fig:growth_hist} plots the growth of the halo mass, total 
gas mass, cold gas mass in the halo, stellar mass, and gas fractions 
for three representative galaxies, \run{Juliet}, \run{m12i}, and 
\run{Batman}.  For completeness, we plot the entirety of our sample in 
Figure~\ref{fig:growth_hist_all}.  The trends that we identify regarding 
the relative amount of star-forming gas using the three representative 
galaxies in Figure~\ref{fig:growth_hist} are robust -- at higher disk 
fractions, the galaxies almost uniformly reach their peak $\mcoldgas$ 
at later times and typically maintain that level of star forming gas 
for several gigayears.

\bsp	
\end{document}